\documentclass{aa}  
\bibpunct{(}{)}{;}{a}{}{,}
\usepackage{graphicx,natbib,url,twoopt}
\usepackage[T1]{fontenc}
\usepackage[labelfont=bf,font=footnotesize]{caption}
\usepackage{subcaption}
\usepackage[]{overpic}
\usepackage{makecell}
\usepackage{txfonts}
\usepackage{colortbl}
\usepackage[table,xcdraw]{xcolor}
\usepackage{amsmath} 
\usepackage{booktabs}
\usepackage{lipsum}
\usepackage{multicol}
\usepackage{float}
\usepackage{soul}
\def\logg{\hbox{$\log{g}$}}         
\def\rhk{\hbox{$\log{(R^{'}_{HK})}$}}         
\def\ha{\hbox{$H_{\alpha}$}}         
\def\ms{\hbox{\,m\,s$^{-1}$}}         
\def\cms{\hbox{\,cm\,s$^{-1}$}}       
\def\m2s2{\hbox{\,m$^{2}$\,s$^{-2}$}} 
\def\Msun{\hbox{$\mathrm{M}_{\odot}$}}             

\def\Mearth{\hbox{$\mathrm{M}_{\oplus}$}}             

\def\teff{\hbox{$T_{\rm eff}$}}
\def\feh{\hbox{[Fe/H]}}

\usepackage{natbib,twoopt}
\usepackage{hyperref} 
\hypersetup{breaklinks=true}
\bibpunct{(}{)}{;}{a}{}{,}             
\makeatletter
  \newcommandtwoopt{\citeads}[3][][]{\href{http://adsabs.harvard.edu/abs/#3}%
    {\def\hyper@linkstart##1##2{}%
     \let\hyper@linkend\@empty\citealp[#1][#2]{#3}}}
  \newcommandtwoopt{\citepads}[3][][]{\href{http://adsabs.harvard.edu/abs/#3}%
    {\def\hyper@linkstart##1##2{}%
     \let\hyper@linkend\@empty\citep[#1][#2]{#3}}}
  \newcommandtwoopt{\citetads}[3][][]{\href{http://adsabs.harvard.edu/abs/#3}%
    {\def\hyper@linkstart##1##2{}%
     \let\hyper@linkend\@empty\citet[#1][#2]{#3}}}
  \newcommandtwoopt{\citeyearads}[3][][]%
    {\href{http://adsabs.harvard.edu/abs/#3}
    {\def\hyper@linkstart##1##2{}%
     \let\hyper@linkend\@empty\citeyear[#1][#2]{#3}}}
\makeatother

\begin{document}

\title{The HARPS search for southern extra-solar planets\thanks{Based on observations made with HARPS spectrograph on the 3.6-m ESO telescope at La Silla Observatory, Chile.}}

\subtitle{XLVI: 12 super-Earths around the solar type stars HD39194, HD93385, HD96700, HD154088, and HD189567}


\author{N. Unger\inst{\ref{i:geneva}}
\and D. S\'egransan\inst{\ref{i:geneva}}
\and D. Queloz\inst{\ref{i:geneva},\ref{i:cambridge}}
\and S. Udry\inst{\ref{i:geneva}}
\and C. Lovis\inst{\ref{i:geneva}}
\and C. Mordasini\inst{\ref{i:bern}}
\and E. Ahrer\inst{\ref{i:warwick},\ref{i:habi-warwick}}
\and W. Benz\inst{\ref{i:bern}}
\and F. Bouchy\inst{\ref{i:geneva}}
\and J.-B. Delisle\inst{\ref{i:geneva}}
\and R. F. Díaz\inst{\ref{i:icas}}
\and X. Dumusque\inst{\ref{i:geneva}}
\and G. Lo Curto\inst{\ref{i:eso}}
\and M. Marmier\inst{\ref{i:geneva}}
\and M. Mayor\inst{\ref{i:geneva}}
\and F. Pepe\inst{\ref{i:geneva}}
\and N. C. Santos\inst{\ref{i:porto}, \ref{i:porto2}}
\and M. Stalport\inst{\ref{i:geneva}}
\and R. Alonso\inst{\ref{i:tenerife}}
\and A. Collier Cameron \inst{\ref{i:standrews}}
\and M. Deleuil \inst{\ref{i:marseille}}
\and P. Figueira \inst{\ref{i:porto}}
\and M. Gillon \inst{\ref{i:liege}}
\and C. Moutou \inst{\ref{i:toulouse}}
\and D. Pollacco \inst{\ref{i:warwick},\ref{i:habi-warwick}}
\and E. Pompei \inst{\ref{i:eso}}
}
          

\institute{D\'epartement d'Astronomie, Universit\'e de Gen\`eve, Chemin Pegasi 51, CH-1290 Versoix, Suisse \label{i:geneva}
\and Astrophysics Group, Cavendish Laboratory, JJ Thomson Avenue, CB3 0HE Cambridge, UK\label{i:cambridge}
\and Physikalisches Institut Universität Bern, Sidlerstrasse 5, 3012, Bern, Switzerland \label{i:bern}
\and Department of Physics, University of Warwick, Gibbet Hill Road, CV4 7AL Coventry, UK \label{i:warwick}
\and Centre for Exoplanets and Habitability, University of Warwick, Gibbet Hill Road, CV4 7AL Coventry, UK \label{i:habi-warwick}
\and International Center for Advanced Studies (ICAS) and ICIFI (CONICET), ECyT-UNSAM, Campus Miguelete, 25 de Mayo y Francia, (1650) Buenos Aires, Argentina \label{i:icas}
\and European Southern Observatory, Alonso de Cordova 3107, Vitacura, Santiago, Chile \label{i:eso}
\and Instituto de Astrofísica e Ciências do Espaço, Universidade do Porto, CAUP, Rua das Estrelas, 4150-762 Porto, Portugal \label{i:porto}
\and Departamento de Física e Astronomia, Faculdade de Cièncias, Universidade do Porto, Rua do Campo Alegre, 4169-007 Porto, Portugal \label{i:porto2}
\and Instituto de Astrofísica de Canarias, 38200 La Laguna, Tenerife, Spain \label{i:tenerife}
\and School of Physics and Astronomy, University of St Andrews, North Haugh, St Andrews, Fife KY16 9SS, UK \label{i:standrews}
\and Aix Marseille Univ, CNRS, CNES, LAM, Marseille, France \label{i:marseille}
\and Astrobiology Research Unit, Université de Liège, Allée du 6 Août 19C, B-4000 Liège, Belgium \label{i:liege}
\and Univ. de Toulouse, CNRS, IRAP, 14 avenue Belin, 31400 Toulouse, France \label{i:toulouse}
}

 \date{Received; accepted}

 
\abstract
{
We present precise radial-velocity measurements of five solar-type stars observed with the HARPS Echelle spectrograph mounted on the 3.6-m telescope in La Silla (ESO, Chile). With a time span of more than 10 years and a fairly dense sampling, the survey is sensitive to low mass planets down to super-Earths on orbital periods up to 100 days.}
{Our goal was to search for planetary companions around the stars HD\,39194, HD\,93385, HD\,96700, HD\,154088, and HD\,189567 and use Bayesian model comparison to make an informed choice on the number of planets present in the systems based on the radial velocity observations. These findings will contribute to the pool of known exoplanets and better constrain their orbital parameters.}
%
{
A first analysis was performed using the DACE (Data \& Analysis Center for Exoplanets)  online tools to assess the activity level of the star and the potential planetary content of each system. We then used Bayesian model comparison on all targets to get a robust estimate on the number of planets per star. We did this using the nested sampling algorithm \textsc{PolyChord}. For some targets, we also compared different noise models to disentangle planetary signatures from stellar activity. Lastly, we ran an efficient MCMC (Markov chain Monte Carlo) algorithm for each target to get reliable estimates for the planets' orbital parameters.}
{We identify 12 planets within several multiplanet systems. These planets are all in the super-Earth and sub-Neptune mass regime with minimum masses ranging between 4 and 13 \Mearth\ and orbital periods between 5 and 103 days. Three of these planets are new, namely HD\,93385 b, HD\,96700 c, and HD\,189567 c.}
{}

\keywords{ stars: planetary systems -- stars: individual: HD39194, HD93385, HD96700, HD154088, HD189567 -- technique: spectroscopy -- technique: radial velocities}

\maketitle
%

\section{Introduction}

The radial-velocity (RV) planet search survey using the High Accuracy Radial velocity Planet Searcher (HARPS) spectrograph installed on the ESO-3.6m telescope at La Silla, Chile (\citeads{2000SPIE.4008..582P}; \citeads{2003Msngr.114...20M}) has contributed to the detection of hundreds of exoplanets, most of them being Super-Earths and hot Neptunes (see detections for solar-type stars in e.g., \citeads{2011arXiv1109.2497M}; \citeads{2011A&A...528A.112L}; \citeads{2013A&A...551A..59L}; \citeads{2016A&A...585A.134D}; \citeads{2019A&A...622A..37U}; \citeads{2021MNRAS.503.1248A}; for M-dwarfs in e.g., \citeads{2013A&A...549A.109B}, \citeyearads{2013A&A...556A.110B}; \citeads{2015A&A...575A.119A}, \citeyearads{2017A&A...602A..88A}; or \citeads{2009A&A...496..513M}). 

The original sample of HARPS targets are a subsample of the historical CORALIE \citepads{2000A&A...356..590U} volume-limited sample and they have been selected to maximize the detectability of low-mass (down to a few Earth masses) exoplanets around solar-type stars (from late-F to late-K dwarfs). For example, only stars with chromospheric activity indexes lower than $\rhk = -4.75$ and low-rotation rates were selected. These targets were observed for 6 years between 2003 and 2009 as part of the Guaranteed Time Observations (GTO) (PI: M. Mayor) and later on continued as ESO Large Programs. \footnote{GTO program 072.C-0488 and Large programs 082.C-0842, 091.C-0936, 183.C-0972, 192.C-0852, and 198.C-0836}

Using the high-precision sample from HARPS together with more than 10 years of data from CORALIE available at the time, \citetads{2011arXiv1109.2497M} found that around 50\% of solar-type stars contain at least one planetary companion with a period <100 days and mass <30\Mearth. In addition, over 70\% of these are multiplanet systems. The \textit{Kepler} mission (\citeads{2010Sci...327..977B}; \citeads{2016RPPh...79c6901B}) has since discovered thousands of new planets which corroborate \citepads{2012ApJS..201...15H} and refine \citepads{2019AJ....158..109H} the statistic found by \citetads{2011arXiv1109.2497M}.

Spectrographs such as HARPS have reached a precision where planets of a few Earth masses can be detected, but these usually have low amplitude radial velocity signatures (0.5 - 0.7 \ms). Retrieving these signals is even more challenging because of stellar noise. Activity on the surface of the star can produce RV signals at various timescales \citepads{2016A&A...593A...5D}. Stellar granulation and pulsations occur on short timescales of up to several hours, while starspots and plages occur on timescales on the order of the rotation of the star, and magnetic activity cycles can last several years. For quiet stars, this intrinsic noise reaches levels of about 1 \ms, comparable to the RV semi-amplitude of the planets we want to find. Many of these sources of noise are also correlated in time, which introduces additional challenges to the modeling of the RVs.

Another challenge in the detection and characterization of planetary signals is the irregular sampling of the measurements. This prevented us from using a purely data-driven approach and instead we had to make assumptions about the properties of the noise, which can be correlated in time. There are efforts to obtain well sampled radial velocities probing a wider range of frequencies. For instance, \citetads{2011A&A...534A..58P} managed to find planets with RV semi-amplitudes as low as $\sim\!$~0.5 \ms, or the future HARPS3 spectrograph \citepads{2016SPIE.9908E..6FT} which will improve on the observing techniques used in \citetads{2011A&A...534A..58P}.

The development of new and better instruments is also important to improve precision. Instruments such as ESPRESSO (\citeads{2010SPIE.7735E..0FP}, \citeyearads{2014AN....335....8P}, \citeyearads{2021A&A...645A..96P}), EXPRES \citepads{2016SPIE.9908E..6TJ}, or NEID \citepads{2016SPIE.9908E..7HS} are designed to have great stability and reach a precision of $\sim\!\!$~15~\cms\ or lower. This level of precision enables the possibility for a better characterization of the stellar noise in quiet stars and detect Earth-like planets.

To analyze these noisy RV datasets, we used advanced statistical tools. Markov Chain Monte Carlo (MCMC) techniques have been used for decades now (e.g., \citeads{2005AJ....129.1706F}, \citeyearads{2006ApJ...642..505F}) to estimate the posterior probabilities of each parameter in a model. This works very well for parameter estimation, but to determine the best model we use Bayesian model comparison, which has proven to be a useful tool for model selection (e.g., \citeads{2014MNRAS.437.3540F}; \citeads{2016A&A...585A.134D}; \citeads{2016A&A...588A..31F}, \citeyearads{2020A&A...635A..13F}). 

Bayes theorem provides direct relative probabilities between different models to decide which one is the best. To do this, we need to calculate the so-called Bayesian evidence, which is no easy task since this involves the calculation of a high dimensional integral. The challenges of Bayesian model comparison were clearly shown by \citetads{2020AJ....159...73N}, where they found a high variance in results from different algorithms in models containing three or more planets. To overcome this difficulty, we use \textsc{PolyChord} \citepads{2015MNRAS.453.4384H}, a state-of-the-art nested sampling algorithm designed to perform well in high dimensional problems. We tested \textsc{PolyChord} on the simulated datasets of \citetads{2020AJ....159...73N} and found its results closely matching those of the other nested samplers. In future works we aim to also use other techniques such as Bayesian model averaging, which was recently implemented as a detection criterion for exoplanet detection with radial velocities by \citetads{2021arXiv210506995H}.


In this article, we present the data analysis and the orbital solutions of five planetary systems announced in \citetads{2011arXiv1109.2497M}, namely: HD\,39194, HD\,93385, HD\,96700, HD\,154088, and HD\,189567. With more than five additional years of data, we were able to further constrain these planetary systems and confirm three new planets that were not found at the time.

This paper is organized as follows. In Sect. \ref{sec:stellar-characteristics}, we present how the HARPS data is obtained and the stellar host characteristics; in Sect. \ref{sec:model-methods}, we describe the methods and models used for the data analysis; in Sect. \ref{sec:model-comparison}, we introduce the principles of Bayesian model comparison and \textsc{PolyChord}; in Sect. \ref{sec:data-analysis}, we go through the analysis of each star and present the results; in Sect. \ref{sec:discussion} we discuss possible formation and evolution paths for these planetary systems, and finally in Sect. \ref{sec:conclusions} we conclude the article.


\section{HARPS data and stellar characteristics} \label{sec:stellar-characteristics}

\begin{table*}[h]
\centering
\caption{Characteristics of the host stars. $^\dagger$ 95\% variability}
\label{tab:stellar-characteristics}
\begin{tabular}{@{}lrccccc@{}}
\toprule
Parameters & Units & HD\,39194     & HD\,93385     & HD\,96700     & HD\,154088    & HD\,189567     \\ \midrule
Spectral Types        &           & K0V         & G2/G3V      & G0V         & K0IV-V      & G3V          \\
$B$       & {[}mag]   & 8.841        & 8.08        & 7.11        & 7.411        & 6.71         \\
$V$       & {[}mag]   & 8.075        & 7.486       & 6.50        & 6.584        & 6.07         \\ 
$B-V$     & {[}mag]   & 0.766        & 0.594       & 0.61        & 0.827       & 0.64         \\ 
$M_V$     & {[}mag]   & 5.964        & 4.30        & 4.476       & 5.27        & 4.75         \\ 
$G$       & {[}mag]   & 7.852        & 7.338       & 6.352       & 6.363       & 5.899        \\
$\pi$     & {[}mas]   & $37.831\pm0.038$    & $23.04\pm0.03$    & $39.324\pm0.044$    & $57.71\pm0.05$  & $55.81\pm0.04$\\
$L$       & {[}L$_{\odot}$]         & 0.39        &  1.61       & 1.38        & 0.74        & 2.11          \\
\teff     & {[}K]     & $5205\pm23$        & $5977\pm18$        & $5845\pm13$        & $5374\pm43$        & $5726\pm15$         \\
\logg     & {[}cgs]   & $4.53\pm0.05$        & $4.42\pm0.02$        & $4.39\pm0.02$        & $4.37\pm0.07$        & $4.41\pm0.01$         \\
\feh      & {[}dex]   & $-0.61\pm0.02$       & $+0.02\pm0.01$       & $-0.18\pm0.01$       & $+0.28\pm0.03$       & $-0.24\pm0.01$        \\
\rhk  &           & -4.95 (0.04)$^\dagger$ & -4.99 (0.02)$^\dagger$ & -4.95 (0.02)$^\dagger$ & -5.05 (0.06)$^\dagger$ & -4.92 (0.02)$^\dagger$ \\
$M$   & {[}\Msun] & $0.67\pm0.04$           & $1.04\pm0.01$          & $0.89\pm0.01$          & $0.91\pm0.02$          & $0.83\pm0.01$         \\
age       & {[}Gyr]   & $10\pm3$ & $3.3\pm1.3$ & $8.6\pm0.7$ & $8\pm2$ & $11.0\pm0.5$ \\ \bottomrule
\end{tabular}
\end{table*}


Radial velocities have been obtained with the HARPS high-resolution spectrograph installed on the 3.6m ESO telescope at La Silla Observatory \citepads{2003Msngr.114...20M}. Every night calibrations are performed with a ThAr lamp and simultaneous calibrations are obtained during observation with a second fiber. From 2003 to 2013 these simultaneous calibrations were done with the same ThAr lamp and later from 2013 onward with a Fabry–Pérot interferometer. The data reduction is done on site with the latest HARPS data reduction software, which extracts the spectra, calibrates it and obtains a cross correlation function (CCF) with a stellar template. Other stellar parameters are also derived from the HARPS CCF like the Full Width Half Maximum (FWHM) and the Bisector Inverse Slope (BIS) (\citeads{2000A&A...359L..13Q}, \citeyearads{2001A&A...379..279Q}; \citeads{2002A&A...388..632P}; \citeads{2007A&A...468.1115L}; \citeads{2009A&A...507..487M}, \citeyearads{2009A&A...493..639M}).


The bulk physical characteristic of stars with planets  presented in this paper are summarized in Table \ref{tab:stellar-characteristics}. The spectroscopic atmospheric parameters \teff, \feh\ and  \logg\  are derived from the work of \citetads{2008A&A...487..373S}. The magnitudes and colors are extracted from the \textit{Tycho-2} catalog \citepads{2000A&A...355L..27H}, except for HD~189567 where the photometry comes from Vizier \citepads{2002yCat.2237....0D}. G band photometry and parallaxes are extracted from the \textit{Gaia} Data Release 2 \citepads{2018A&A...616A...1G}. The absolute magnitude $M_V$ is derived from these quantities.  Stellar masses and ages are estimated using PARAM-1.5 (\citeads{2006A&A...458..609D}; \citeads{2014MNRAS.445.2758R}; \citeyearads{2017MNRAS.467.1433R}) - an online tool\footnote{\url{http://stev.oapd.inaf.it/cgi-bin/param}} that performs a Bayesian interpolation within the stellar evolutionary grid produced by the Padova and Trieste Stellar evolution code \citepads{2012MNRAS.427..127B}. However, the uncertainties on the stellar masses are probably underestimated.

The chromospheric activity index \rhk\ has been measured on the HARPS spectra according to methods described in \citetads{2000A&A...361..265S} and implemented on HARPS by \citetads{2011arXiv1107.5325L}. For each star, the mean chromospheric activity \rhk\ is listed together with the measurement of its intrinsic scattering (in parenthesis) expressed as the 95\% variability.


\section{Model and methodology} \label{sec:model-methods}

For the data analysis, we used the RV module of the Data \& Analysis Center for Exoplanets (DACE) web-platform \footnote{\url{https://dace.unige.ch/radialVelocities/}}, which provides tools for data visualization and analysis. The formalism of the data analysis implemented in DACE is described in \citetads{2016A&A...590A.134D} and Ségransan et al. (in revision). We first performed a quick analysis with DACE to obtain a preliminary model for each target and then used Bayesian model comparison to confirm how many planets there are in the system (see Sect. \ref{sec:model-comparison}).

The RV model we used in this paper consists of five parts: a systemic velocity offset, $\gamma_i$; a polynomial drift, $\mathrm{drift}(t)$; the effect of long term magnetic cycle, $\mathrm{mag}(t)$; the Keplerian curves, $\mathrm{kep}(t)$; and a Gaussian white noise, $\epsilon$. At any given $k$-th radial velocity data point our model predicts:

\begin{equation}
    v(t_k) = \gamma_i + \mathrm{mag}(t_k) + \mathrm{drift}(t_k) + \mathrm{kep}(t_k) + \epsilon_k \enspace .
\end{equation}


\subsection{Systemic radial velocity}

The systemic radial velocity offset $\gamma_i$ is independent for each available instrument $i$. Even though we only used HARPS data in this paper, HARPS underwent a fiber change in May 2015 \citepads{2015Msngr.162....9L}, which introduced an offset in the radial velocity measurements. It was observed that this offset is not constant for all stars and depends on the width of the cross-correlation function (CCF) \citepads{2015Msngr.162....9L}, the wider the CCF the higher the offset in the RV before and after the fiber change. Because of this RV offset we consider data from these two periods as taken from separate instruments, which we denote as H03 and H15, respectively.


\subsection{Drifts and magnetic cycles}

First, we looked at the periodogram \citepads{2008MNRAS.385.1279B} of the RV data in search of any long term signal (>1000 day period). To model these long term effects we used a polynomial drift of up to order three:

\begin{equation} \label{eq:drift}
    \mathrm{drift}(t) = \alpha_1 \,\tau + \alpha_2 \,\tau^2 + \alpha_2 \,\tau^3 \enspace ,
\end{equation}
where $\alpha_1$, $\alpha_2$ and $\alpha_3$ are the linear, quadratic and cubic terms respectively. We represented the time $\tau$ in years to avoid $\alpha_i$ being too large. 

We also used the detrending technique described in \citetads{2018A&A...614A.133D}, where we used the time series of one of the activity indicators (e.g., FWHM or \rhk) as the model of these long term signals. This time series can be smoothed out by either using a Gaussian or an Epachenikov kernel \citep{epanechnikov69} to only keep the low (or high) frequency components. The specific kernel and timescale we used is indicated in the corresponding results table for each target. Then to make sure that the proportionality factor is in units of \ms\, we re-scaled and centered the data to have zero mean and a semi-amplitude of one. The magnetic cycle model is represented in the following way:

\begin{equation} \label{eq:activity}
    \mathrm{mag}(t) = A \cdot \mathrm{smoothed\;activity} \enspace ,
\end{equation}
where $A$ is the proportionality factor between the activity and the radial velocity of the star.

We looked for a correlation between the time series of the RV and the time series of some activity indicators, namely the FWHM, BIS, \ha, and \rhk. If we saw any correlation, or a periodic signal that is present in both datasets, then we used the technique described above to remove that signal from the RV data.


\subsection{Keplerians} \label{sec:keplerians}

We then started an iterative process of searching for the most significant peaks in the periodogram of the radial velocities time-series. We fitted a Keplerian curve (see Eq. \ref{eq:keplerians}) at the period of the most significant peak. We considered a peak to be significant if the false alarm probability (FAP) is below 1\% \citepads{2008MNRAS.385.1279B}. The process was then repeated using the periodogram of the residuals of the previous fit. We did this until there are no more significant periodic signals in the residuals of the radial velocity data. The Keplerian model is given by:

\begin{equation} \label{eq:keplerians}
    \mathrm{kep}(t) = \sum_{j=0}^N K_j \left[\cos \left(\nu_j(t) + \omega_j \right) + e_j \cos (\omega_j) \right] \enspace .
\end{equation}

For each planet $j$, we have: $K_j$ the semi-amplitude of the periodic function, $\nu_j$ the true anomaly, $\omega_j$ the argument of periastron, and $e_j$ the eccentricity. The calculation of the true anomaly ($\nu$) requires the orbital period $P_j$, the mean longitude $\lambda_{0j}$ at a given reference epoch, and the eccentricity $e_j$. To do this calculation, one has to solve the Kepler equation, which is a transcendental equation.





\subsection{Noise}

We added a Gaussian white noise term (also called jitter) to each data point. This jitter term ($\sigma_{\mathrm{J}_i}$) was added quadratically to the uncertainties provided by the HARPS data reduction software ($\sigma_k$):

\begin{equation}
    \sigma^2 = \sigma_k^2 + \sigma_{\mathrm{J}_i}^2 \enspace .
\end{equation}

This additional noise is calculated independently for each instrument $i$, which in our case are H03 and H15.

The planets we analyzed in this article have high enough amplitudes to bypass a full covariance modeling to mitigate the noise. These are mostly quiet stars and if there is no clear sign of stellar activity in the RV and/or indicators, training a Gaussian Process only on the RVs would result in absorbing the signal of one of the planets. Then even if the Gaussian Process is a better fit than the Keplerian, it does not necessarily mean that the signal is stellar activity. That is why we opted for fairly simple models in these high significance detections.


\section{Bayesian model comparison and \textsc{PolyChord}} \label{sec:model-comparison}

After this initial analysis with DACE (Sect. \ref{sec:model-methods}) and identifying all significant planetary signals using the FAP, we estimated the Bayesian evidence for all models with up to four planets. The a posteriori bayesian analysis shows evidence of at most three planets in these systems. We always analyzed models up to at least one more planet than we think is true for completeness and to confirm that there are indeed no more planets.

This analysis helps to validate or disprove the presence of the signals. One of the limitations of the iterative process described in Sect. \ref{sec:keplerians} is that the signals are fitted sequentially by removing one and looking for the next. This can have problems by falling into local maxima of the Likelihood while a better solution could be found by fitting all planets at the same time.


With the value of the evidence for each planet model we then compared these values using the Jeffrey's scale (see Table~\ref{tab:jeffscale}) to decide which model is preferred. Additional considerations may have been taken such as the convergence of the orbital parameters to decide on the best model (see Sect. \ref{sec:data-analysis}).

\begin{table}[h] 
    \centering
    \caption{Slightly modified version of the original Jeffreys scale \citepads{1939thpr.book.....J} for deciding how conclusive a model is over another when comparing their Bayesian evidence. The modification is only to make it easier to work with the logarithm of $\mathcal{O}$.}
    \label{tab:jeffscale}
    \begin{tabular}{ccc}
    \hline$\ln \mathcal{O}$ & Odds & Remark \\
    \hline$<1.0$ & $\lesssim 3: 1$ & Inconclusive \\
    1.0 & $\sim 3: 1$ & Weak evidence \\
    3.0 & $\sim 20: 1$ & Moderate evidence \\
    5.0 & $\sim 150: 1$ & Strong evidence \\
    >5.0 & $> 150: 1$ & Decisive \\
    \hline
    \end{tabular}
\end{table}

\begin{table*}[t]
\centering
\caption{List of priors used for each parameter. RV$_{\mathrm{min}}$ and RV$_{\mathrm{max}}$ are the minimum and maximum radial velocity measured, respectively, for each target. All keplerians added to the models have the same priors.}
\label{tab:priors}
\begin{tabular}{@{}lccl@{}}
\toprule
\textbf{Parameter}           & \textbf{Units}       & \textbf{Prior Distribution}                        & \textbf{Description}                       \\ \midrule
$K$                   & \ms         & Log Uniform [0.1, 20]           & Semi-amplitude                                \\
$P^{-1}$              & days$^{-1}$ & Uniform [10$^{-3}$, 1.5$^{-1}$] & Orbital frequency                             \\
$e$                   &             & Beta [0.867, 3.03]\tablefootmark{$\dagger$}  & Eccentricity                                  \\
$\omega$              & rad         & Uniform [0, $2\pi$]             & Argument of periastron                         \\
$\lambda_{0} = M_{0}+\omega$ & rad                  & Uniform [0, $2\pi$]                                & Mean longitude at a given reference epoch.  \\ \midrule
$\gamma$                     & \ms                  & Uniform [RV$_{\mathrm{min}}$, RV$_{\mathrm{max}}$] & Constant velocity offset                   \\
$\sigma_{\mathrm{J}}$ & \ms         & Uniform [0, 20]                 & Additional white noise (Jitter)               \\
$A$                   & \ms         & Uniform [-20, 20]               & Linear scaling factor with activity indicator \\
$\alpha_i$                   & m s$^{-1}$ yr$^{-i}$ & Uniform [-10, 10]                                  & \textit{i}-th term of the polynomial drift \\
Epoch                   & BJD & Fixed at 2\,455\,500                                  & Reference epoch \\\bottomrule
\end{tabular}
\tablefoot{\tablefoottext{$\dagger$}\ \!\!\citetads{2013MNRAS.434L..51K}
}
\end{table*}

\subsection{Bayesian inference} \label{sec:bayes-inference}

A detailed description of Bayesian inference is presented in Appendix \ref{app:bayes-inference}, but we introduce some key equations here.  The Bayesian evidence $\mathcal{Z}$ is defined as the weighted average of the Likelihood $\mathcal{L}(\theta)$ over the prior parameter space $\pi(\theta)$:

\begin{equation} \label{eq:evidence-main}
    \mathcal{Z} = \int \mathcal{L}(\theta) \pi(\theta) \; d\theta \enspace .
\end{equation}

This integral has as many dimensions as parameters in the model. That fact makes the calculation of the integral very challenging in high dimensional models. To compare two models $\mathcal{M}_1$ and $\mathcal{M}_2$, we define the odds ratio \citepads{2010blda.book.....G} which is a measure of evidence for (or against) one model over another. It takes the following form after applying the natural logarithm: 

\begin{equation} \label{eq:oddsratio}
    \ln{\mathcal{O}_{12}} = \ln{\frac{\mathcal{Z}_{1}\pi_{1}}{\mathcal{Z}_{2}\pi_{1}}} = \ln{\mathcal{Z}_{1}} - \ln{\mathcal{Z}_{2}} + \ln{\frac{\pi_{1}}{\pi_{2}}} \enspace .
\end{equation}

In Eq. \ref{eq:oddsratio}, $\pi_1$ and $\pi_2$ are the prior probabilities for each model. We compare models with different number of planets and we do not have any good physical reasons to assume a specific distribution for the number of planets orbiting a star. If we were to use the current known distribution on the number of planets in multiplanet systems, we would be heavily influenced by our observational and instrumental biases. That is why we use an uninformative uniform prior, or $\pi_i = \pi_j$, which reduces Eq. \ref{eq:oddsratio} to the difference of $\ln{\mathcal{Z}}$ between both models.

The log odds ratio then simply becomes the difference of the $\ln \mathcal{Z}$ values. We then compare this value with Table \ref{tab:jeffscale} to decide on the best model.


\subsection{\textsc{PolyChord}} \label{sec:polychord}


As we saw in the previous section, the Bayesian evidence is a high dimensional integral which makes its computation very difficult and therefore sophisticated algorithms had to be developed.

Because of this difficulty, one may wish to avoid this integral by using approximations such as the Bayesian Information Criterion or Chib's approximation \citep{chib01}. But these are only rough estimates and do not work well for multimodal posterior distributions. So more advanced techniques are needed for a robust calculation of the Bayesian evidence. Specifically Nested Sampling \citepads{2004AIPC..735..395S} has shown to be a reliable technique \citepads[see][]{2020AJ....159...73N} and several implementations exist. For example, diffusive nested sampling, DNEST (\citeads{2009arXiv0912.2380B}, \citeyearads{2016arXiv160603757B}); multimodal nested sampling, MULTINEST (\citeads{2009MNRAS.398.1601F}, \citeyearads{2011ascl.soft09006F}); or \textsc{PolyChord} \citepads{2015MNRAS.453.4384H}, which we used in this work. More recent implementations came out like JAXNS \citepads{2020arXiv201215286A} that aims to improve performance using the JAX framework and UltraNest \citepads{2021JOSS....6.3001B} with a focus on the correctness of the evidence estimation. We aim to test and use these implementations in future works.

\textsc{PolyChord} is a next-generation nested sampling algorithm that calculates the Bayesian evidence (Eq.~\ref{eq:evidence-main}) and was developed specifically to work well with high dimensional models, that is models with a lot ($>\!20$) of free parameters. We implemented \textsc{PolyChord} using the Python wrapper provided by the developers. More technical notes about \textsc{PolyChord} and the tuning parameter we used can be found in Appendix \ref{app:polychord}.


\subsection{Priors} \label{sec:priors}

A complete list with the priors we used for all \textsc{PolyChord} runs is presented in Table~\ref{tab:priors}. We explain the reasoning for some of them.

In Eq. \ref{eq:keplerians} we showed that the five main parameters for a Keplerian curve are the semi-amplitude $K$, orbital period $P$, eccentricity $e$, argument of periastron $\omega$ and the mean longitude at a reference epoch $\lambda_0$. These last two, $\omega$ and $\lambda_0$, define the orientation of the orbit which is arbitrary and thus we set the prior uniform for all angles. 

For the eccentricity we used a distribution derived by \citetads{2013MNRAS.434L..51K}. They find that the distribution of eccentricities can be described well by a Beta distribution with parameters $a=0.867$ and $b=3.03$. This result is based on nearly 400 exoplanets found using the RV technique. We repeated this analysis including more than 300 new exoplanets discovered since that study (709 in total) and found that the Beta distribution derived by \citeads{2013MNRAS.434L..51K} still holds remarkably well (within the original uncertainties).

For the orbital frequency we chose a uniform prior from $10^{-3}$ to $1.5^{-1}$ day$^{-1}$, or equivalently from 1.5 to 1000 days for the orbital period. This choice is based on the fact that we expect the width of the posterior distribution to be equal when plotted with the frequency instead of the period. We did not consider periods shorter than 1.5 days to avoid aliases around 1 day. In Sect. \ref{sec:data-analysis} we indicate all the instances where aliases close to 1 day appear in the periodogram. By avoiding these periods we obtain a better convergence of the Nested Sampling runs.

In the periodograms we do not see any long period signals ($\gtrsim\!\!1000$ days) that could be caused by planets. Extending the parameter space greatly increases the computational intensity of nested sampling, especially for the period since the Likelihood function is highly multimodal along the period. So we decided to cut the period of the keplerians at 1000 days to save on computational time.


\subsection{MCMC}

Once we selected the best model from the Bayesian model comparison analysis, we ran an efficient MCMC algorithm (\citeads{2014MNRAS.441..983D}; \citeyearads{2016A&A...590A.134D}) to obtain posterior distributions for each parameter. Even though \textsc{PolyChord} also provides samples from the posterior, an MCMC algorithm is more efficient and optimized for this purpose. We performed all MCMC runs with 2\,000\,000 iterations and discarded the first 500\,000 as the burn-in phase.


\section{Data Analysis} \label{sec:data-analysis}


\subsection{HD\,39194}

HD\,39194 is an early K type star located at 26.4 pc. It is a chromospherically quiet star with the \rhk\ showing a low peak to peak amplitude ($\lesssim$ 0.1). The \rhk\ also shows a long term drift but no significant short term variability. However, the activity indicator \ha\ shows a periodic feature at ~34.5 days, which may be caused by the rotation period of the star.

Since 2003, HARPS has gathered 273 nightly binned radial velocities for HD\,39194. In the periodogram of the radial velocity, we can immediately see a very significant periodic signal at 14 days. After fitting a Keplerian at this period and looking at the periodogram of the residuals, we can see another significant periodic signal at 5.6 days. Peaks close to 1 day are also visible, but these are the one day aliases of the 14 and 5.6 day signals. Repeating this process once more, we find another signal at 33.8 days. Fig. \ref{fig:HD39194&HD93385-FAP} shows the procedure of iteratively fitting a Keplerian at the most significant peak of the periodogram and repeating the process for the residuals. 

\begin{figure}[h] 
    \centering
    \includegraphics[width=\columnwidth]{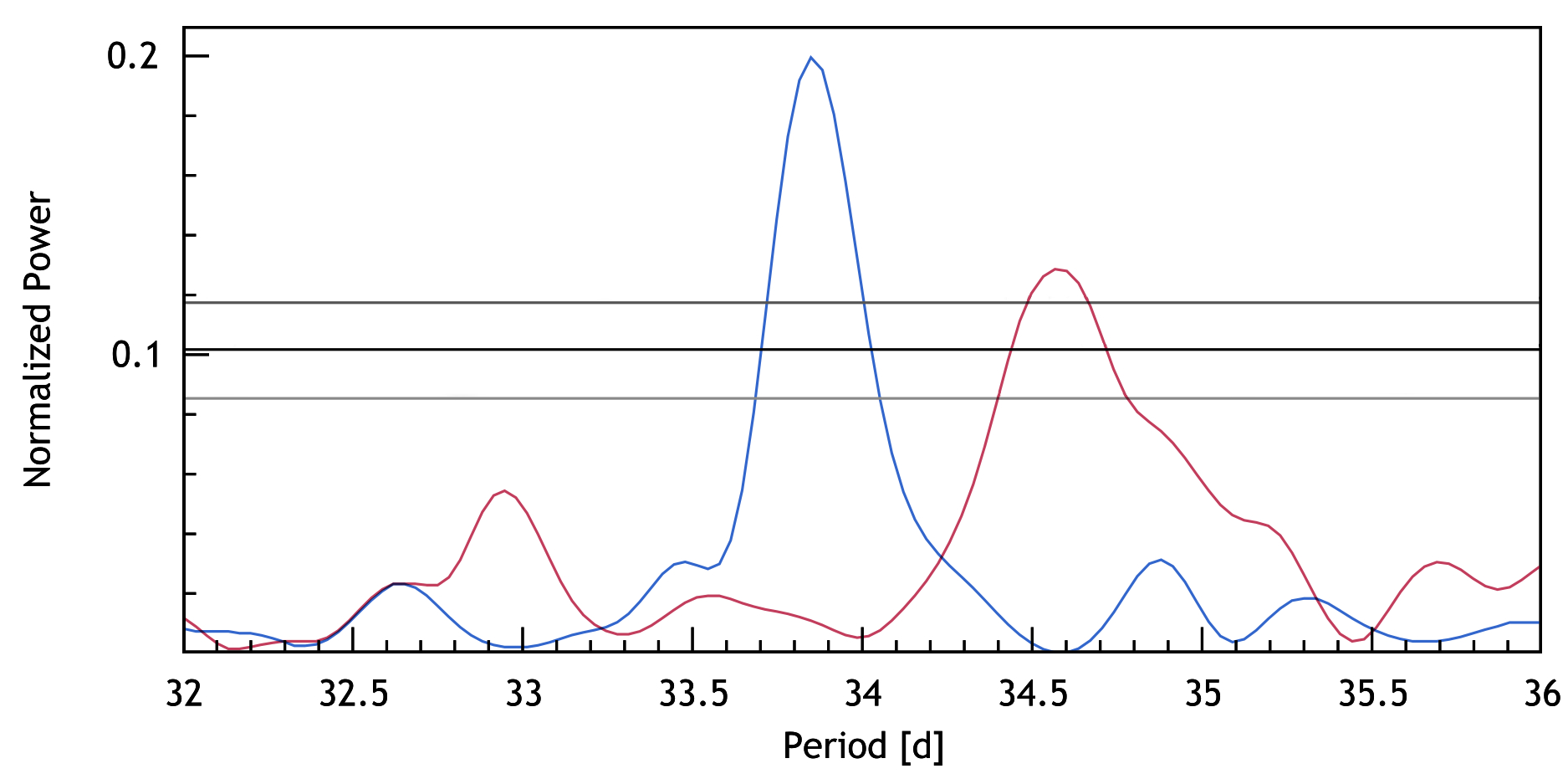}
    \caption{Periodogram of the activity indicator \ha\ (in red), after removing a long term drift of $\sim$~8000 days, together with the periodogram of the residuals of the RV time-series (in blue) after removing the 14 and 5 day signals. We can see that both peaks are independent. False alarm probability (FAP) thresholds are shown as horizontal lines for FAP=$10\%, 1\%$ and $0.1\%$.}
    \label{fig:HD39194-Halpha}
\end{figure}

This last signal is very close to the 34.5-day signal we found in the \ha. However, both signals are independent in frequency space, so they are unlikely to stem from the same origin. In Fig. \ref{fig:HD39194-Halpha} both periodograms are shown on top of each other to visualize the difference.

After fitting these three Keplerians, we can see two additional peaks at 370 and 2000 days. The former may be caused by the stitching effect where periods close to one year can appear due to a wavelength calibration error introduced by the stitching of the CCD blocks in the HARPS camera (see \citeads{2015ApJ...808..171D}; \citeads{2019A&A...622A..37U}; \citeads{2019A&A...629A..27C}). We are using data that was corrected for this using the technique from \citetads{2015ApJ...808..171D} which greatly reduced the power of this signal at 370 days, but after the correction there is still some residual power. Still, this signal gets significantly reduced when we add a linear parameter that scales with \rhk\ and smoothed using an Epanechnikov kernel at a timescale of 1.5 years, which makes us suspect that it may be activity related after all. 

For the 2000 day signal, we chose to remove it by fitting a third-order polynomial drift. A planetary origin for this signal is unlikely with the current data because a keplerian fit results in a high eccentricity ($\sim\!0.9$) and only one complete period is observed. So many more observations would be needed to confirm this as a planet.

The final model we chose for HD\,39194 consists of three Keplerians, with a linear parameter scaling with \rhk\ and a third-order polynomial drift. With this model, we proceeded to estimate the Bayesian evidence for models including from zero up to four planets. The results are presented in Table \ref{tab:odds_ratios}. 

\begin{figure*}
    \centering
    \begin{minipage}{\textwidth}
        \centering
        \begin{multicols}{3}
            \begin{overpic}[width=0.32\textwidth]{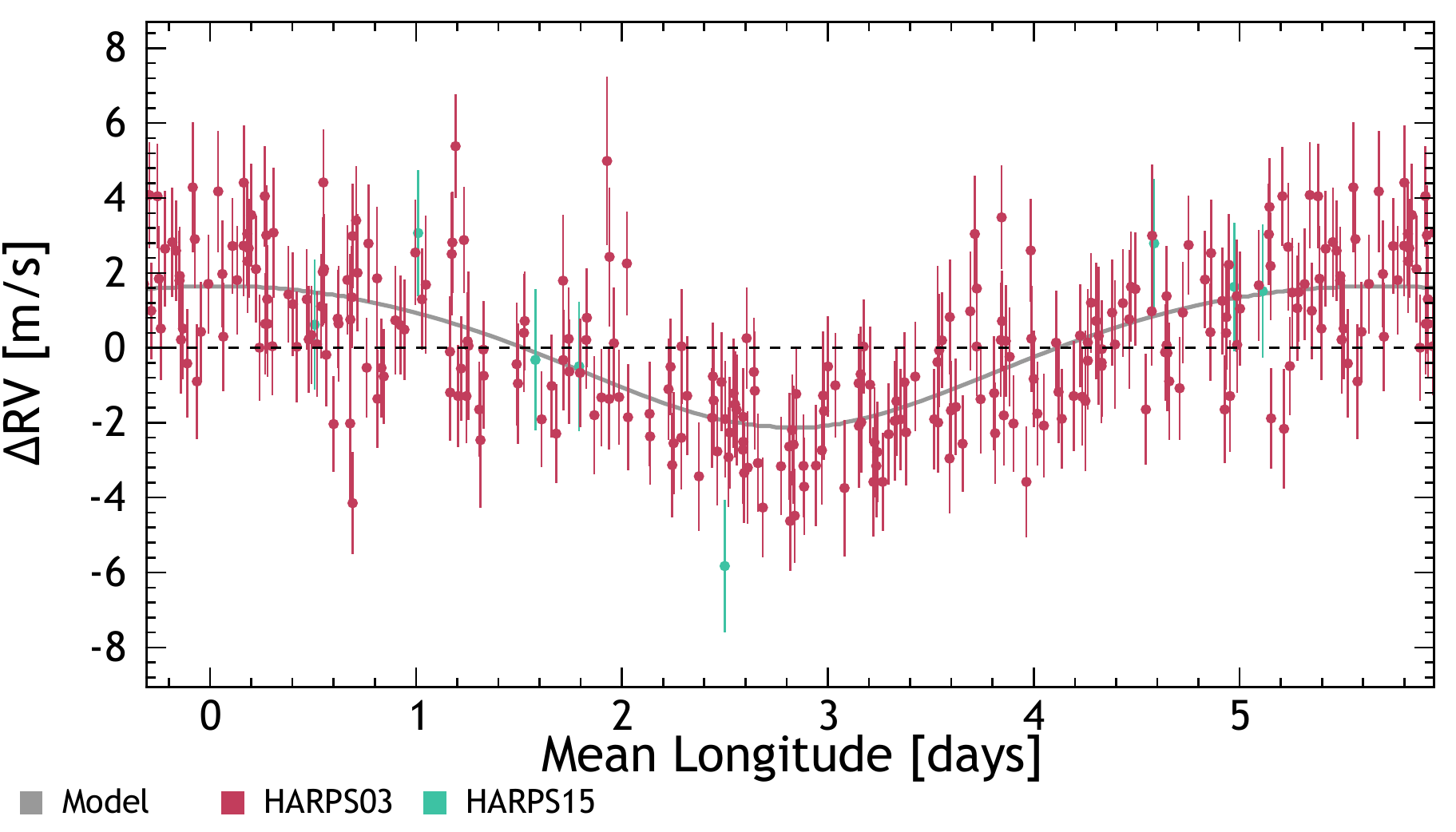}
            \end{overpic}
            \columnbreak
            
            \begin{overpic}[width=0.32\textwidth]{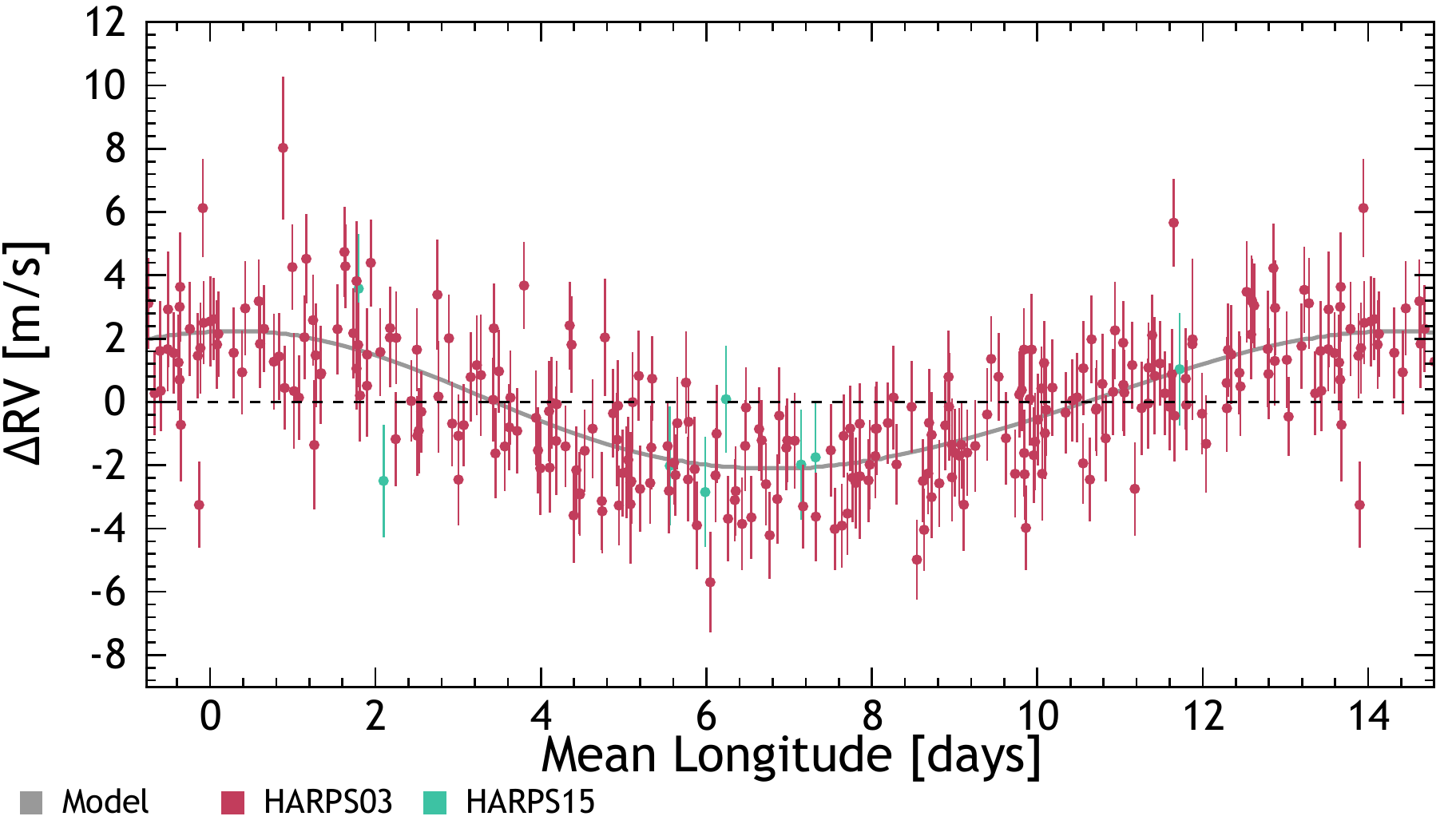}
            \end{overpic}
            \columnbreak
            
            \begin{overpic}[width=0.32\textwidth]{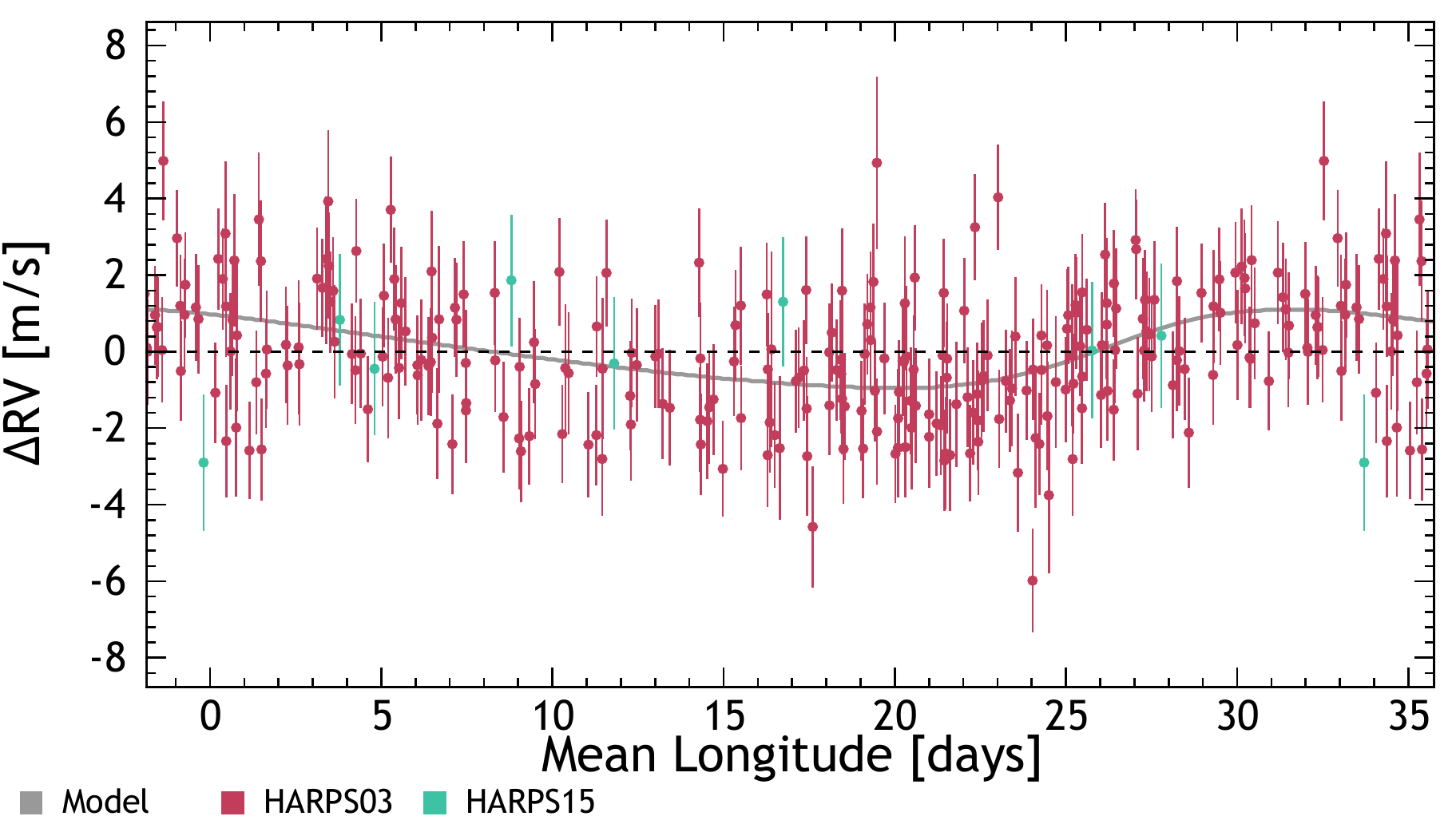}
            \end{overpic}
            
        \end{multicols}
    \end{minipage}
    \caption{Phase folded plots for each planet in HD\,39194. From left to right: planets b, c and d with orbital periods of 5.64, 14.03 and 33.9 days, respectively.}
    \label{fig:HD39194-phasefold}
\end{figure*}

We see a significant increase of evidence from the zero planet model up to the three planet model. Then, the four planet model has an advantage of $\ln(\mathcal{O}) = 2.0 \pm 1.8$ which is only weak evidence in its favor (see Table \ref{tab:jeffscale}). Furthermore, there is no clear period candidate for a fourth planet in the posterior. This leads us to conclude that HD\,39194 has three planets at periods of 5.63, 14.03, and 33.91 days with minimum masses of 4.0, 6.3, and 4.0 \Mearth\, respectively. The phase-folded RVs can be seen in Fig. \ref{fig:HD39194-phasefold}. 

\begin{figure*}
    \centering
    \includegraphics[width=0.7\textwidth]{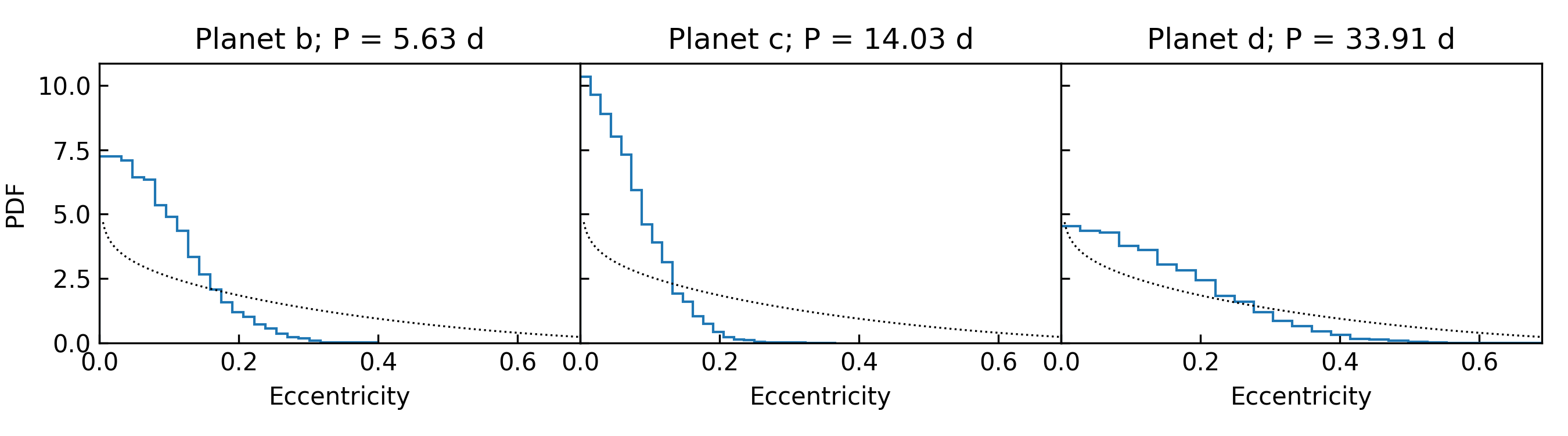}
    \caption{Posterior distribution of the eccentricities for the planetary companions of HD\,39194. The dotted line represents the eccentricity prior.}
    \label{fig:HD39194_eccs}
\end{figure*}

We then ran an MCMC using the same model described above for better parameter estimation of the orbital elements. The full results are presented in Table \ref{tab:HD39194-posterior}. The histogram of the MCMC samples for the eccentricities are shown in Fig. \ref{fig:HD39194_eccs}. All planets seem to have mostly circular orbits. Planets b and c have posteriors that are more concentrated than the prior distribution, indicating that the current data constrain the planet eccentricities beyond the prior level, but only mildly for planet d.

Compared to the planet parameters derived by \citeads{2011arXiv1109.2497M}, we find a reduced semi-amplitude for planet d from 1.49 to 1.04 \ms, which in turn also reduced its minimum mass from 5.13 to 4.0 \Mearth.

\begin{table*}[]
\centering
\caption{Relative difference of the logarithm of the evidence values for all five stars and each planet model. The odds ratios are all taken with respect to the model with the proposed number of planets for that star. Those cells are also colored in gray. The uncertainties on the odds ratios are calculated as the standard deviation of the evidence value of three identical \textsc{PolyChord} runs of each model.}
\label{tab:odds_ratios}
\begin{tabular}{@{}lccccc@{}}
\toprule
\textbf{Star} & \multicolumn{5}{c}{\textbf{Models}}                                                                                                                       \\ \midrule
 &
  \begin{tabular}[c]{@{}c@{}}0 planets\\ $\ln(\mathcal{O})$\end{tabular} &
  \begin{tabular}[c]{@{}c@{}}1 planet\\ $\ln(\mathcal{O})$\end{tabular} &
  \begin{tabular}[c]{@{}c@{}}2 planets\\ $\ln(\mathcal{O})$\end{tabular} &
  \begin{tabular}[c]{@{}c@{}}3 planets\\ $\ln(\mathcal{O})$\end{tabular} &
  \begin{tabular}[c]{@{}c@{}}4 planets\\ $\ln(\mathcal{O})$\end{tabular} \\ \midrule
HD\,39194         & $-99.8 \pm 0.3$  & $-66.1 \pm 1.2$                       & $-14.4 \pm 1.3$                       & \cellcolor[HTML]{DADADA}$0.0 \pm 2.3$ & $+2.0 \pm 1.8$ \\
HD\,93385         & $-61.2 \pm 0.1$  & $-41.3 \pm 0.6$                       & $-18.7 \pm 1.9$                       & \cellcolor[HTML]{DADADA}$0.0 \pm 1.5$ & $-0.5 \pm 2.3$ \\
HD\,96700         & $-130.5 \pm 0.2$ & $-59.0 \pm 1.6$                       & $-10.3 \pm 0.8$                       & \cellcolor[HTML]{DADADA}$0.0 \pm 0.9$ & $-1.0 \pm 2.8$ \\
HD\,154088        & $-18.6 \pm 0.2$  & \cellcolor[HTML]{DADADA}$0.0 \pm 0.6$ & $+0.1 \pm 0.2$                        & $+0.5 \pm 1.3$                        & $-1.3 \pm 0.9$ \\
HD\,189567        & $-74.3 \pm 0.3$  & $-26.2 \pm 0.9$                       & \cellcolor[HTML]{DADADA}$0.0 \pm 2.9$ & $+1.9 \pm 1.4$                        & $+1.7 \pm 2.7$ \\ \bottomrule
\end{tabular}
\end{table*}

\begin{table}[]
\centering
\caption{Posterior values of all parameters used for HD39194. We give the values of the mode and 1$\sigma$ confidence interval of the posterior distribution from the MCMC. $^{(\dagger)}$ The linear parameter was adjusted on \rhk\ smoothed with an Epanechnikov kernel with a 1.5 yr timescale. $^{(\ddagger)}$ Eccentricity does not differ significantly from zero; the 68\% and 95\% upper limits are reported. The argument of periastron $\omega$ is therefore unconstrained.}
\label{tab:HD39194-posterior}
\resizebox{\hsize}{!}{%
\begin{tabular}{@{}lcccc@{}}
\toprule
\multicolumn{5}{c}{\textbf{HD39194}}                                                                                                \\ \midrule
Parameter              & \multicolumn{1}{c}{Units}               & \multicolumn{1}{l}{} &                     &                     \\ \midrule
\multicolumn{5}{c}{\textit{Offset and drift}}                                                                                       \\ \midrule
$\gamma_{\rm H03}$ & \ms                                     & \multicolumn{3}{c}{14\,175.6$\pm$1.2}                              \\
$\gamma_{\rm H15}$ & \ms                                     & \multicolumn{3}{c}{14\,181.7$\pm$1.6}                              \\
$\alpha_1$             & m s$^{-1}$ yr$^{-1}$                    & \multicolumn{3}{c}{0.47$\pm$0.19}                                \\
$\alpha_2$             & m s$^{-1}$ yr$^{-2}$                    & \multicolumn{3}{c}{-0.05$\pm$0.04}                               \\
$\alpha_3$             & m s$^{-1}$ yr$^{-3}$                    & \multicolumn{3}{c}{0.013$\pm$0.005}                              \\
A$^{(\dagger)}$               & \ms         & \multicolumn{3}{c}{-4.1$\pm$1.6}                                 \\ \midrule
\multicolumn{5}{c}{\textit{Noise}}                                                                                                  \\ \midrule
$\sigma_{\rm H03}$ & \ms                                     & \multicolumn{3}{c}{1.22$\pm$0.08}                                \\
$\sigma_{\rm H15}$ & \ms                                     & \multicolumn{3}{c}{1.9$\pm$1.1}                                  \\
$\sigma_{(O-C)}$       & \ms                                     & \multicolumn{3}{c}{1.46}                                         \\ \midrule
\multicolumn{5}{c}{\textit{Keplerians}}                                                                                             \\ \midrule
                       &                                         & \textbf{HD39194\,b}  & \textbf{HD39194\,c} & \textbf{HD39194\,d} \\ \midrule
$P$                    & day                                     & 5.6368$\pm$0.0004    & 14.030$\pm$0.003    & 33.91$\pm$0.03      \\
$K$                    & \ms                                     & 1.86$\pm$0.13        & 2.19$\pm$0.14       & 1.04$\pm$0.14       \\
$e$                    &                & <0.105; <0.207 $^{(\ddagger)}$ & <0.078; <0.154 $^{(\ddagger)}$ & <0.174; <0.333 $^{(\ddagger)}$       \\
$\omega$               & deg                                     & - $^{(\ddagger)}$          & - $^{(\ddagger)}$          & - $^{(\ddagger)}$         \\ 
$\lambda_0$            & deg                                     & -62$\pm$4            & 113$\pm$4           & 144$\pm$8           \\ \midrule
$T_{\rm Periastron}$            & BJD                                     & 2\,455\,503.8$\pm$1.4    & 2\,455\,499$\pm$3       & 2\,455\,516$\pm$7       \\
$m\,\sin i$            & \Mearth                                 & 4.0$\pm$0.3          & 6.3$\pm$0.5         & 4.0$\pm$0.6         \\
$a$                    & AU                                      & 0.056$\pm$0.001    & 0.103$\pm$0.002     & 0.185$\pm$0.0033    \\ \midrule
Ref. Epoch                  & BJD                                & \multicolumn{3}{c}{2\,455\,500}                  \\ \bottomrule 
\end{tabular}}
\end{table}

\subsubsection*{Potential 5:2 mean motion resonance between planets b and c}

The two inner planets have a period ratio $P_c/P_b \sim 2.49$. This is close enough to $2.5$ for us to wonder if this inner pair lies in the 5:2 mean motion resonance (MMR).

In order to explore this possibility, we computed a chaoticity map in the neighborhood of the 5:2 MMR. As such, we defined $121\times121$ system's configurations spread on a grid of the initial eccentricity of planet c, $e_c$, and the period ratio of the inner pair, $P_c/P_b$. All the other parameters were initially fixed at their median value from the posterior of the MCMC exploration, and all the planetary orbits were inferred coplanar and aligned to the line-of-sight ($i$=90 deg). The period ratio is the main parameter that influences the proximity of the planets' pair to the 5:2 MMR. Other parameters such as $e_b$ and the inclinations influence as well the resonance width, and indirectly the proximity of the pair to the resonance. We opted to explore the eccentricity of planet c, $e_c$, as the second parameter because the resonance shape in the eccentricity - period ratio subspace has a well-defined V-shape. This choice is still somewhat arbitrary, and the influence of other parameters on the resonance could also be explored. Nevertheless, such an exhaustive resonance study is beyond the scope of this paper.

The orbital evolution of each configuration was numerically computed over 30 kyr with REBOUND\footnote{REBOUND is an open-source software package dedicated to N-body integrations: \url{http://rebound.readthedocs.org}}, making use of the adaptive time-step high order N-body integrator IAS15 (\citeads{2012A&A...537A.128R}; \citeads{2015MNRAS.446.1424R}). A correction for general relativity was considered via the library REBOUNDx\footnote{\url{https://reboundx.readthedocs.org}} \citepads{2020MNRAS.491.2885T}, following the developments of \citeads{1975ApJ...200..221A}. The level of chaos was then estimated with the NAFF fast chaos indicator (\citeads{1992PhyD...56..253L}; \citeyearads{1993PhyD...67..257L}), based on the diffusion of the planetary mean motions $n$. For each planet $i$, the NAFF basically computes the diffusion rate $\frac{\Delta n_i}{n_0}$, where $n_0$ is the initial mean motion of the considered planetary orbit, and $\Delta n_i$ is the difference of the averaged mean motions over the two halves of the integration. The maximal diffusion rate among the three planets was retained. 

\begin{figure}
    \centering
    \includegraphics[scale=0.7]{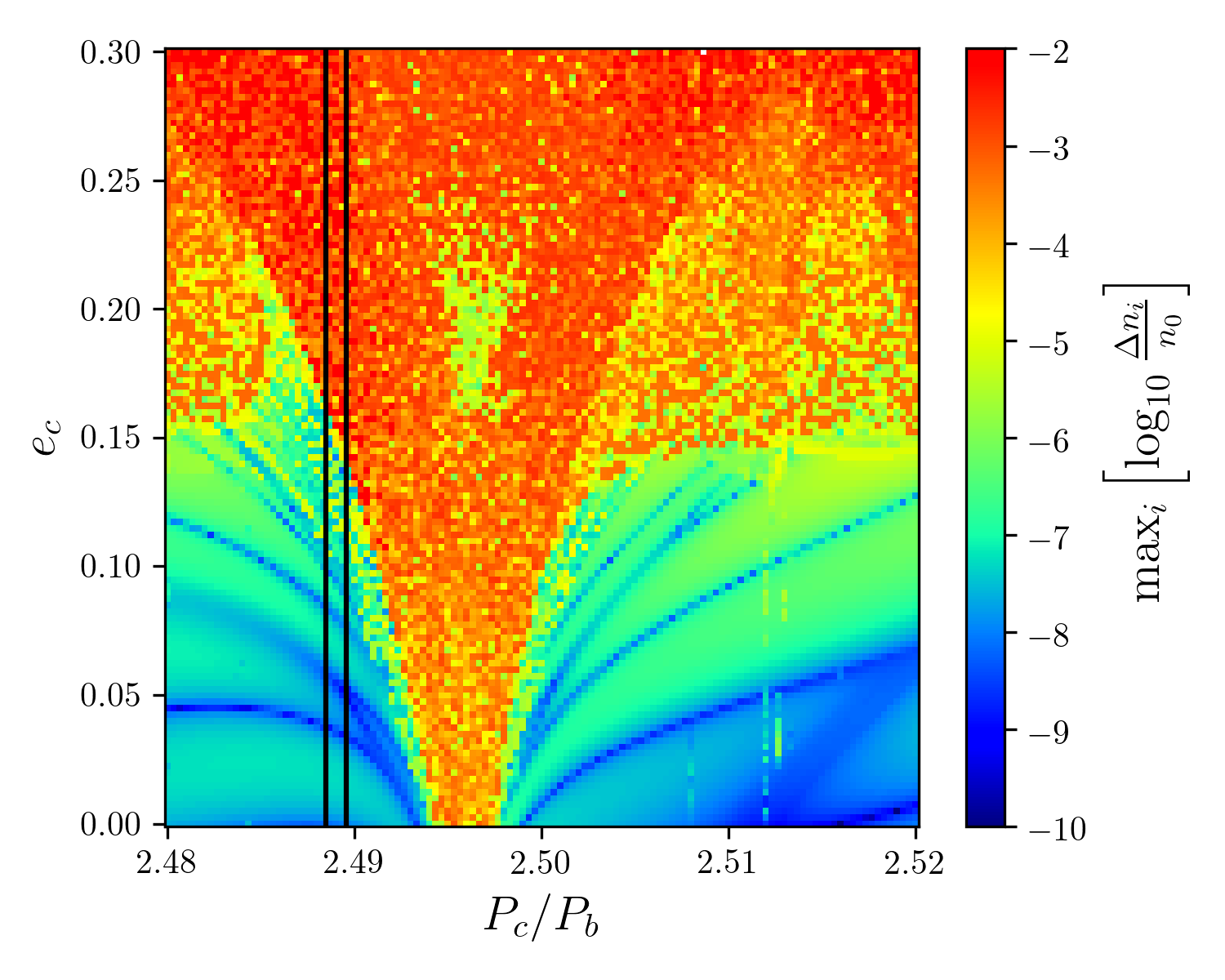}
    \caption{Chaoticity map of HD\,39194 around the 5:2 MMR of the inner planet pair. We explore the eccentricity of planet c on the vertical axis, and the period ratio $P_c/P_b$ on the horizontal axis, via a total set of 14641 configurations. A color is assigned to each configuration based on the NAFF chaos indicator.}
    \label{HD39194_StMap}
\end{figure}

The resulting chaoticity map is shown in Fig. \ref{HD39194_StMap}, with the color code depicting the level of chaos. The redder, the more diffusive are the orbital mean motions and therefore the more chaotic is the configuration. The two vertical lines depict the 1$\sigma$ confidence interval on $P_c$, reported on $P_c/P_b$. It is obvious from this figure that the inner planet pair in HD\,39194 lies outside of the 5:2 MMR, given the $1\sigma$ upper bound on $e_c$ of 0.1. The resonance width is dependent on some parameters such as the planetary masses or eccentricities. However, we do not expect our conclusions to change among the parameter values allowed by the MCMC exploration and with orbital inclinations such that $sin\,i \sim 1$.

Furthermore, we note that the configurations with $e_c > 0.15$ are very unlikely, given their high level of chaos. However, we take this observation with caution since only two parameters, $e_c$ and $P_c/P_b$, were explored in the parameter space. To provide proper bounds on the eccentricity, studying the influence of the other parameters is essential. In any case, we notice that no constraint from stability is added on top of the MCMC results. \\ 

\subsection{HD\,93385} \label{sec:HD93385}

\begin{table}[]
\centering
\caption{Posterior values of all parameters used for HD93385. We give the values of the mode and the 1$\sigma$ confidence interval of the posterior distribution from the MCMC. $^{(\ddagger)}$ Eccentricity does not differ significantly from zero; the 68\% and 95\% upper limits are reported. The argument of periastron $\omega$ is therefore unconstrained.}
\label{tab:HD93385-posterior}
\resizebox{\columnwidth}{!}{%
\begin{tabular}{@{}lcccc@{}}
\toprule
\multicolumn{5}{c}{\textbf{HD93385}}                                                                                  \\ \midrule
Parameter              & \multicolumn{1}{c}{Units} & \multicolumn{1}{l}{} &                     &                     \\ \midrule
\multicolumn{5}{c}{\textit{Offset and drift}}                                                                         \\ \midrule
$\gamma_{\rm H03}$ & \ms                       & \multicolumn{3}{c}{47\,576.2$\pm$0.1}                            \\
$\gamma_{\rm H15}$ & \ms                       & \multicolumn{3}{c}{47\,594.0$\pm$2.7}                              \\ \midrule
\multicolumn{5}{c}{\textit{Noise}}                                                                                    \\ \midrule
$\sigma_{\rm H03}$ & \ms                       & \multicolumn{3}{c}{1.45$\pm$0.09}                                \\
$\sigma_{\rm H15}$ & \ms                       & \multicolumn{3}{c}{2.4$\pm$1.3}                                  \\
$\sigma_{(O-C)}$       & \ms                       & \multicolumn{3}{c}{1.57}                                         \\ \midrule
\multicolumn{5}{c}{\textit{Keplerians}}                                                                               \\ \midrule
                       &                           & \textbf{HD93385\,b}  & \textbf{HD93385\,c} & \textbf{HD93385\,d} \\ \midrule
$P$                    & day                       & 7.3426$\pm$0.0012    & 13.180$\pm$0.003    & 45.85$\pm$0.05      \\
$K$                    & \ms                       & 1.36$\pm$0.15        & 1.87$\pm$0.15       & 1.51$\pm$0.16       \\
$e$                    &                           & <0.161; <0.295 $^{(\ddagger)}$        & <0.107; <0.200 $^{(\ddagger)}$       & $0.09_{-0.05}^{+0.15}$       \\
$\omega$               & deg                       & - $^{(\ddagger)}$            & - $^{(\ddagger)}$          & $-55\pm41$          \\
$\lambda_0$            & deg                       & 199.3$\pm$6.9        & 228$\pm$4.7         & 134.2$\pm$4.1       \\ \midrule
$T_{\rm Periastron}$            & BJD                       & 2\,455\,504.5$\pm$1.1    & 2\,455\,499$\pm$3       & 2\,455\,522.3$\pm$7.4   \\
$m\,\sin i$            & \Mearth                   & 4.2$\pm$0.5          & 7.1$\pm$0.6         & 8.7$\pm$0.9         \\
$a$                    & AU                        & 0.0756$\pm$0.0013    & 0.112$\pm$0.002     & 0.2565$\pm$0.0043   \\ \midrule
Ref. Epoch                  & BJD                                & \multicolumn{3}{c}{2\,455\,500}                  \\ \bottomrule 
\end{tabular}}
\end{table}

HD\,93385 is a quiet star with low chromospheric activity levels and has been regularly observed since 2006, collecting a total of 240 nightly binned radial velocities. The activity indicator \rhk\ shows a peak to peak amplitude of 0.04. We see a long term feature of ~2800 days, but this is not visible in the RV data.

In the periodogram of the RV data, we find a clear significant peak at 13.18 days. Other significant peaks are present close to 1, 7.3, 45, and 52 days. The peak close to 1 day is the one day alias of this 13.18 day signal. Fitting a Keplerian at this period and looking at the periodogram of the residuals results in another significant signal at 45.84 days (with its one year alias right next to it at 52 days). Repeating this once more, we see a significant signal at 7.34 days. After fitting this last Keplerian, we do not see any more periodic signals in the residuals with a FAP lower than 10\%. This procedure and the corresponding periodogram of the residuals after each step can be seen in Fig. \ref{fig:HD39194&HD93385-FAP}. 

\begin{figure*} 
    \centering
    \begin{minipage}{\textwidth}
        \centering
        \begin{multicols}{3}
            \begin{overpic}[width=0.32\textwidth]{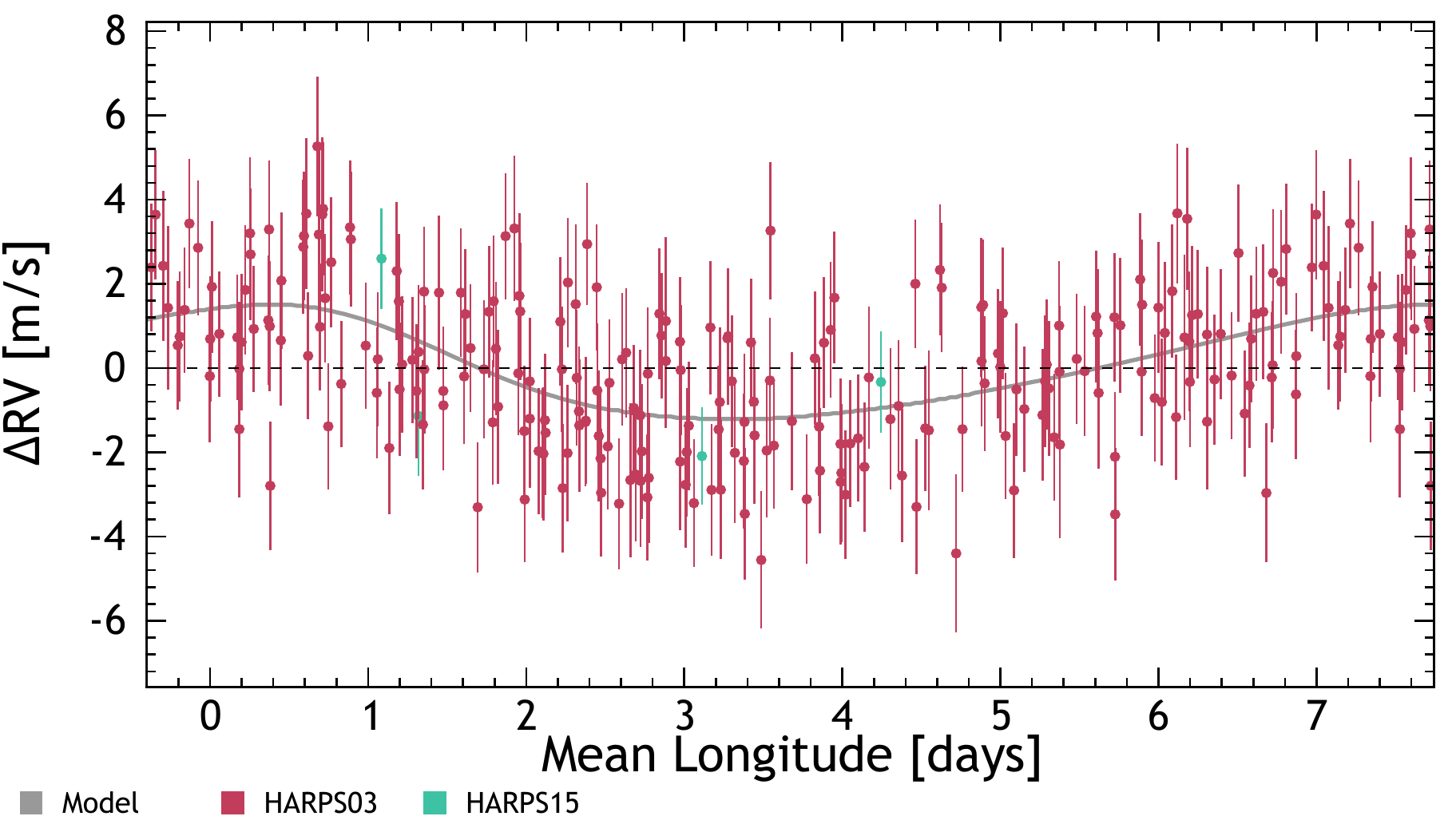}
            \end{overpic}
            \columnbreak
            
            \begin{overpic}[width=0.32\textwidth]{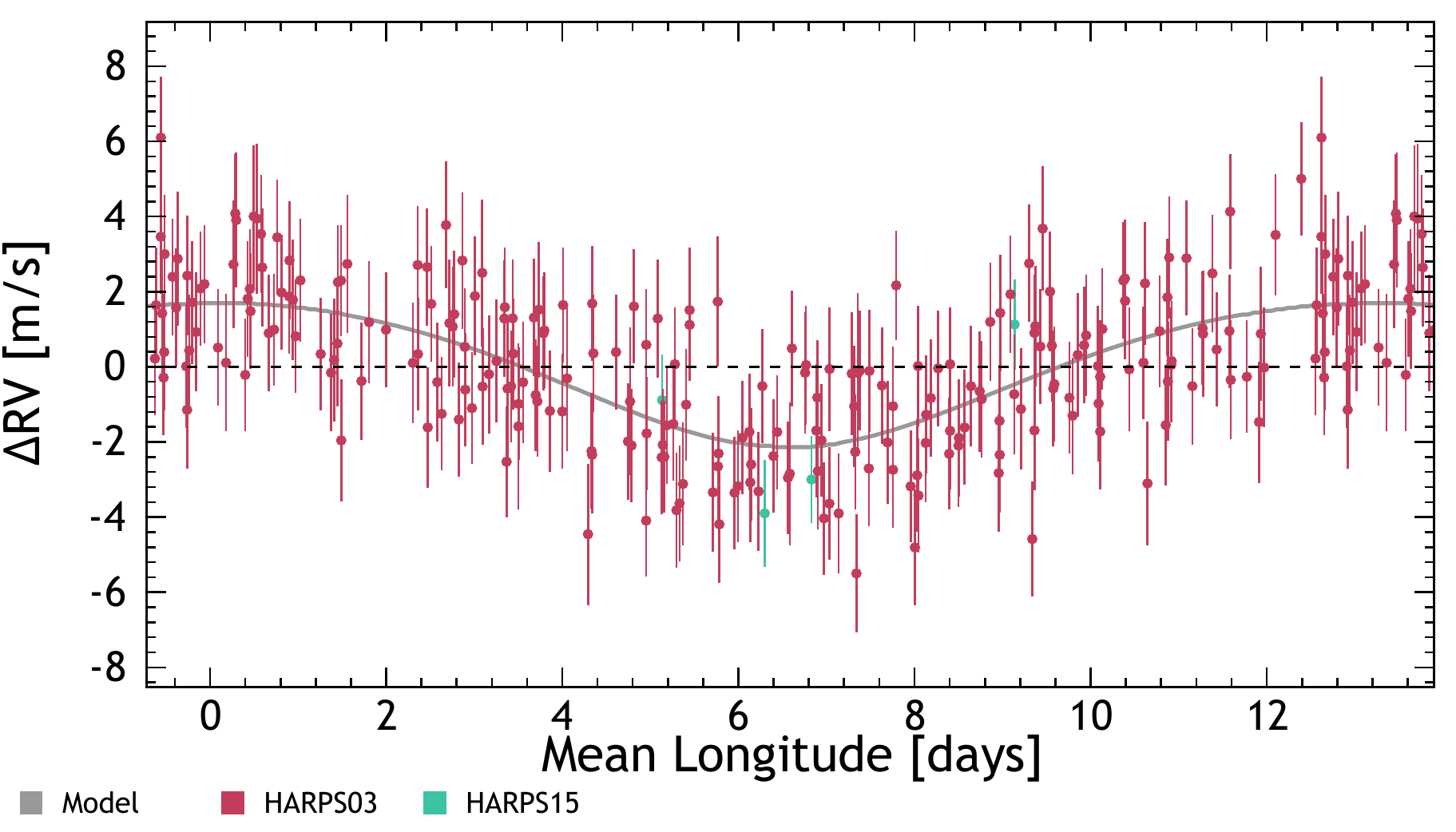}
            \end{overpic}
            \columnbreak
            
            \begin{overpic}[width=0.32\textwidth]{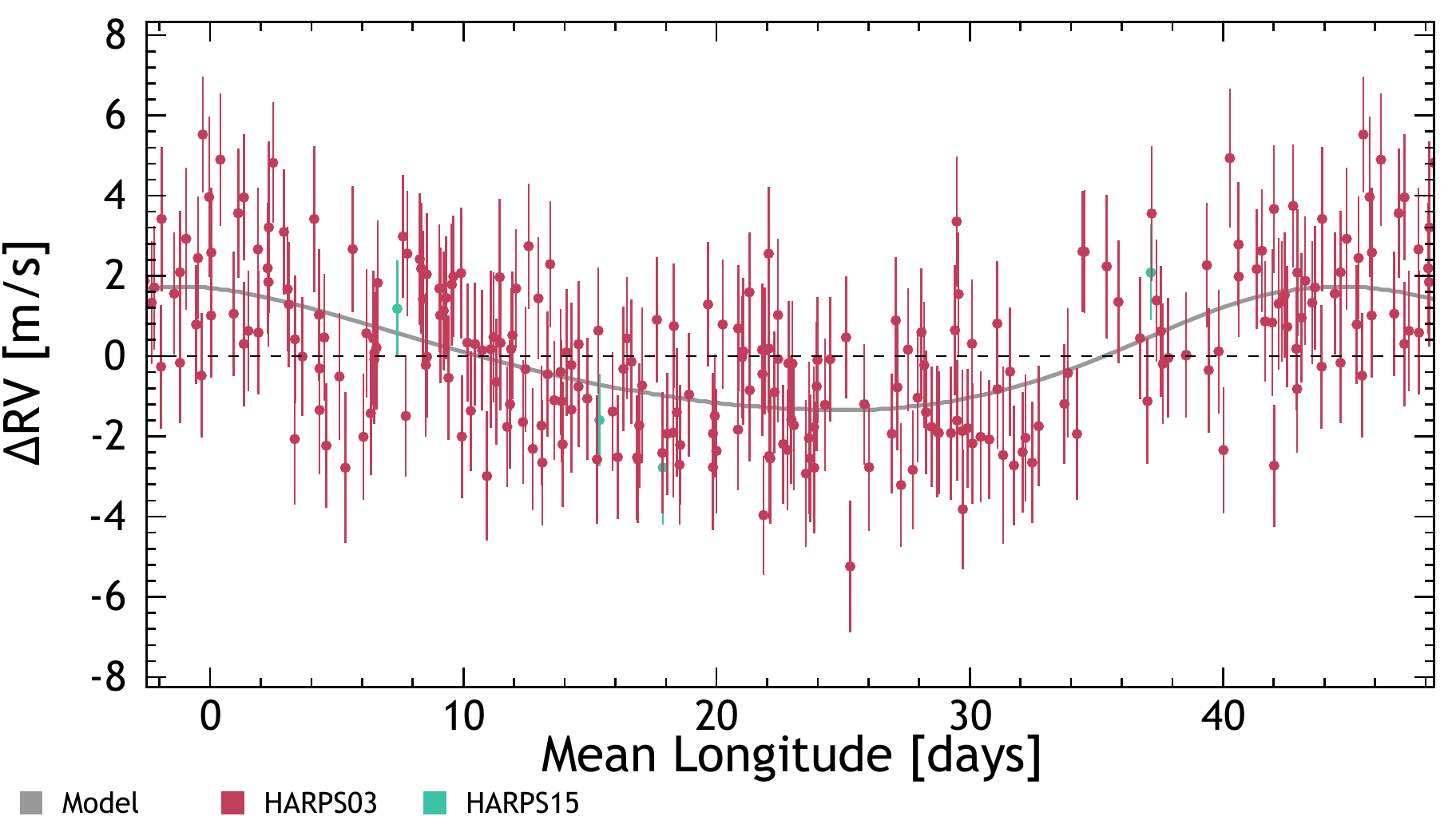}
            \end{overpic}
            
        \end{multicols}
    \end{minipage}
    \caption{Phase folded plots for each planet in HD\,93385. From left to right: planets b, c and d with orbital periods of 7.34, 13.18 and 45.8 days, respectively.}
    \label{fig:HD93385-phasefold}
\end{figure*}

We then estimated the Bayesian evidence using \textsc{PolyChord} for models containing up to four planets. The results are shown in Table \ref{tab:odds_ratios}. We see a significant increase in evidence up to the model with three planets and a slight decrease in the four-planet model. The Bayes factor of the three-planet model compared to the four-planet model, is only $0.5 \pm 2.3$, which is inconclusive. Additionally, we observe that the posterior of the fourth Keplerian orbital period does not converge to a unique value. We cannot confirm the presence of a fourth planet with the current data set, which leaves HD\,93385 with three significant planetary signals at periods of 7.34, 13.18, and 45.84 days and minimum masses of 4.2, 7.1, and 8.7 \Mearth\, respectively. The phase-folded RVs can be seen in Fig. \ref{fig:HD93385-phasefold}.

\begin{figure*}
    \centering
    \includegraphics[width=0.7\textwidth]{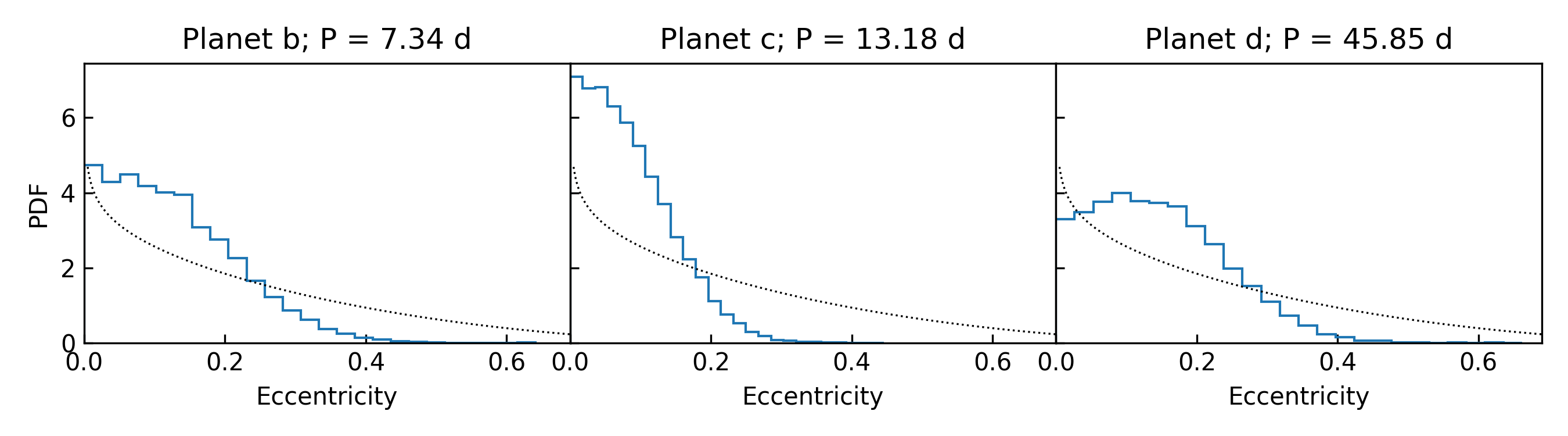}
    \caption{Posterior distribution of the eccentricities for the planetary companions of HD93385. The dotted line represents the eccentricity prior.}
    \label{fig:HD93385_eccs}
\end{figure*}

The orbital elements and their uncertainties, estimated from running an MCMC, are listed in Table \ref{tab:HD93385-posterior}. In Fig. \ref{fig:HD93385_eccs} we show the posterior distribution for the eccentricities of HD93385's companions. Planets b and c do not differ much from zero and are more constrained than the prior, while planet d peaks at around $e=0.1$ but with a weak convergence.

The semi-amplitude's we derive for planets c and d are $\boldsymbol \sim$15\% lower from the ones originally derived by \citeads{2011arXiv1109.2497M}. Planet c was found to have a semi-amplitude of 2.21 \ms\ while we find 1.87 \ms, and planet d's semi-amplitude was calculated at 1.82 \ms\ and we find 1.51 \ms. These changes also reduced their minimum masses from 8.36 to 7.1 \Mearth\ for planet c, and from 10.12 to 8.7 \Mearth\ for planet d.

It is interesting to note that the 7.34-day planet was not reported in \citetads{2011arXiv1109.2497M}. We find this periodic signal with a very high significance level (FAP $\sim 10^{-8.8}$), which we think is a result of the additional ${\sim} 10$ years of RV data that we now have since the publication of \citetads{2011arXiv1109.2497M}. Indeed, when we redo this analysis using only data taken before 2011, the 7.34-day signal is there but with a lower significance at a FAP level of $\sim \! 8 \% $, above the 1\% threshold that \citeads{2011arXiv1109.2497M} used to look for planets.

\subsection{HD\,96700}

HARPS has been observing HD\,96700 since 2003 with some measurements taken during commissioning of the instrument. We removed these data points because of uncertain operation conditions. After commissioning of the instrument only 6 data points were taken until 2008 where the star began to be observed regularly. To reduce aliases in the periodogram we discard these 6 data points as well and only use data taken from 2008 onward, which leaves us with 235 nightly binned radial velocity measurements.

We started by looking at the activity indicator \rhk, and after removing a long term drift with a period of ~3900 days we notice a 41 day signal. This is relevant because the estimated stellar rotation period for HD\,96700 using \citetads{2008ApJ...687.1264M} is around 20.1 days, roughly half of the periodic signal we see in \rhk.

The most significant peak in the periodogram of the RV data is at 8.1 days. Other significant peaks can be seen at 0.9, 1, 1.1 and 103 days. The peaks close to one day are the one day aliases of the 8.1 and 103 day signals. After fitting a Keplerian at an 8.1 day period and looking at the periodogram of the residuals we find the 103.5 day peak as the most significant. Also present is a peak at 80 days which is the one year alias of the 103 day peak. Repeating the process once more by fitting a keplerian with a period of 103 days we see a peak at 19.88 days with a FAP level of $10^{-5.7}$. This period is suspiciously close to the estimated rotation period of the star at 20.1 days.

Even though this signal could be related to the rotation period of the star, we do not see any evidence of this in the activity indicators. The 41-day period signal we see in \rhk\ could actually be the real rotation period, and what we see in the RV is a harmonic of the rotation, due to, for example, star spots opposite to each other. Still, by comparing the phase of this 19.88 day signal in the first and second half of the $\sim\!10$ year dataset, we see that the phases in both periods differ by less than 6 degrees, and the amplitude differs by less than 0.1~\ms, which are within their respective uncertainties. This makes the signal consistent in phase and amplitude for at least 10 years.

We calculated the Bayesian evidence for two additional models to try and get some more insight into this 20 day signal. First we tried using the FF' model \citepads{2012MNRAS.419.3147A} as it was used in \citeads{2021MNRAS.503.1248A}. This model uses the flux of the star $\Psi(t)$ and its time derivative $\dot\Psi(t)$ to account for spot coverage and RV variability. We used the FWHM as a flux proxy of the intensity of the star. We made this choice because we do not have photometric measurements available for HD96700 and it has been shown that the CCF FWHM can track the photometry very closely (e.g. \citeads{2020A&A...639A..77S}).

The second model we tried is assuming that the noise is correlated and modeled it by including a quasi-periodic kernel in the covariance matrix following the formalism of \citetads{2020A&A...638A..95D}. We chose a Gaussian prior for the period of the quasi-periodic kernel, centered at the signal found in the RV (19.88 days) and a standard deviation of 2 days to allow for slight differences in the final period. The kernel also includes an exponential decay and the decay time scale was set with a log-uniform prior between 1 hour and 500 days.

\begin{table}[h]
\centering
\caption{Bayesian model comparison of different noise models for HD\,96700. The Base model contains only keplerians and Gaussian white noise, FF' is modeled following \citetads{2012MNRAS.419.3147A} and the Correlated Noise model includes a quasi-periodic kernel in the covariance matrix. All evidence values are reported relative to the evidence of the base model with 3 planets.}
\label{tab:HD96700-noise}
\resizebox{\columnwidth}{!}{%
\begin{tabular}{@{}lccccc@{}}
\toprule
\textbf{HD\,96700} & \multicolumn{5}{c}{Number of planets}                                                \\ \midrule
Model   & 0                & 1               & 2              & 3               & 4            \\ \midrule
Base    & $-130.5 \pm 0.2$ & $-59.0\pm1.6$   & $-10.3\pm0.8$  & \cellcolor[HTML]{C0C0C0}$0.0\pm0.9$     & $-1.0\pm2.8$ \\
FF'     & $-136.7 \pm 0.5$ & $-64.3 \pm 1.4$ & $-9.7 \pm 1.6$ & $-13.3 \pm 1.5$ &              \\
Correlated Noise & $-115.3 \pm 0.1$ & $-38 \pm 3.5$ & $+6.6 \pm 1.7$ & $+6.88 \pm 1.4$ & $+6.1 \pm 1.2$ \\ \bottomrule
\end{tabular}}
\end{table}

In Table \ref{tab:HD96700-noise} we present the evidence values for these models, together with the base model of just keplerians and Gaussian white noise. All values are presented relative to the base 3-planet model (colored in gray). Both the FF' and correlated noise models try to model the ~20 day signal without a keplerian, so the 2-planet models of these noise models have to be compared with the 3-planet model of the base model.

We get contradicting results where the FF' model is significantly disfavored, while the correlated noise model is favored by a difference of $\ln(\mathcal{O}) = 6.6$. If the rotation period of the star is indeed around 40 days, we would expect to detect other harmonics in the radial velocity data, at the very least we would expect to see the fundamental frequency. Such series of signals are not observed in the radial velocity periodogram and it does not suggest any other periods either. In addition, the fact that this 19.88 day signal is consistent in phase and amplitude for nearly 10 years makes it unlikely to stem from any activity related effect. In Sect. \ref{sec:discussion} we analyze possible formation and evolution paths for this system and find that this planetary architecture is compatible with the latest planetary formation and evolution models. With the data and information we have at this moment we thus conclude that this 19.88 day signal in the RV is better explained by a planetary companion. 

\begin{figure*} 
    \centering
    \begin{minipage}{\textwidth}
        \centering
        \begin{multicols}{3}
            \begin{overpic}[width=0.32\textwidth]{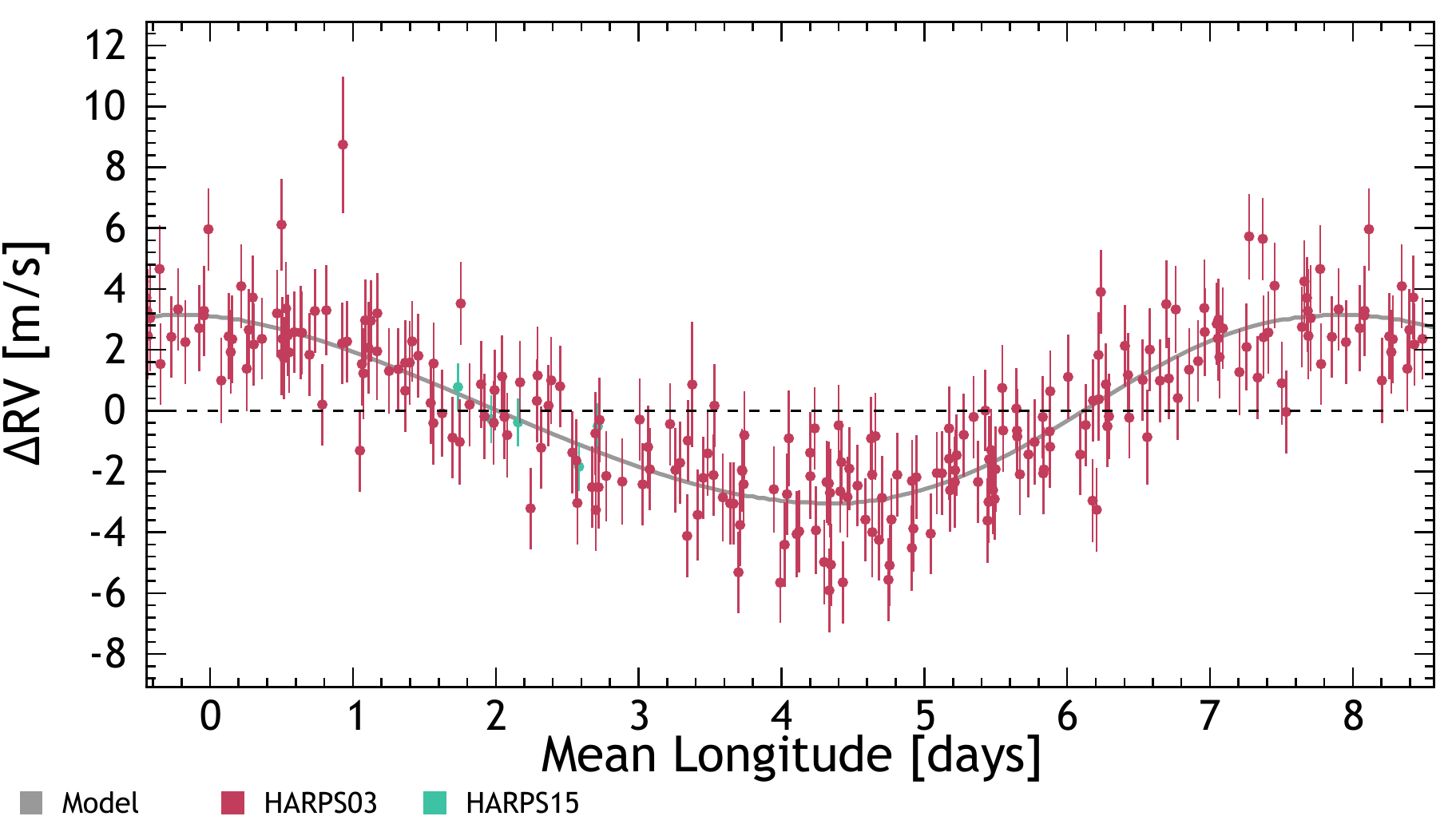}
            \end{overpic}
            \columnbreak
            
            \begin{overpic}[width=0.32\textwidth]{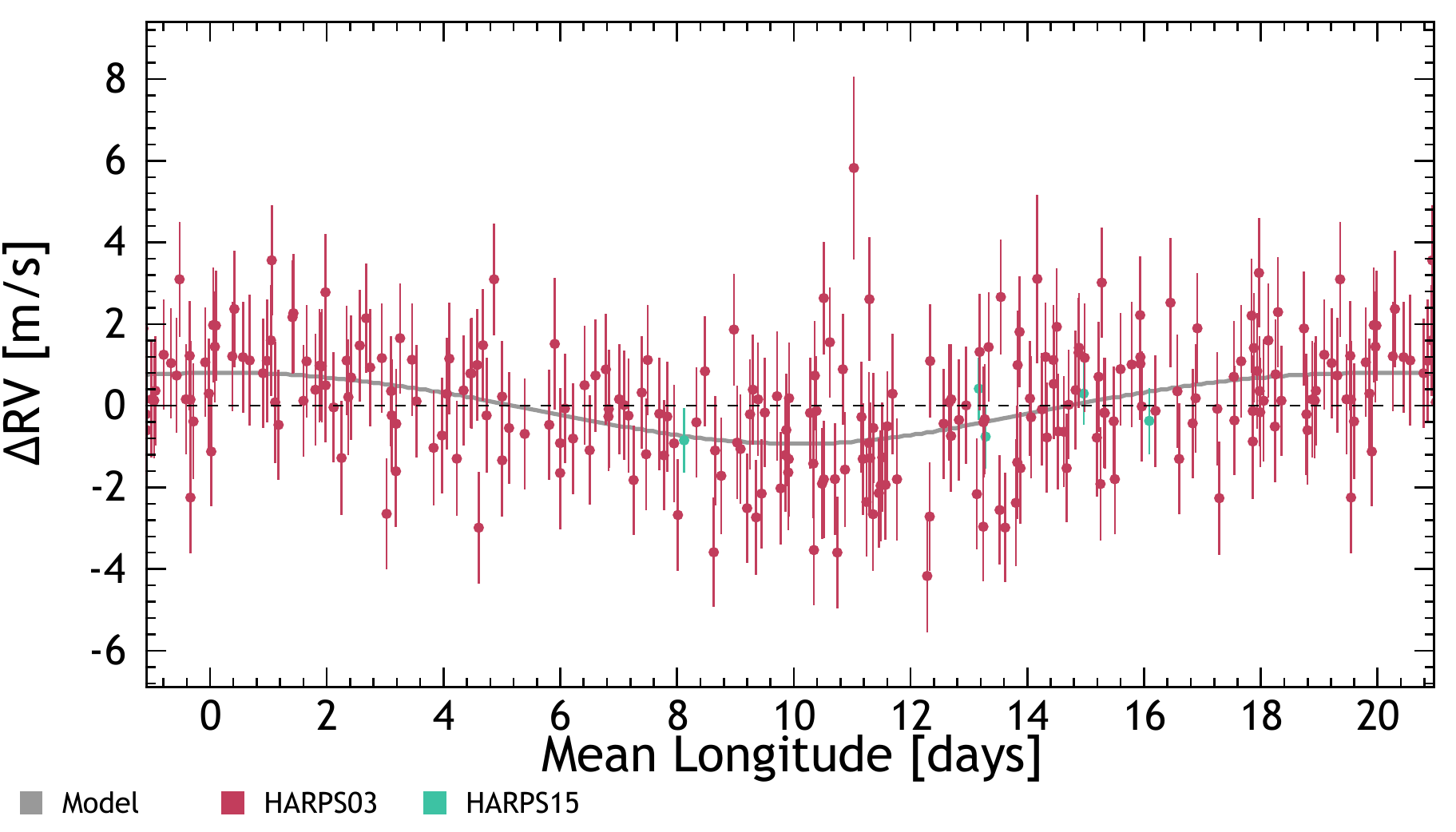}
            \end{overpic}
            \columnbreak
            
            \begin{overpic}[width=0.32\textwidth]{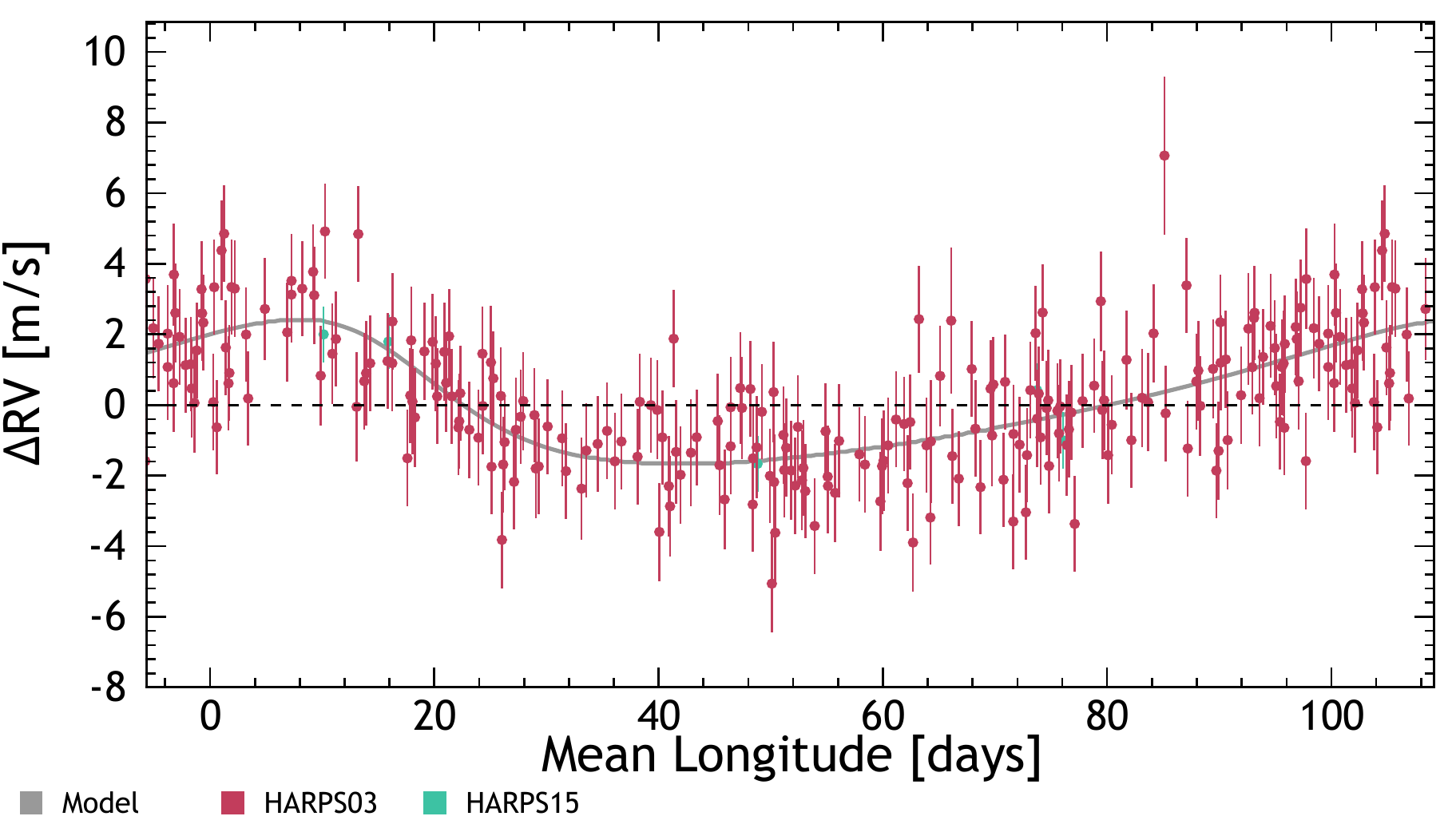}
            \end{overpic}
            
        \end{multicols}
    \end{minipage}
    \caption{Phase folded plots for each planet in HD\,96700. From left to right: planets b, c and d with orbital periods of 8.12, 19.88 and 103.5 days, respectively.}
    \label{fig:HD96700-phasefold}
\end{figure*}

\begin{figure*}
    \centering
    \includegraphics[width=0.7\textwidth]{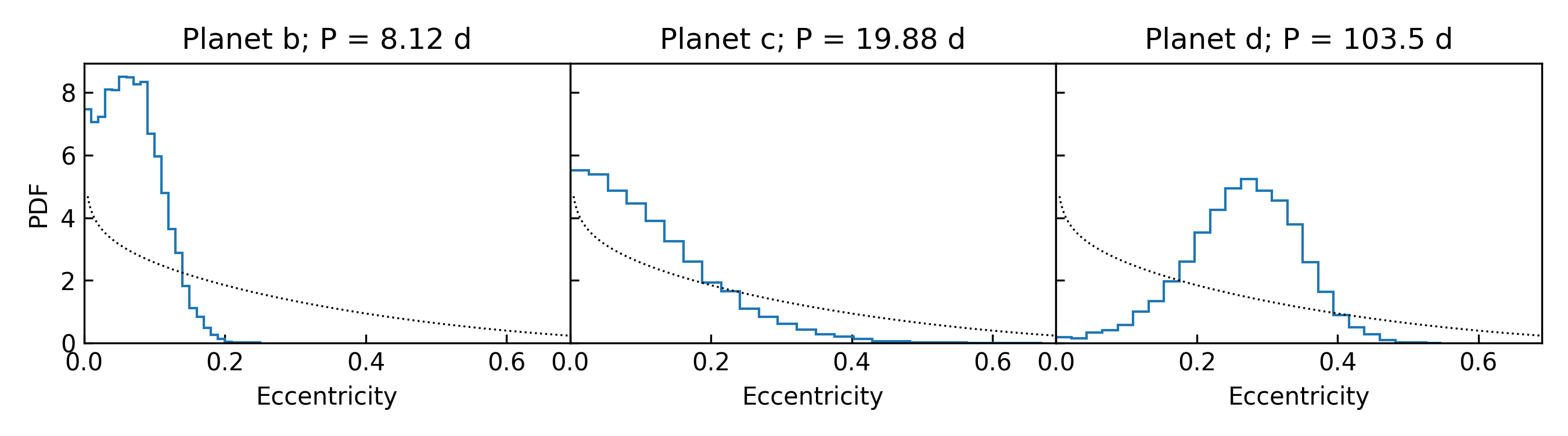}
    \caption{Posterior distribution of the eccentricities for the planetary companions of HD\,96700. The dotted line represents the eccentricity prior.}
    \label{fig:HD96700_eccs}
\end{figure*}

In conclusion, for HD\,96700 we detect three planetary companions with orbital periods of 8.12, 19.88, and 103.5 days and minimum masses of 8.9, 3.5, and 12.7 \Mearth\, respectively. The phase-folded RVs can be seen in Fig. \ref{fig:HD96700-phasefold}. In Table \ref{tab:HD96700-posterior} we present the MCMC posterior estimates of all the model parameters for the Base model with three keplerians. The posterior for the eccentricities are shown in Fig. \ref{fig:HD96700_eccs}. The orbits of planets b and c are compatible with circular while planet d has a clear non zero eccentricity at $e_d=0.27\pm0.08$. In addition, the planet parameters we find for planets b and d are the same than the ones found by \citeads{2011arXiv1109.2497M}.

\begin{table}[]
\centering
\caption{Posterior values of all parameters used for HD96700. We give the values of the mode and the 1$\sigma$ confidence interval of the posterior distribution from the MCMC. $^{(\ddagger)}$ Eccentricity does not differ significantly from zero; the 68\% and 95\% upper limits are reported. The argument of periastron $\omega$ is therefore unconstrained.}
\label{tab:HD96700-posterior}
\resizebox{\columnwidth}{!}{%
\begin{tabular}{@{}lcccc@{}}
\toprule
\multicolumn{5}{c}{\textbf{HD96700}}                                                                                  \\ \midrule
Parameter              & \multicolumn{1}{c}{Units} & \multicolumn{1}{l}{} &                     &                     \\ \midrule
\multicolumn{5}{c}{\textit{Offset and drift}}                                                                         \\ \midrule
$\gamma_{\rm H03}$ & \ms                       & \multicolumn{3}{c}{12\,862.45$\pm$0.1}                             \\
$\gamma_{\rm H15}$ & \ms                       & \multicolumn{3}{c}{12\,879.7$\pm$0.8}                              \\ \midrule
\multicolumn{5}{c}{\textit{Noise}}                                                                                    \\ \midrule
$\sigma_{\rm H03}$ & \ms                       & \multicolumn{3}{c}{1.25$\pm$0.07}                                \\
$\sigma_{\rm H15}$ & \ms                       & \multicolumn{3}{c}{0.6$\pm$0.5}                                  \\
$\sigma_{(O-C)}$       & \ms                       & \multicolumn{3}{c}{1.3}                                          \\ \midrule
\multicolumn{5}{c}{\textit{Keplerians}}                                                                               \\ \midrule
                       &                           & \textbf{HD96700\,b}  & \textbf{HD96700\,c} & \textbf{HD96700\,d} \\ \midrule
$P$                    & day                       & 8.1245$\pm$0.0006    & 19.88$\pm$0.01      & 103.5$\pm$0.1       \\
$K$                    & \ms                       & 3.06$\pm$0.13        & 0.9$\pm$0.1         & 1.94$\pm$0.15       \\
$e$                    &                           & <0.085; 0.138 $^{(\ddagger)}$ & <0.144; <0.293 $^{(\ddagger)}$       & 0.27$\pm$0.08       \\
$\omega$               & deg                       & - $^{(\ddagger)}$          & - $^{(\ddagger)}$          & 42$\pm$18           \\
$\lambda_0$            & deg                       & 196$\pm$3            & -34$\pm$9           & 144$\pm$8           \\ \midrule
$T_{\rm Periastron}$            & BJD                       & 2\,455\,501.3$\pm$1.3    & 2\,455\,513$\pm$5       & 2\,455\,500$\pm$5       \\
$m\,\sin i$            & \Mearth                   & 8.9$\pm$0.4          & 3.5$\pm$0.4         & 12.7$\pm$1.0        \\
$a$                    & AU                        & 0.0777$\pm$0.0013    & 0.141$\pm$0.002     & 0.424$\pm$0.007     \\ \midrule
Ref. Epoch                  & BJD                                & \multicolumn{3}{c}{2\,455\,500}                  \\ \bottomrule 
\end{tabular}}
\end{table}

\subsection{HD\,154088}

\begin{table}[]
\centering
\caption{Posterior values of all parameters used for HD154088. We give the values of the mode and the 1$\sigma$ confidence interval of the posterior distribution from the MCMC. $^{(\dagger)}$ Linear parameter fitted to \rhk\ with a Gaussian kernel at a 0.5 yr timescale. $^{(\ddagger)}$ Eccentricity does not differ significantly from zero; the 68\% and 95\% upper limits are reported. The argument of periastron $\omega$ is therefore unconstrained.}
\label{tab:HD154088-posterior}
\resizebox{0.6\columnwidth}{!}{%
\begin{tabular}{@{}lcccc@{}}
\toprule
\multicolumn{5}{c}{\textbf{HD154088}}                                                        \\ \midrule
Parameter              & \multicolumn{1}{c}{Units} & \multicolumn{1}{l}{}      &       &     \\ \midrule
\multicolumn{5}{c}{\textit{Offset and drift}}                                                \\ \midrule
$\gamma_{\rm H03}$ & \ms                       & \multicolumn{3}{c}{14\,298.12$\pm$0.16}   \\ 
A $^{(\dagger)}$      & \ms                       & \multicolumn{3}{c}{2.46$\pm$0.31}       \\  \midrule
\multicolumn{5}{c}{\textit{Noise}}                                                           \\ \midrule
$\sigma_{\rm H03}$ & \ms                       & \multicolumn{3}{c}{1.60$\pm$0.09}       \\
$\sigma_{(O-C)}$       & \ms                       & \multicolumn{3}{c}{1.3}                 \\ \midrule
\multicolumn{5}{c}{\textit{Keplerians}}                                                      \\ \midrule
                       &                           & \multicolumn{3}{c}{\textbf{HD154088\,b}} \\ \midrule
$P$                    & day                       & \multicolumn{3}{c}{18.56$\pm$0.01}      \\
$K$                    & \ms                       & \multicolumn{3}{c}{1.7$\pm$0.2}         \\
$e$                    &                           & \multicolumn{3}{c}{<0.193; <0.344 $^{(\ddagger)}$}       \\
$\omega$               & deg                       & \multicolumn{3}{c}{- $^{(\ddagger)}$}          \\ 
$\lambda_0$            & deg                       & \multicolumn{3}{c}{239$\pm$6}           \\ \midrule
$T_{\rm Periastron}$            & BJD                       & \multicolumn{3}{c}{2\,455\,498.2$\pm$3.5}   \\
$m\,\sin i$            & \Mearth                   & \multicolumn{3}{c}{6.6$\pm$0.8}         \\
$a$                    & AU                        & \multicolumn{3}{c}{0.134$\pm$0.002}     \\ \midrule
Ref. Epoch                  & BJD                                & \multicolumn{3}{c}{2\,455\,500}                  \\ \bottomrule 
\end{tabular}}
\end{table}

HD\,154088 is a K dwarf at a distance of 17.8 pc from Earth. HARPS has been observing this star regularly since 2008 until 2015, when the measurements start to become more sparse. There are three RV points taken in 2006, but we disregarded these because of the long time separation until the next measurements in 2008. This time gap can introduce unwanted harmonics in the periodogram. After the fiber change of HARPS in 2015, only one RV point was taken in 2017. We also discarded this point for the same reason mentioned before. We end up with 183 nightly binned radial velocities for HD\,154088.

The periodogram of the RV time series shows a peak at 18.5 days and a long term magnetic cycle at $\sim\!3000$ days. The magnetic cycle is easily removed by adding a linear parameter proportional to the time series of \rhk. After fitting the magnetic cycle and one Keplerian, no more significant peaks appear in the periodogram.

The results from the Bayesian model comparison analysis (see Table \ref{tab:odds_ratios}) shows a significant increase in evidence from zero to one planet. It then plateaus giving similar evidence values for the models with one, two, and three planets, followed by a slight decrease for four planets. The evidence for models with more than one planet are inconclusive compared to the one planet model and we do not see any new period candidate for an additional planet in the PolyChord posteriors. 

\begin{figure}
    \centering
    \includegraphics[width=0.6\columnwidth]{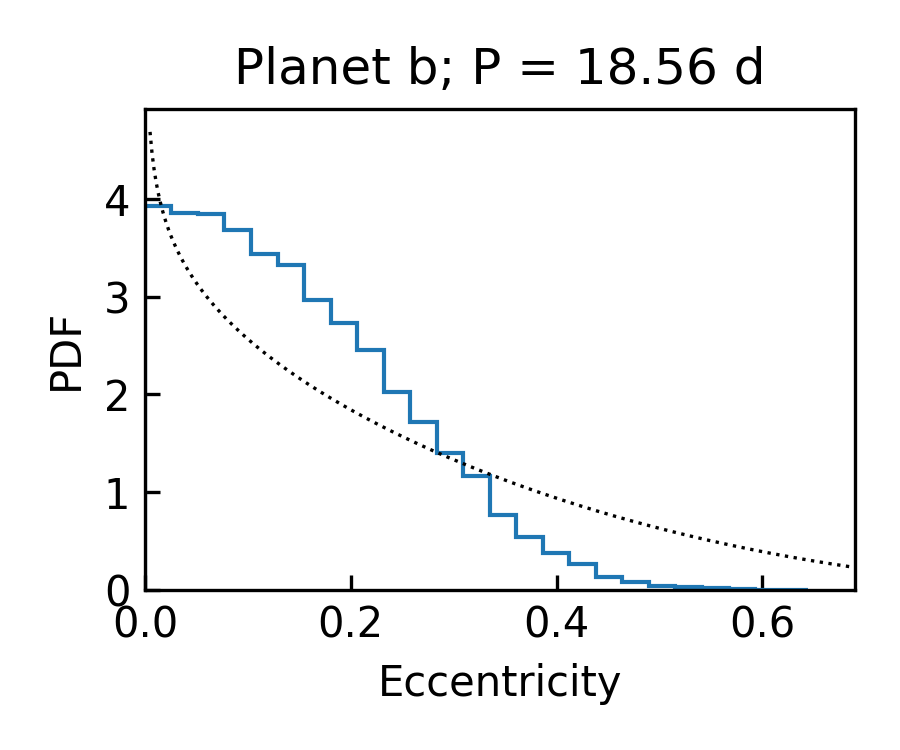}
    \caption{Posterior distribution of the eccentricities for the planetary companion of HD154088. The dotted line represents the eccentricity prior.}
    \label{fig:HD154088_eccs}
\end{figure}

We then conclude that HD\,154088 has one planet with a 18.56 day period and a minimum mass of 6.6 \Mearth. The phase-folded RVs can be seen in Fig. \ref{fig:HD189567-phasefold} and the full orbital parameters derived by running an MCMC can be seen in Table \ref{tab:HD154088-posterior}. Fig. \ref{fig:HD154088_eccs} shows the posterior distribution of the eccentricity of HD\,154088b. The planet parameters we derived in this article for HD\,154088b are very similar to the ones found by \citeads{2011arXiv1109.2497M}, only for the eccentricity do we find a lower value. They reported an eccentricity of 0.38, while we find an eccentricity compatible with 0 and an upper bound of 0.34 at the 95\% level.

\begin{figure} 
    \centering
    \includegraphics[width=0.8\columnwidth]{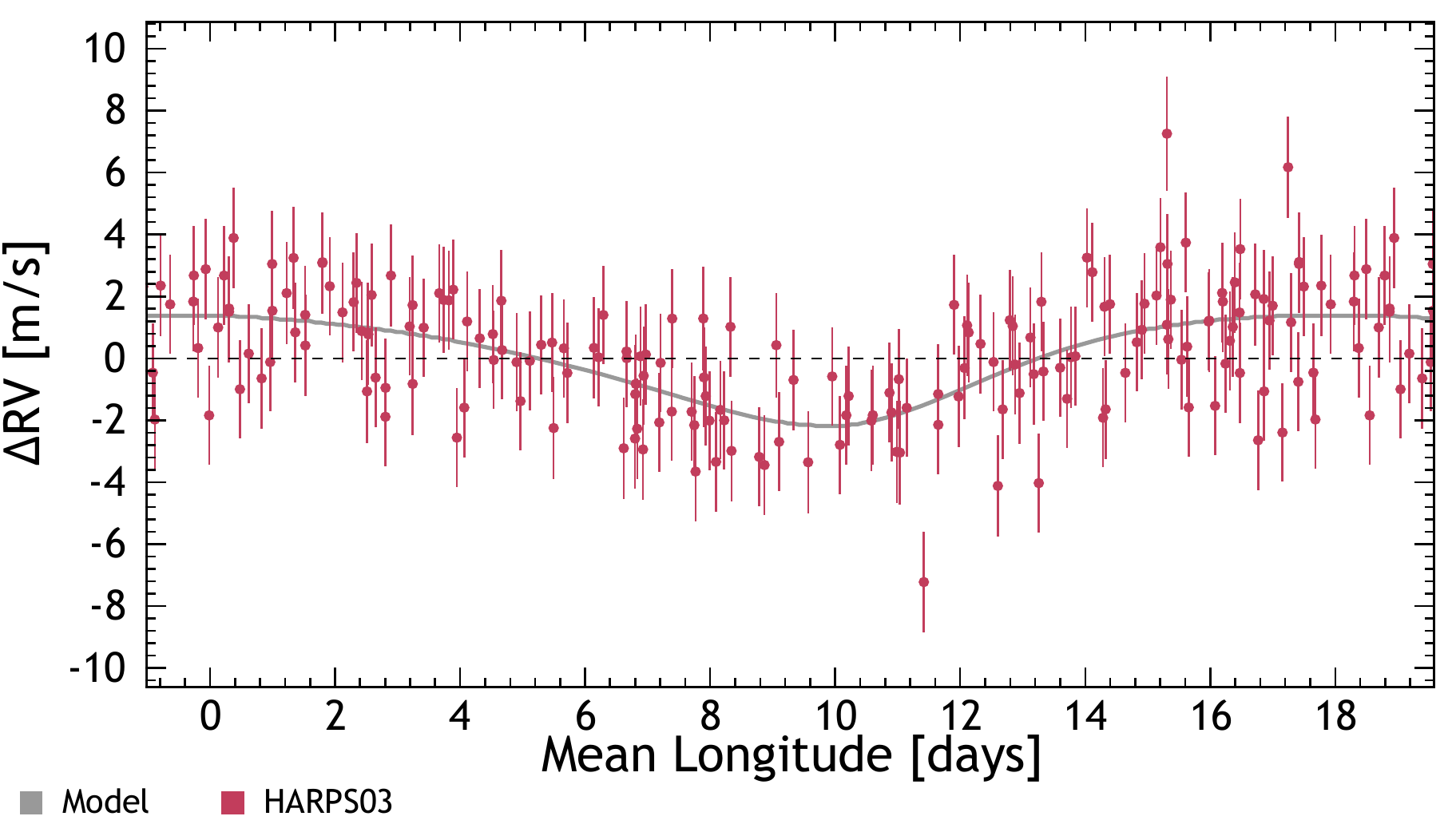}
    \caption{Phase folded plot for the single planet in HD\,154088 with an orbital period of 18.56 days.}
    \label{fig:154088-phasefold}
\end{figure}

\subsection{HD\,189567}

\begin{table}[]
\centering
\caption{Posterior values of all parameters used for HD189567. We give the values of the mode and the 1$\sigma$ confidence interval of the posterior distribution from the MCMC. $^{(\dagger)}$ Linear parameter fitted on the FWHM using an Epanechnikov kernel with a high pass filter at a timescale of 1.5 yr. $^{(\ddagger)}$ Eccentricity does not differ significantly from zero; the 68\% and 95\% upper limits are reported. The argument of periastron $\omega$ is therefore unconstrained.}
\label{tab:HD189567-posterior}
\resizebox{0.9\columnwidth}{!}{%
\begin{tabular}{@{}lccc@{}}
\toprule
\multicolumn{4}{c}{\textbf{HD189567}}                                                                             \\ \midrule
Parameter                               & \multicolumn{1}{c}{Units} & \multicolumn{1}{l}{} &                      \\ \midrule
\multicolumn{4}{c}{\textit{Offset and drift}}                                                                     \\ \midrule
$\gamma_{\rm H03}$                  & \ms                       & \multicolumn{2}{c}{-10\,475.76$\pm$0.18}    \\
$\gamma_{\rm H15}$                  & \ms                       & \multicolumn{2}{c}{-10\,463.8$\pm$1.4}        \\
$\alpha_1$                              & m s$^{-1}$ yr$^{-1}$       & \multicolumn{2}{c}{0.24$\pm$0.10}           \\
$\alpha_2$                              & m s$^{-1}$ yr$^{-2}$       & \multicolumn{2}{c}{0.06$\pm$0.02}           \\
$\alpha_3$                              & m s$^{-1}$ yr$^{-3}$       & \multicolumn{2}{c}{-0.007$\pm$0.004}        \\
A $^{(\dagger)}$ & \ms                       & \multicolumn{2}{c}{3.54$\pm$0.36}           \\ \midrule
\multicolumn{4}{c}{\textit{Noise}}                                                                                \\ \midrule
$\sigma_{\rm H03}$                  & \ms                       & \multicolumn{2}{c}{1.97$\pm$0.09}           \\
$\sigma_{\rm H15}$                  & \ms                       & \multicolumn{2}{c}{1.0$\pm$0.9}             \\
$\sigma_{(O-C)}$                        & \ms                       & \multicolumn{2}{c}{1.84}                    \\ \midrule
\multicolumn{4}{c}{\textit{Keplerians}}                                                                           \\ \midrule
                                        &                           & \textbf{HD189567\,b} & \textbf{HD189567\,c} \\ \midrule
$P$                                     & day                       & 14.288$\pm$0.002     & 33.688$\pm$0.025     \\
$K$                                     & \ms                       & 2.53$\pm$0.18        & 1.6$\pm$0.2          \\
$e$                                     &                           & <0.101; <0.189 $^{(\ddagger)}$  & 0.16$\pm$0.09        \\
$\omega$                                & deg                       & - $^{(\ddagger)}$           & 219$\pm$58           \\ 
$\lambda_0$                             & deg                       & -57$\pm$5            & 196$\pm$7            \\ \midrule
$T_{\rm Periastron}$                             & BJD                       & 2\,455\,509$\pm$3        & 2\,455\,502$\pm$5        \\
$m\,\sin i$                             & \Mearth                   & 8.5$\pm$0.6          & 7.0$\pm$0.9          \\
$a$                                     & AU                        & 0.111$\pm$0.002      & 0.197$\pm$0.003      \\ \midrule
Ref. Epoch                  & BJD                                & \multicolumn{2}{c}{2\,455\,500}                  \\ \bottomrule 
\end{tabular}}
\end{table}

\begin{figure*}[t] 
    \centering
    \begin{minipage}{\textwidth}
        \centering
        \begin{multicols}{2}
            \begin{overpic}[width=0.40\textwidth]{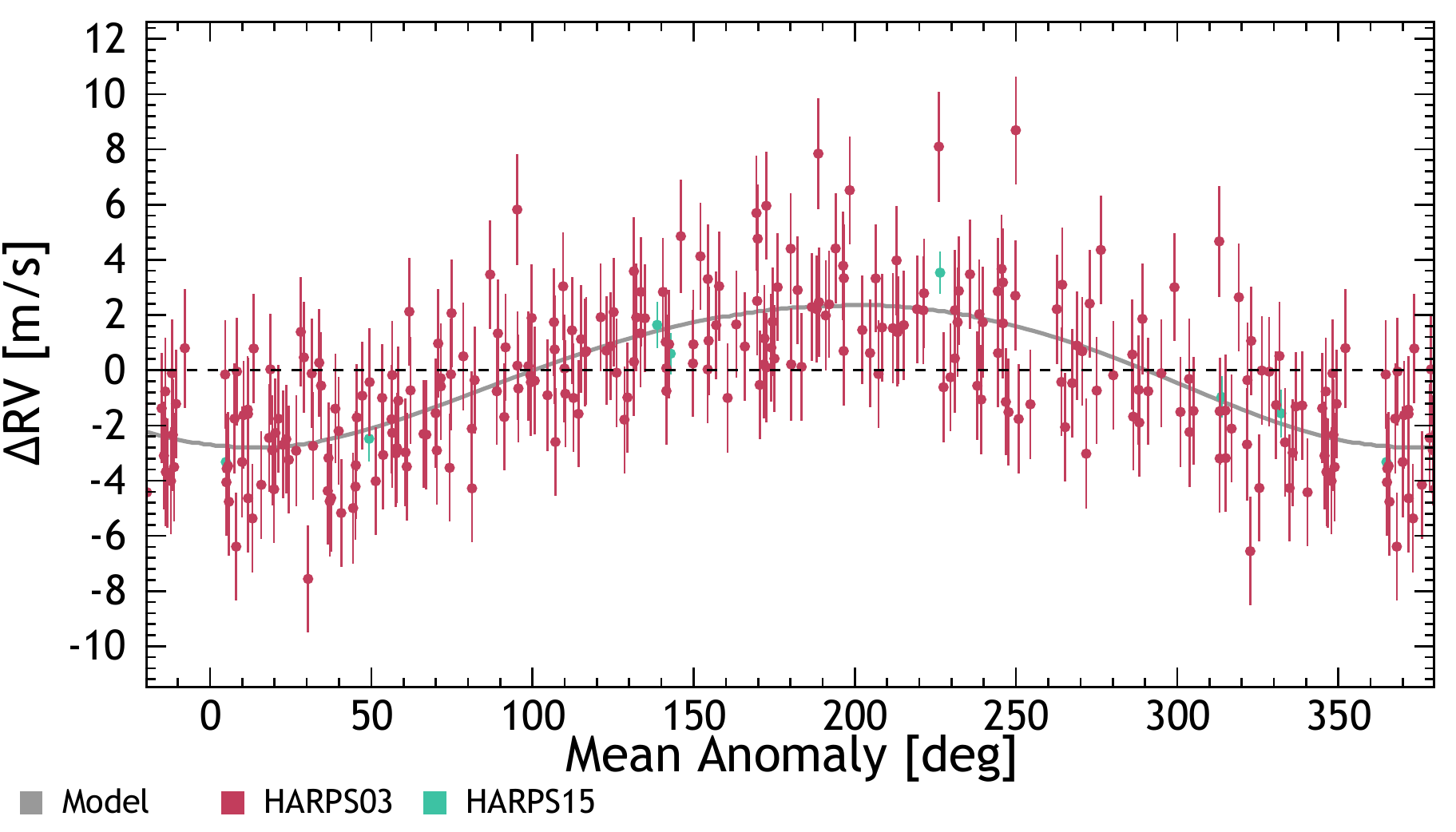}
            \end{overpic}
            \columnbreak
            
            \begin{overpic}[width=0.40\textwidth]{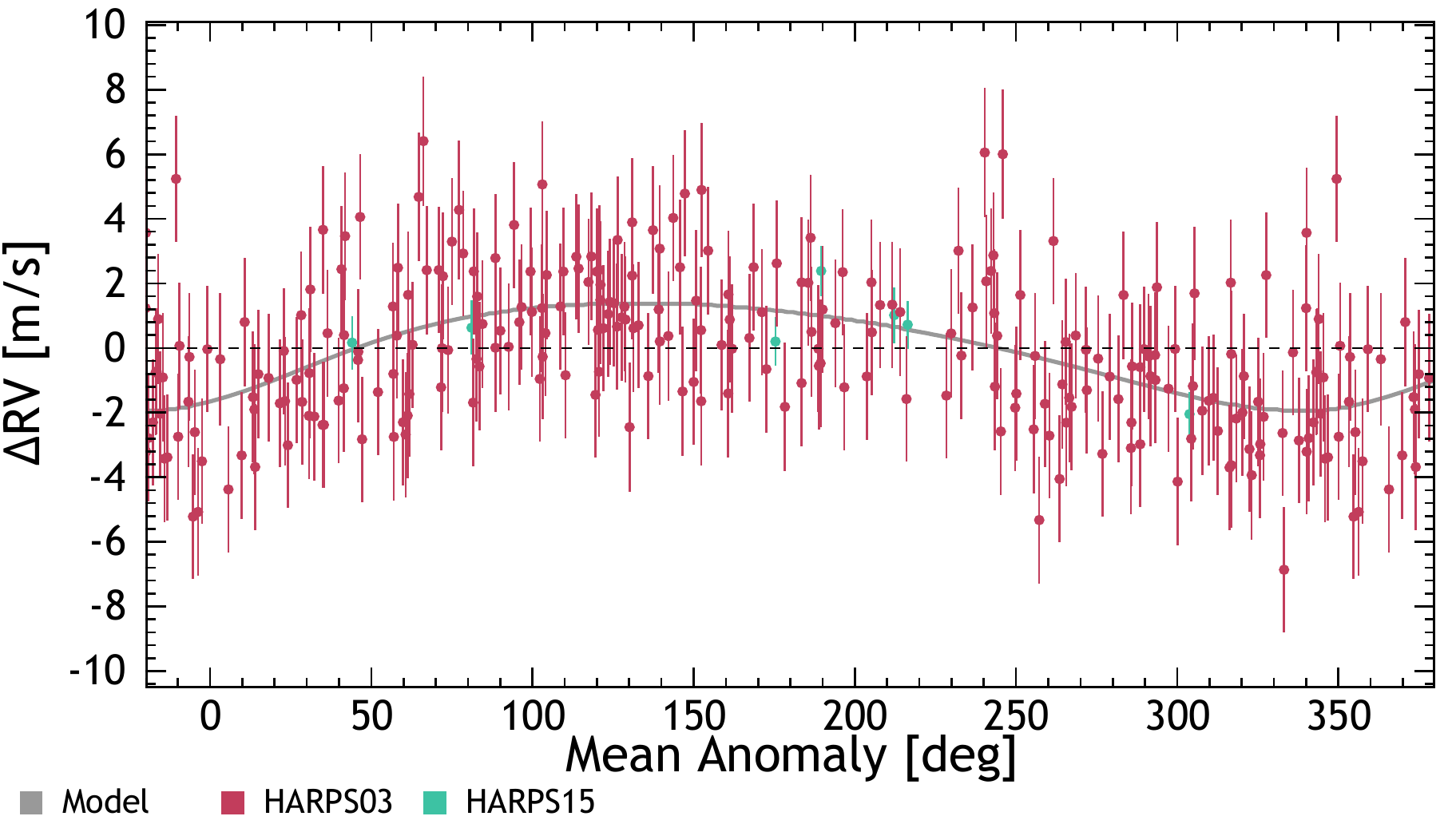}
            \end{overpic}
            
        \end{multicols}
    \end{minipage}
    \caption{Phase folded plots for each planet in HD\,189567. From left to right: planets b and c with orbital periods of 14.28 and 33.68 days, respectively.}
    \label{fig:HD189567-phasefold}
\end{figure*}

\begin{figure}
    \centering
    \includegraphics[width=0.9\columnwidth]{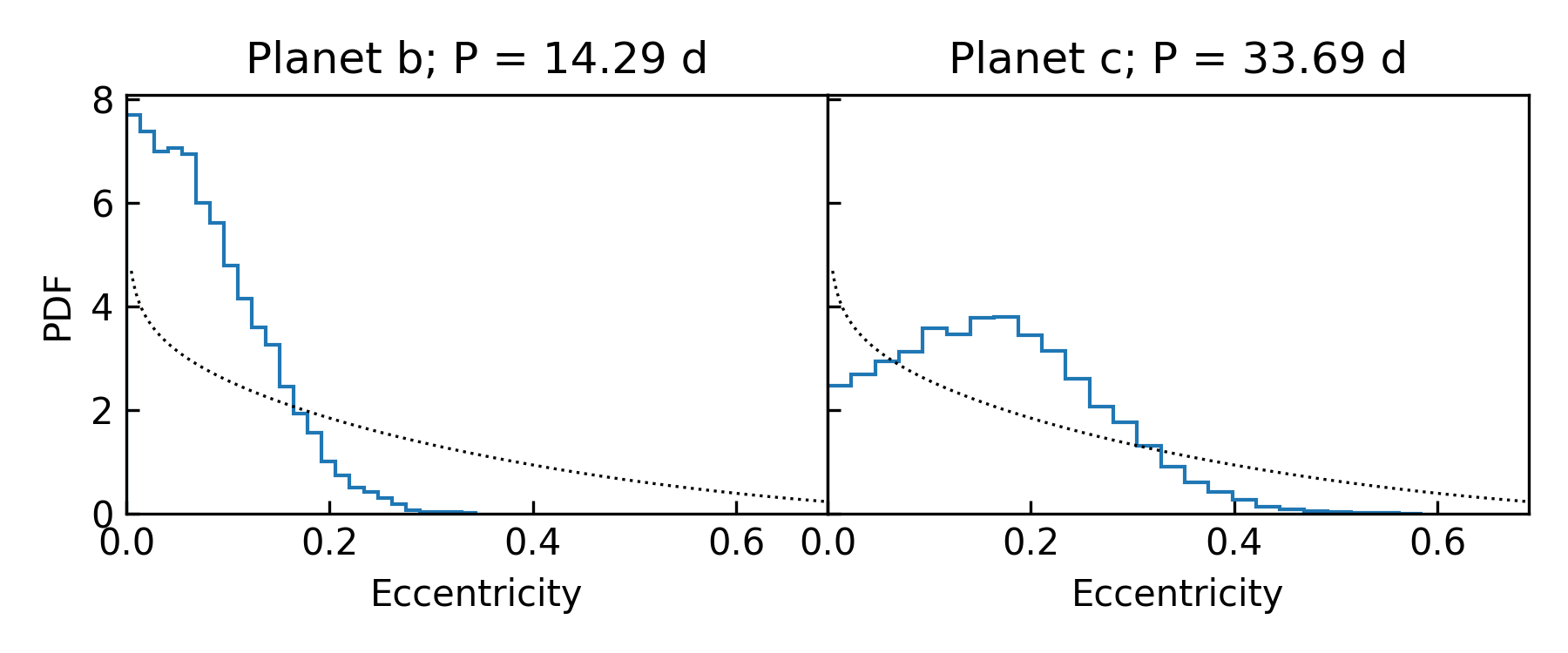}
    \caption{Posterior distribution of the eccentricities for the planetary companions of HD189567. The dotted line represents the eccentricity prior.}
    \label{fig:HD189567_eccs}
\end{figure}

HD\,189567 has been regularly observed since 2004 collecting more than 15 years of data. We removed four data points with low signal to noise ratio\footnote{We removed the spectra taken taken at these dates: BJD - 2\,400\,000 = 53545.842524, 53549.847373, 55370.738814, 56410.801187}. The activity indicator \rhk\ shows a periodic signal at 38.8 days, which is likely the star's rotation period. Also, the FWHM has a peak at 61.9 days.

Looking at the periodogram of the radial velocities time series, we immediately see a significant peak at 14.2 days. After fitting one Keplerian, the periodogram of the residuals shows a few more significant peaks at 1, 33.6, 37, 61, and 2600 days. The peak at 1 day is the one day alias of the 61 day signal, while the 37 day peak is the one year alias of the 33.7 day signal. The 61 day peak is precisely the period we saw in the FWHM, so we can suspect that this is an effect of stellar activity. To remove the 61-day peak, we fit a linear parameter with the FWHM and observe that we can improve the fit by applying a high pass filter using an Epanechnikov kernel at a timescale of 1.5 years to only keep the high frequencies. Then we add a second-order polynomial drift to remove the $\sim\!$~2600 day signal which is an artifact of the sampling of the data.

After fitting the linear parameter with the FWHM and the polynomial drift we are only left with a clear significant peak at 33.7 days with its one day (1 day) and one year (37 day) aliases at lower significance. We fit a keplerian at this 33.7 day period and are left with residuals with an RMS of 1.8 \ms\ and a peak at 18.6 days. This signal is weak though at a FAP level of $\sim\!2.5\%$.

The Bayesian model comparison analysis (see Table \ref{tab:odds_ratios}) shows that there is a substantial ($\ln(\mathcal{O}) > 20$) increase in evidence up to the model with two planets. Then it plateaus for three and four planets with a $\ln(\mathcal{O})$ of a bit less than 2, only weak evidence in their favor. This shows that the 18.6-day signal is not significantly detected.

HD\,189567 has then two planets at periods of 14.3 and 33.7 days and minimum masses of 8.8 and 7.2 \Mearth. The phase-folded RVs can be seen in Fig. \ref{fig:HD189567-phasefold} and the posterior distribution of the orbital parameters obtained with an MCMC can be seen in Table \ref{tab:HD189567-posterior}. Fig. \ref{fig:HD189567_eccs} shows the posterior distribution of the eccentricities of the planets from HD\,189567. Planet b's solution is compatible with a circular orbit, while planet c has a non zero eccentricity at $e=0.16\pm0.09$. Compared to the results obtained by \citeads{2011arXiv1109.2497M}, we find a lower semi-amplitude for planet b, from 3.02 to 2.53 \ms\ which reduced the minimum mass from 10.03 to 8.5 \Mearth. HD\,189567c is a new planet confirmation from this article.


\section{Discussion} \label{sec:discussion}

The new planetary systems presented in this paper have masses between the ones of Earth and Neptune. Planets in this mass range are often identified as Super-Earths or Sub-Neptunes. Out of the five stars of this paper, four of them host more than one planet. We also did not detect any massive planets like Jupiter in these systems. Any planet with a period smaller than 5 years and a mass larger than 20 \Mearth\ would have been detected at the precision of HARPS (i.e., semi-amplitudes of $\gtrsim\!1\ms$).

It is interesting to study these planetary systems in the context of synthetic planet populations. We made use of the Bern model of planet formation and evolution (\citeads{2005A&A...434..343A}; \citeads{2009A&A...501.1139M}, \citeyearads{2012A&A...547A.111M}, \citeyearads{2018haex.bookE.143M}) which uses the core accretion paradigm to synthesize planetary systems around solar-type stars. Specifically we used the New Generation Planetary Population Synthesis (NGPPS) from \citetads{2020arXiv200705561E}. 

\begin{figure}
    \centering
    \includegraphics[width=\columnwidth]{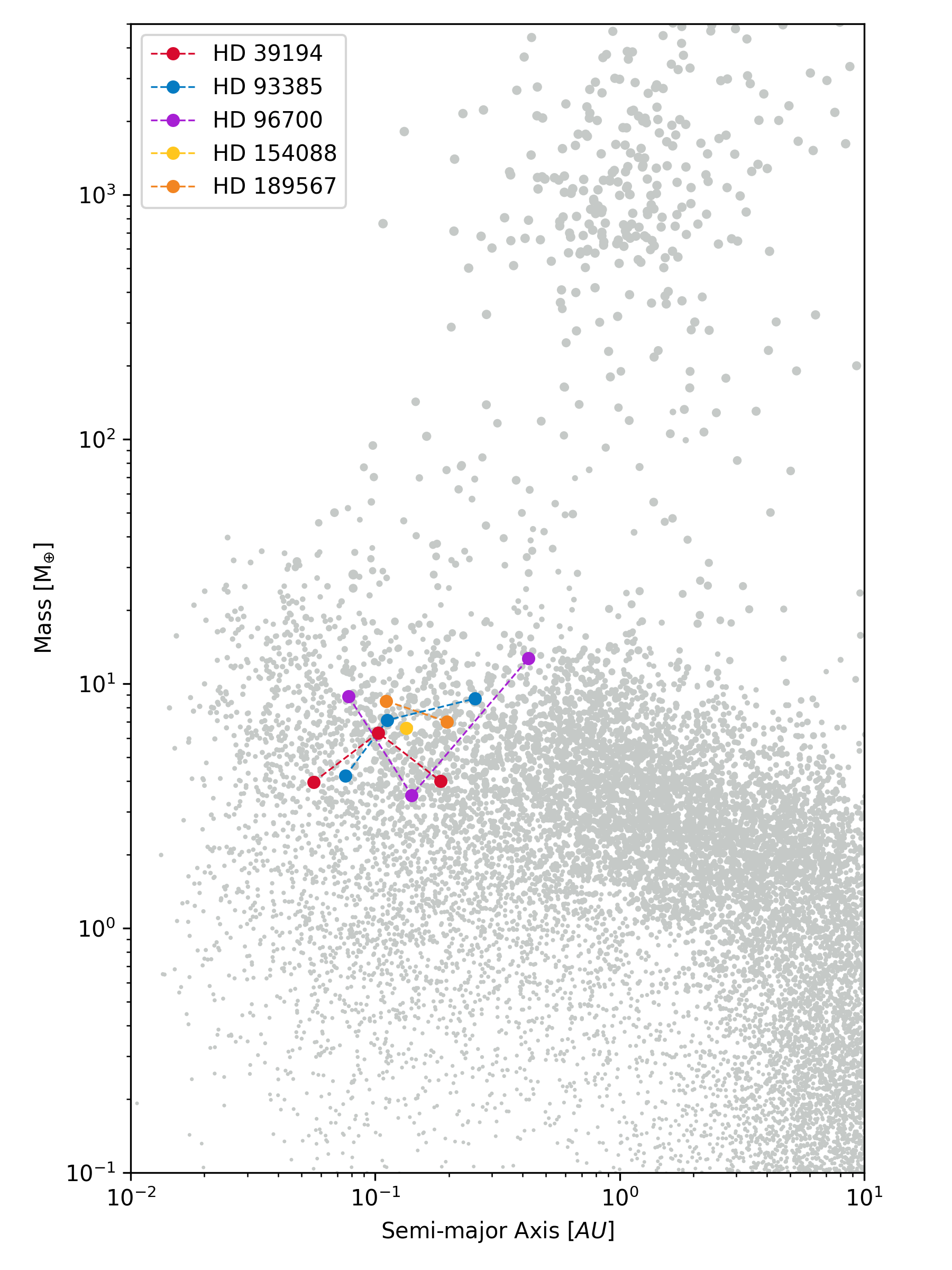}
    \caption{Mass versus semi-major axis diagram comparing the New Generation Planet Population Synthesis \citepads{2020arXiv200705561E} (in gray) to the planets presented in this paper (in color). For synthetic planets the size of the dot is proportional to the planet radius in a linear scale. For the mass of the planets presented in this paper we use their minimum masses.}
    \label{fig:ng76}
\end{figure}

In Fig. \ref{fig:ng76} we show the mass versus semi-major axis diagram of the synthetic planet population NG76. This specific synthetic population from the NGPPS was generated from 1000 systems around a 1 $M_\odot$ star, each starting with 100 lunar-mass embryos and simulated up to a time of 5 Gyr. Each gray dot on the plot is one of the planets present after the 5 Gyr simulation. On top we placed in color the planets presented in this article (using their minimum masses). These are all in the mid-range mass area (4 to 13 \Mearth) of close-in ($<0.5$ AU) Super-Earth planets.

Given the synthetic planet population of the NGPPS, we were interested in explaining how these planets could have been formed. From the five systems analyzed in this article we selected HD\,39194 and HD\,96700 to analyze their possible planet formation and evolution paths. Both systems present similar but inverted mass architectures, low-high-low mass as a function of orbital distance for HD\,39194 and high-low-high for HD\,96700. The NG76 planet population synthesis consists of 1000 planetary systems from where we selected two systems that have a similar planetary architecture to HD\,39194 and HD\,96700.

\begin{figure*} 
    \centering
    \begin{subfigure}[b]{0.48\textwidth}
        \centering
        \includegraphics[width=\textwidth]{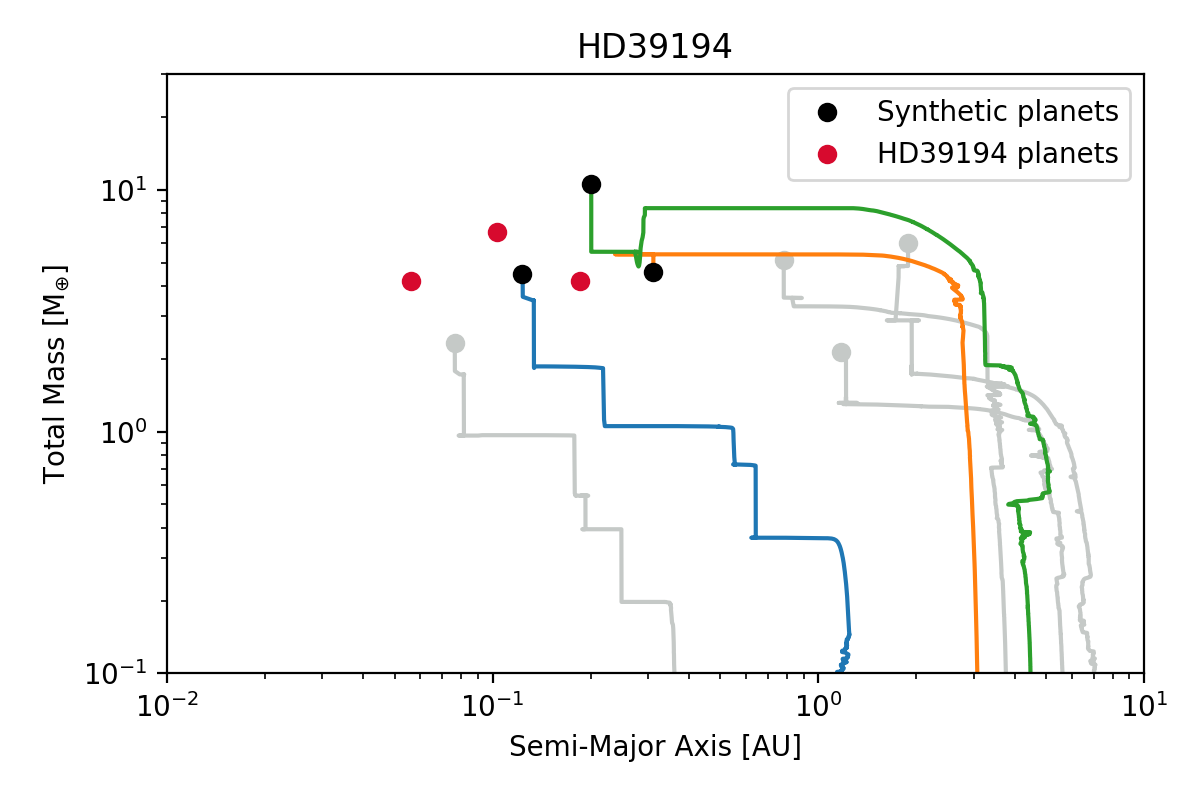}
        \label{fig:synth_HD39194}
    \end{subfigure}
    \hfill
    \begin{subfigure}[b]{0.48\textwidth}
        \centering
        \includegraphics[width=\textwidth]{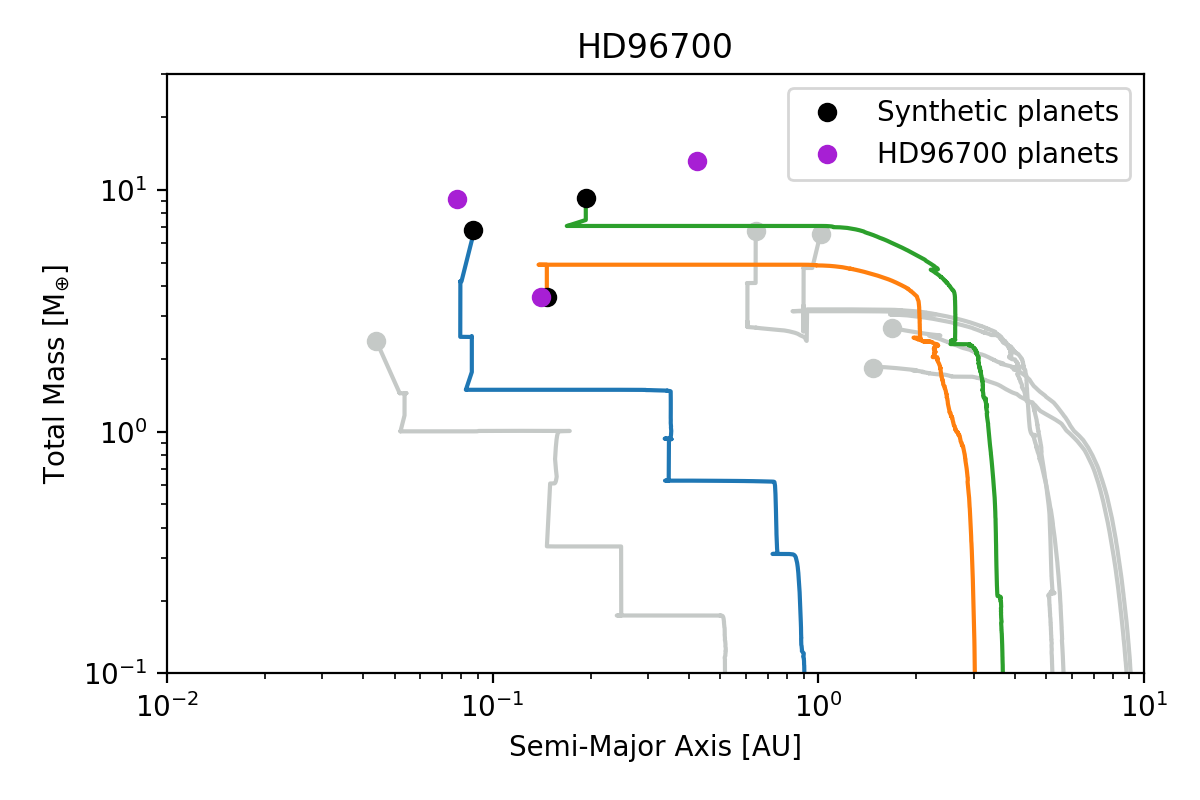}
        \label{fig:synth_HD96700}
    \end{subfigure}
    \vspace{-0.5cm}
    \caption{Planets presented in this paper for HD\,39194 (left) and HD\,96700 (right) compared with similar synthetic planet configurations from the NG76 population of the NGPPS, ID 126 (left) and ID 862 (right). We show in black the synthetic planets that are similar to the ones we found and in gray some of the other relevant synthetic planets of those systems, the rest were removed for clarity. The upward near-vertical steps seen in some tracks correspond to protoplanet-protoplanet collisions (giant impacts).}
    \label{fig:synthetic-vs-real}
\end{figure*}

In Fig. \ref{fig:synthetic-vs-real} we plot the minimum mass and semi-major axis of the real planets together with three planets and their formation tracks from two systems of the NG76 population of the NGPPS. These systems are labeled with ID 126 and 862 within the NG76, respectively. We note that the analysis we do in this section is done using the minimum masses of the real planets. The true masses are probably higher but the general architecture would stay the same if the orbits are coplanar.

There are basically two ways to form these types of planets, either by an initial mass accretion and a later inward migration or by in-situ accretion. From Fig. \ref{fig:synthetic-vs-real} we see a similar evolution in both systems. When the tracks suddenly jump up, it means a giant impact occurred where another less massive proto-planet is accreted. When the tracks go down, it means a loss of hydrogen and helium. This can occur either through envelope impact stripping or XUV-driven photoevaporation \citepads{2018ApJ...853..163J}.

The inner synthetic planets (shown with a blue track) have formed closer in, and both have no volatiles in their core which means that they formed inside the water ice line, which is close to 3 AU. They formed close to 1 AU and accreted solids while migrating in until 0.1 AU. As can be seen seen from the numerous near vertical steps in the blue tracks, these inner synthetic planets grew mainly from the accretion of other protoplanets, that is via giant impacts. The horizontal parts correspond to inward migration through parts of the disk where the planetesimals were already fully accreted by the protoplanets. Thus, no solid accretion occurs. Gas accretion also remains very inefficient because of the long Kelvin-Helmholtz cooling and contraction timescales of such low-mass cores \citepads{2012ApJ...753...66I}.

On the other hand, the outer synthetic planets (shown with orange and green tracks) started beyond 3 AU (just outside the water ice line) with an initial mass accretion of icy planetesimals, then a strong inward migration, and finally a gain or loss in mass. The outer synthetic planets migrate in with roughly similar masses. This mass scale, where migration becomes more important than planetesimal accretion (inward bending from vertical to horizontal tracks), is set by the condition that the growth and migration timescales cross. At lower masses, the planetesimal accretion timescale is shorter than the migration timescale. For higher masses, it is the opposite. This reflects that the oligarchic planetesimal accretion timescale is an increasing function of planet mass, whereas the Type I migration timescale is a decreasing function of it \citepads{2018haex.bookE.143M}.

After the migration, they separate in mass because one accretes another protoplanet and the other loses part of its atmosphere. In the system on the left, all planets still possess some H/He gas at the moment of disk dissipation. After 90 Myr, the envelope of the innermost planet is, however, completely evaporated. The outer two planets in contrast still bear H/He envelopes at 5 Gyr. The consequence is that the planets have significantly different radii, about 1.5, 2.5, and 5.3 $R_\oplus$. This system would thus have planets on both sides of the radius gap (\citeads{2017AJ....154..109F}; \citeads{2020A&A...638A..52M}). 

We can speculate that the formation history for HD\,39194 and HD\,96700 could have been similar to these synthetic systems. The fact that the latest end-to-end models in planet formation and evolution can explain a planetary architecture such as the one we see in HD\,96700 also gives us more confidence in the existence of the 19.88-day planet.

We note that the synthetic systems shown in Fig. \ref{fig:synthetic-vs-real} are just one example we used for comparison. A few other similar synthetic systems are also present in the NGPPS, showing that these are not isolated events.

\section{Conclusions} \label{sec:conclusions}

In this article we characterize the planetary structure of five systems with twelve planets in total. Three systems with three planets each, one with two planets and one with one planet. These planets are all in the Super-Earth and sub-Neptune mass regime with masses between 4 and 13 \Mearth.

These stars have low levels of activity which makes them easier to analyze and to find planetary signals. Using simple linear relations with activity indicators was enough to model some of the present activity. Long term magnetic cycles are easily removed by detrending the RVs with a smoothed time series of an activity indicator such as \rhk. This was very effective for HD\,39194 and HD\,154088. Short term activity effects can also be modeled with the same technique. For example HD\,189567 shows a 61 day signal in the periodogram which we know is caused by activity because we can see the same signal in the FWHM. So a simple detrending of the RVs with the high frequency components of the FWHM removes the 61 day signal, leaving the planetary signal very clear in the periodogram.

We implemented the use of the nested sampling algorithm \textsc{PolyChord} for Bayesian model comparison. Our interest was to confirm the number of planets in each system by estimating the Bayesian evidence of models with different number of planets (from 0 up to 4 planets). This analysis showed us that this technique is very useful to get a clear and robust answer on the planet population of each system. All planets were confirmed with log odds ratios greater than 10 with respect to the models with one fewer planet.

For HD\,96700 we also compared different noise models and saw that some signals are still difficult to model accurately. This shows us that great efforts are still required in the entire pipeline to improve the reliability and accuracy of radial velocity measurements, from spectral extraction and data reduction up to data analysis.

Finally we used the synthetic planet population generated by the Bern model (NGPPS) to explain possible formation paths for HD\,39194 and HD\,96700. We found that the inner planet  probably formed  from inside the water ice-line mainly from giant impacts with other protoplanets, while the outer planets formed beyond the water ice-line with an initial mass accretion dominated by planetesimal accretion and a subsequent Type I inward migration. Additionally, we see that in these synthetic systems one of these outer synthetic planets has a final giant impact where it gains mass through growth of the solid core, while the other suffers from mass reduction by XUV-driven escape. This divergence in mass leads to the different architectural patterns in terms of the mass as a function of orbital distance in multiplanet systems.


\begin{acknowledgements}
This work has been carried out within the framework of the National Centre for Competence in Research PlanetS supported by the Swiss National Science Foundation. We acknowledge the financial support of the SNSF. \\

This publication makes use of the Data \& Analysis Center for Exoplanets (DACE), which is a facility based at the University of Geneva (CH) dedicated to extrasolar planets data visualization, exchange and analysis. DACE is a platform of the Swiss National Centre of Competence in Research (NCCR) PlanetS, federating the Swiss expertise in Exoplanet research. The DACE platform is available at \url{https://dace.unige.ch}.

\end{acknowledgements}

%
\bibliographystyle{aa} 
\bibliography{ads_bibliography.bib,non_ads_bibliography.bib} 
%

\begin{appendix}

\section{Bayesian Inference} \label{app:bayes-inference}

To evaluate the power of a model or hypothesis, we can make use of Bayes formula, which states:

\begin{equation} \label{eq:bayestheorem}
    p\left(\theta_{\mathcal{M}_i} \mid \mathcal{D}, \mathcal{M}_i\right) = \frac{p\left(\mathcal{D} \mid \theta_{\mathcal{M}}, \mathcal{M}_i\right) p\left(\theta_{\mathcal{M}} \mid \mathcal{M}_i\right)}{p(\mathcal{D} \mid \mathcal{M}_i)} \enspace,
\end{equation}
where $\theta_{\mathcal{M}}$ are the parameters of the model.

Bayes theorem relates the probability of the data given the model $p\left(\mathcal{D} \mid \theta_{\mathcal{M}}, \mathcal{M}_i\right) = \mathcal{L}$ also called Likelihood, with the prior probability $p\left(\theta_{\mathcal{M}} \mid \mathcal{M}_i\right) = \mathcal{\pi}$ and the evidence ${p(\mathcal{D} \mid \mathcal{M}_i) = \mathcal{Z}}$ to give us the posterior probability for the parameters $p\left(\theta_{\mathcal{M}_i} \mid \mathcal{D}, \mathcal{M}_i\right)$. In the parameter estimation case, the evidence $\mathcal{Z}$ is just a normalization constant and is defined as the integral of the product of the Likelihood and the prior over the entire parameter space:

\begin{equation} \label{eq:evidence}
    \mathcal{Z} = \int p\left(\mathcal{D} \mid \theta_{\mathcal{M}}, \mathcal{M}_i\right) p\left(\theta_{\mathcal{M}} \mid \mathcal{M}_i\right) d\theta = \int \mathcal{L}(\theta) \pi(\theta) d\theta \enspace .
\end{equation}

This quantity can usually be ignored but it is fundamental in model comparison. To calculate the posterior probability of the model itself, Bayes formula takes the following form:

\begin{equation}\begin{aligned}
    p\left(\mathcal{M}_{i} \mid \mathcal{D}\right) &=\frac{p\left(\mathcal{D} \mid \mathcal{M}_{i}\right) p\left(\mathcal{M}_{i}\right)}{p(\mathcal{D})} \\
    &=\frac{\mathcal{Z}_{i} \pi_{i}}{\sum_{j} \mathcal{Z}_{j} \pi_{j}} \enspace .
\end{aligned}\end{equation}

The posterior probability of two models $\mathcal{M}_1$ and $\mathcal{M}_2$ can then be compared by taking their ratio resulting in what is called the \textit{odds ratio} \citepads{2010blda.book.....G}:

\begin{equation}\begin{aligned} \label{eq:oddsratio_appendix}
\mathcal{O}_{12} &=\frac{p\left(\mathcal{M}_{1} \mid \mathcal{D}\right)}{p\left(\mathcal{M}_{2} \mid \mathcal{D}\right)} =\frac{p\left(\mathcal{D} \mid \mathcal{M}_{1}, I\right) p\left(\mathcal{M}_{1}\right)}{p\left(\mathcal{D} \mid \mathcal{M}_{2}, I\right) p\left(\mathcal{M}_{2}\right)}=\frac{\mathcal{Z}_{1} \pi_{1}}{\mathcal{Z}_{2} \pi_{2}} \\
&=\mathcal{B}_{12} \frac{\pi_{1}}{\pi_{2}} \enspace ,
\end{aligned}\end{equation}
where $\mathcal{B}_{12}$ is ratio of the evidence values between models 1 and 2, also called the Bayes factor. $\pi_{1}/\pi_{2}$ is the prior probability ratio which is usually set to 1 if there is no prior information to suggest that one model is preferred over the other.

Because the Likelihood usually has very low values that can reach the double precision floating point limit, it is common practice to work with the natural logarithm of the Likelihood and thus the logarithm of the evidence $\mathcal{Z}$. We can then rewrite Eq.~\ref{eq:oddsratio} into

\begin{equation}
    \ln{\mathcal{O}_{12}} = \ln{\frac{\mathcal{Z}_{1}\pi_{1}}{\mathcal{Z}_{2}\pi_{1}}} = \ln{\mathcal{Z}_{1}} - \ln{\mathcal{Z}_{2}} + \ln{\frac{\pi_{1}}{\pi_{2}}} \enspace ,
\end{equation}
and make use of the Jeffreys scale (see Table~\ref{tab:jeffscale}) to decide if model 1 is preferred by the data over model 2. A final note is that Bayesian model comparison has a built-in Occam's Razor which automatically penalizes models with too many parameters, only giving them a high probability if the data justifies the complexity of the model. A more detailed description can be found in \citetads[][chap.3]{2010blda.book.....G}.


\subsection{Likelihood}

The log-Likelihood is defined as follows:

\begin{multline}
    \ln \mathcal{L} (\pmb{\theta}) = -\frac{n_{\mathrm{obs}}}{2} \ln(2 \pi) - \frac{1}{2} \ln(|\det \Sigma|) \\ 
    - \frac{1}{2} (\pmb{v} - \pmb{v}_{\mathrm{pred}}(\pmb{\theta}))^T \Sigma^{-1}  (\pmb{v} - \pmb{v}_{\mathrm{pred}}(\pmb{\theta})) \enspace ,
\end{multline}
where $\pmb{\theta}$ is the vector of parameters, $n_{\mathrm{obs}}$ is the total number of observations, $\Sigma$ is the covariance matrix of the data, $\pmb{v}$ is the vector of measurements and $\pmb{v}_{\mathrm{pred}}$ is the predicted model.


\section{Technical notes about \textsc{PolyChord}} \label{app:polychord}

\textsc{PolyChord} \citepads{2015MNRAS.453.4384H} utilizes Slice Sampling \citepads{2000physics...9028N} to find new live points within the iso-likelihood contour which works by using a Markov Chain Monte Carlo (MCMC) procedure. \citetads{2015MNRAS.453.4384H} claim that this procedure is well suited for nested sampling because it samples uniformly and can be adapted to high dimensional Likelihoods.

\textsc{PolyChord} was designed to work well with high dimensional models, however, \citetads{2020AJ....159...73N} showed the difficulty that arises in Bayesian model comparison when high dimensional models are considered. They show the diversity of the several algorithms that exist for this and how they can give different answers to the same problem. We tested \textsc{PolyChord} on the same simulated datasets from \citetads{2020AJ....159...73N} and found similar results to the other nested samplers. It is reassuring, that the nested samplers mostly agreed in their results. So special care has to be taken when calculating the Bayesian evidence to ensure that convergence has been reached.

We implemented \textsc{PolyChord} using the Python wrapper provided by the developers. The Likelihood and priors are mostly written in Python with the exception of the true anomaly calculation (see Sect. \ref{sec:keplerians}) which we wrote in C for optimized run time. The true anomaly is one of the most computationally intense calculations of the Likelihood function.

\subsection{Tuning parameters and uncertainties}

\textsc{PolyChord} has a few tuning parameters, the most important one being the number of live points. In a nutshell, the more live points that are used, the better the sampling will be but this also comes with a higher computational cost. A balance has to be found where the sampling is good enough but the computational cost is not too high. 

In \citetads{2015MNRAS.453.4384H} the authors propose to use a number of live points equal to 25 times the number of free parameters in the model. In all \textsc{PolyChord} runs we used 50 times the number of free parameters in the model as the number of live points. This is double the recommended amount but low amplitude radial velocity problems ($\lesssim\!$ 10 \ms) can have highly multi-modal Likelihoods and experience has shown us that we need a high number of live points to properly explore the entire parameter space. A lower amount of live points leads to a high variance in the estimation of the evidence.

The number of live points is the only tuning parameter that we changed. We kept the rest at their default values as we noticed that these worked well for our purposes. We did however calculate our own estimate for the uncertainty on the value of the evidence. Even though \textsc{PolyChord} provides a value for the error of the evidence we noticed that this value is usually underestimated. By running several identical runs of a particular model we saw that the spread in evidence values we got was larger than the reported error by \textsc{PolyChord}. To take this into account we ran each model three times and took the median and standard deviation of those runs as our estimate of the evidence.

\subsection{Run time}

A few notes about the run time and computational resources used for the \textsc{PolyChord} runs. \textsc{PolyChord} can be parallelized with a primary and secondary core structure using the openMPI architecture \citepads{2004epmu.meet...97G}. We ran all calculations in the \textit{lesta} cluster of the Observatory of Geneva and each simulation was launched on one entire node containing 32 cores of an Intel(R) Xeon(R) Gold 5218 CPU @ 2.30GHz. 

The total run time for each model depends on the number of planets included in the model as this adds an additional five parameters to the model and thus also significantly increases the total number of live points. Simple models (0 and 1 planets) take just a few minutes to complete, while the more complicated models of 4 planets can take up to 15 hours to complete.


\newpage

\section{Periodograms}

\begin{figure*}[] 
\centering
\caption[]{\tabular[t]{p{0.93\textwidth}lp{0.95\textwidth}lp{0.95\textwidth}}\textit{Left:} Periodogram of the RV residuals of HD39194, after sequentially removing, from top to bottom, the instrumental RV offsets (free parameters), the 14.03 d (resp. 5.63 d and 33.91 d) Keplerian signals and a third-order drift plus a linear term with \rhk . \\ \textit{Right:} Periodogram of the RV residuals of HD93385, after sequentially removing, from top to bottom, the instrumental RV offsets (free parameters), the 13.18 d (resp. 45.84 d and 7.34 d) Keplerian signals. \\ False alarm probability (FAP) thresholds are shown as horizontal lines for FAP=10$\%$, 1$\%$ and 0.1$\%$. \endtabular}

\label{fig:HD39194&HD93385-FAP}%
\begin{minipage}{\textwidth}%
    \centering
    \vspace{0.6cm}
    \begin{multicols}{2}%
        \begin{overpic}[width=0.49\textwidth]{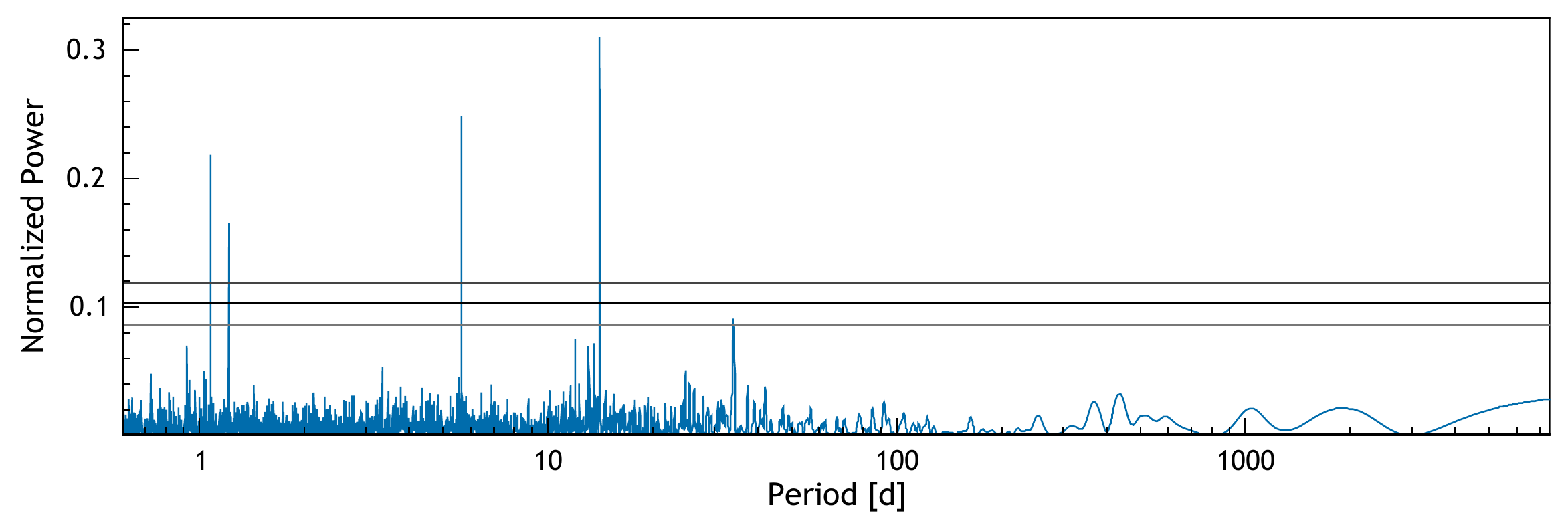}
            \put (45,35) {\bf HD\,39194 }
            \put (24,29) {\tiny $14.03$~d $\displaystyle\rightarrow$  }
            \put (68,28) {\tiny After offset subtraction}
        \end{overpic}\\
        \vspace{0.1cm}
        \begin{overpic}[width=0.49\textwidth]{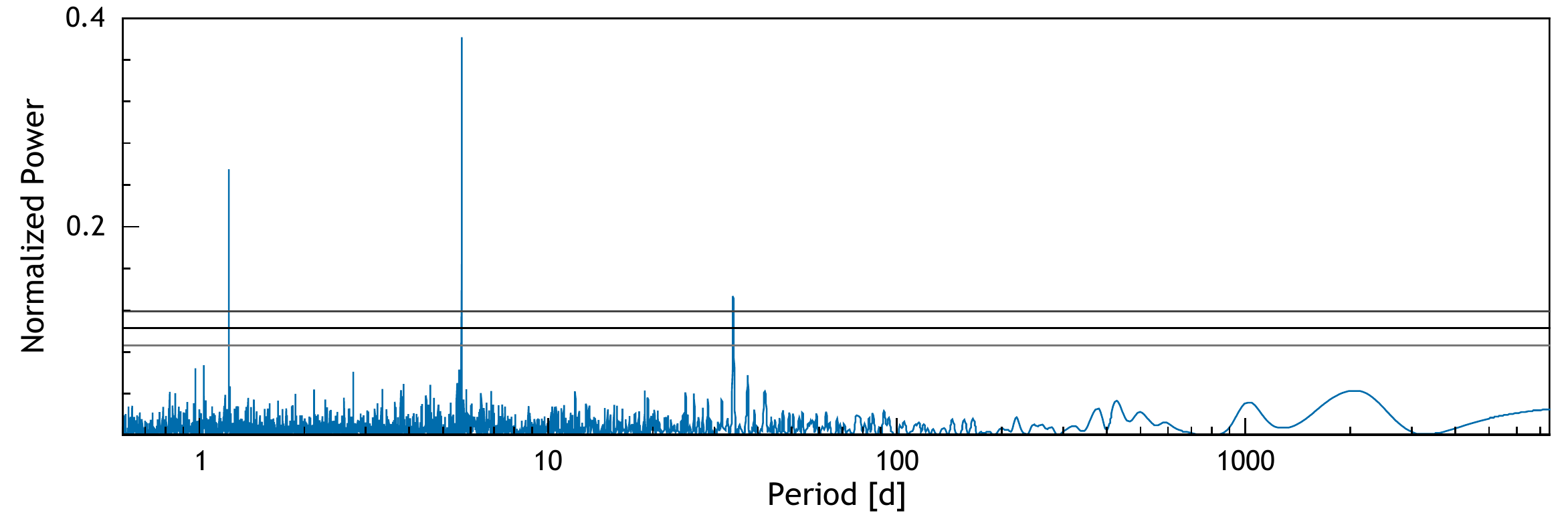}
            \put (17,29) {\tiny $5.63$~d $\displaystyle\rightarrow$ }
            \put (75,26) {\parbox{3cm} {\tiny After 14.03 d \\ signal subtraction}}
        \end{overpic}\\
        \vspace{0.1cm}
        \begin{overpic}[width=0.49\textwidth]{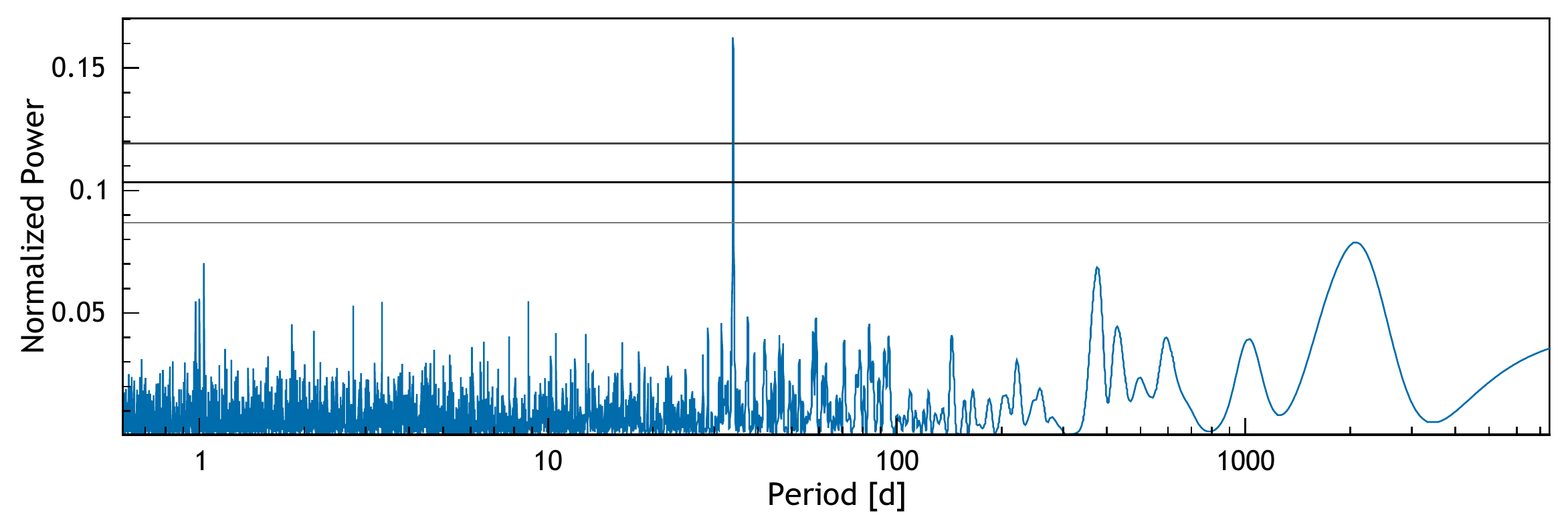}
            \put (32,29) { \tiny $33.91$~d $\displaystyle\rightarrow$ }
            \put (71,27) {\parbox{3cm} {\tiny After 14.03 d, 5.63 d \\ signal subtraction}}
        \end{overpic}
        \vspace{0.1cm}
        \begin{overpic}[width=0.49\textwidth]{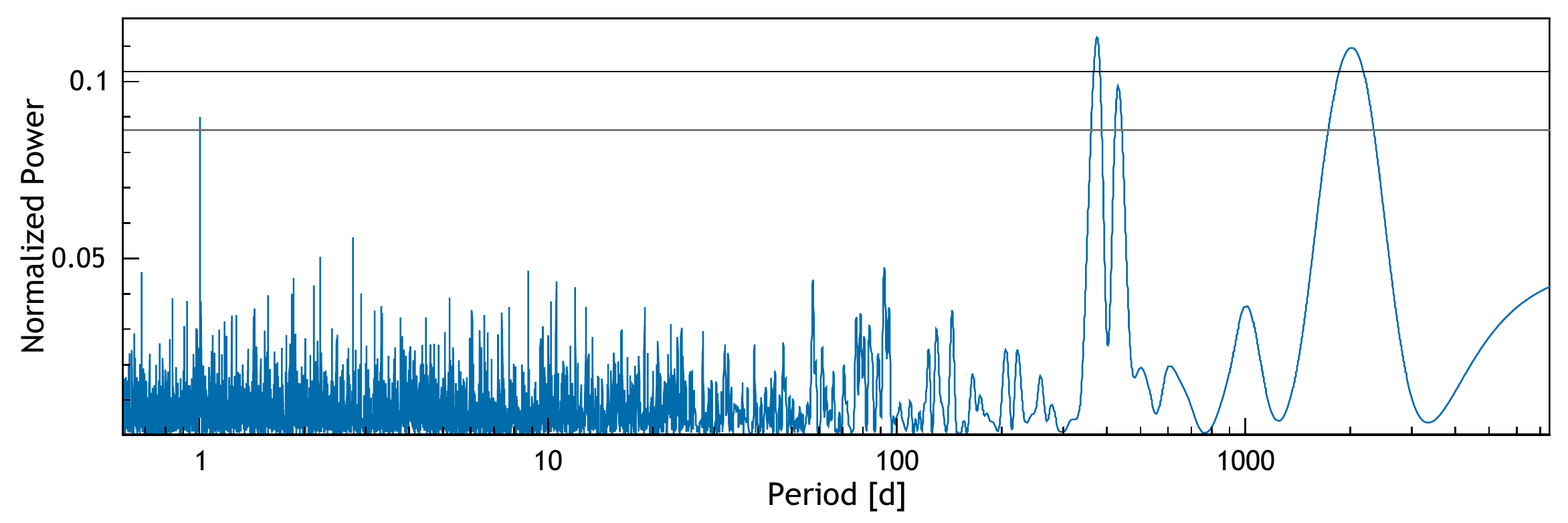}
            \put (15,27.7) {\parbox{3cm} {\colorbox{white}{\tiny After 14.03 d, 5.63 d, }}}
            \put (15,24) {\parbox{3cm} {\colorbox{white}{\tiny 33.91 d signal subtraction }}}
        \end{overpic}
        \begin{overpic}[width=0.49\textwidth]{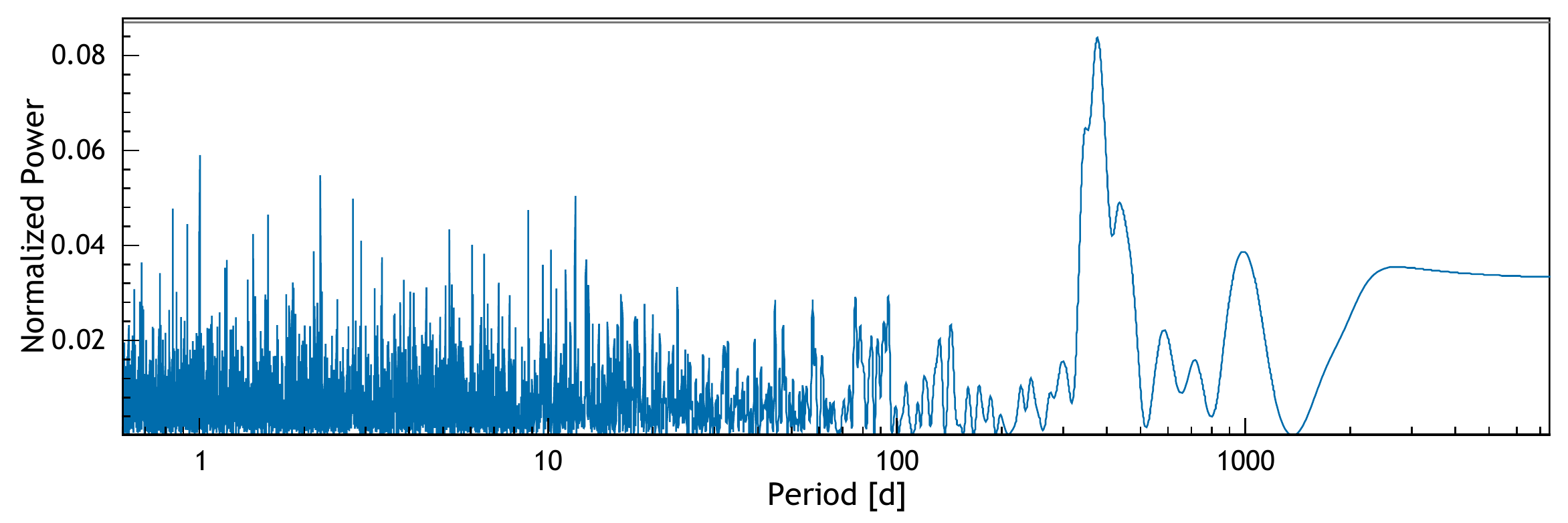}
            \put (15,27.7) {\parbox{3cm} {\colorbox{white}{\tiny After 14.03 d, 5.63 d, 33.91 d and drift }}}
            \put (15,23.7) {\parbox{3cm} {\colorbox{white}{\tiny + lin \rhk\ signal subtraction }}}
        \end{overpic}
        
        \columnbreak
        
        \begin{overpic}[width=0.49\textwidth]{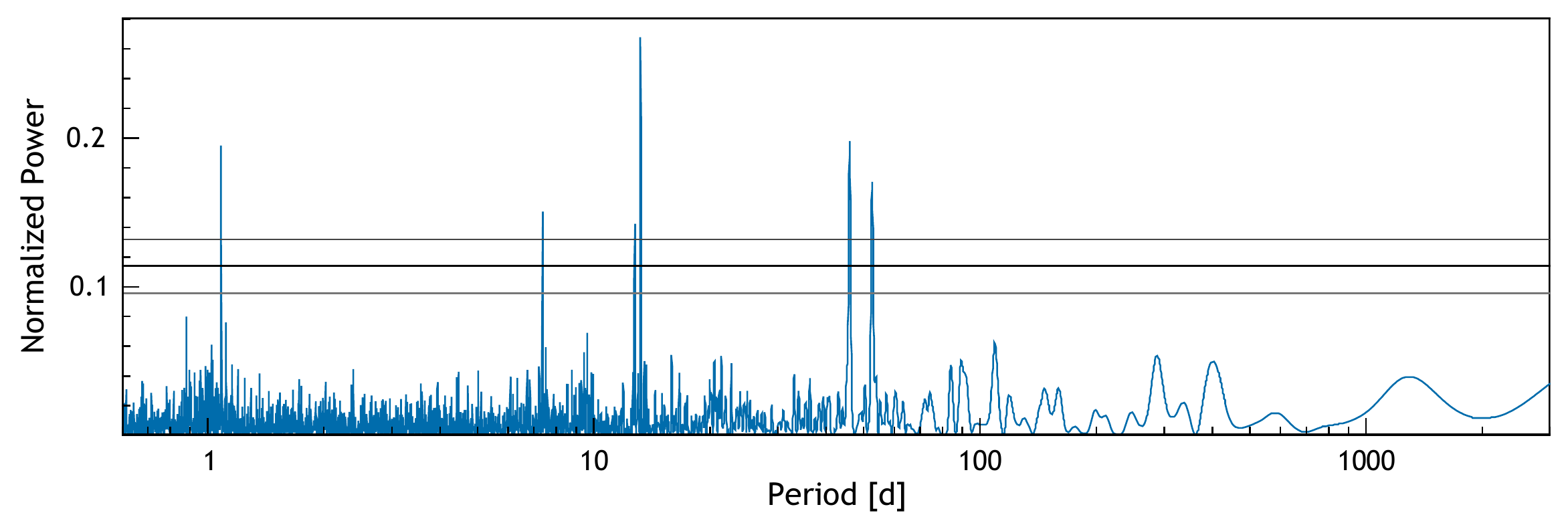}
            \put (45,35) {\bf HD\,93385 }
            \put (26,29) {\tiny $13.18$~d $\displaystyle\rightarrow$}
            \put (68,28) {\tiny After offset subtraction}
        \end{overpic}\\
        \vspace{0.1cm}
        \begin{overpic}[width=0.49\textwidth]{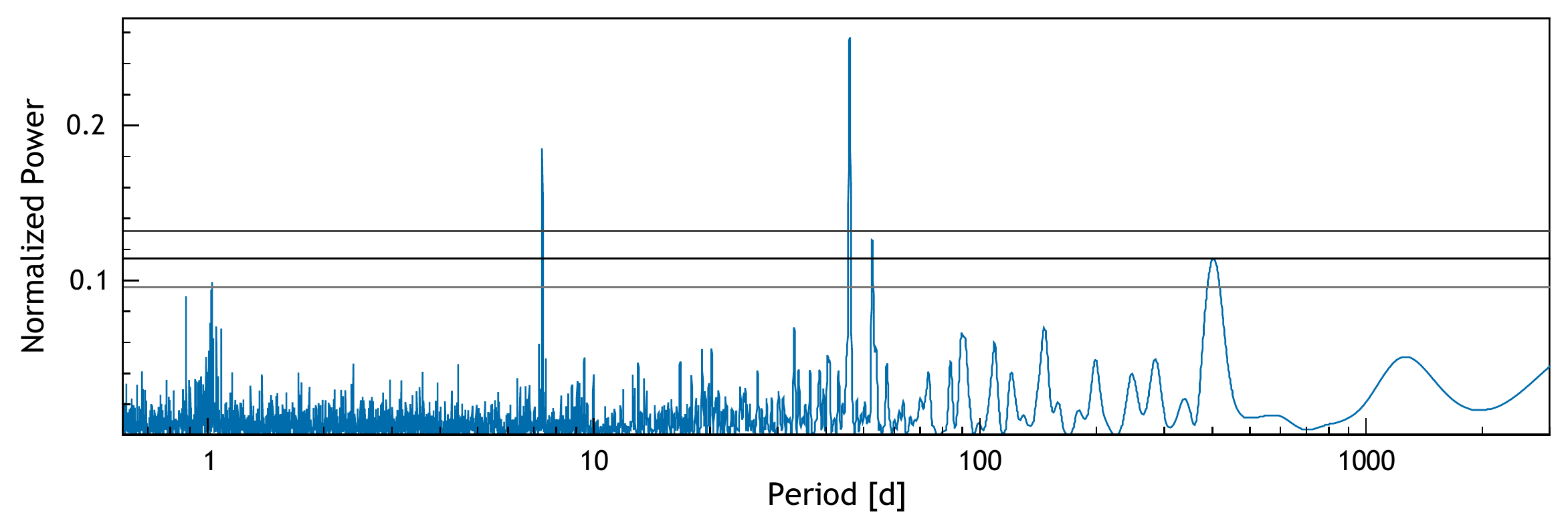}
            \put (39,29) {\tiny $45.84$~d $\displaystyle\rightarrow$ }
            \put (75,26) {\parbox{3cm} {\tiny After 13.18 d \\ signal subtraction}}
        \end{overpic}\\
        \vspace{0.1cm}
        \begin{overpic}[width=0.49\textwidth]{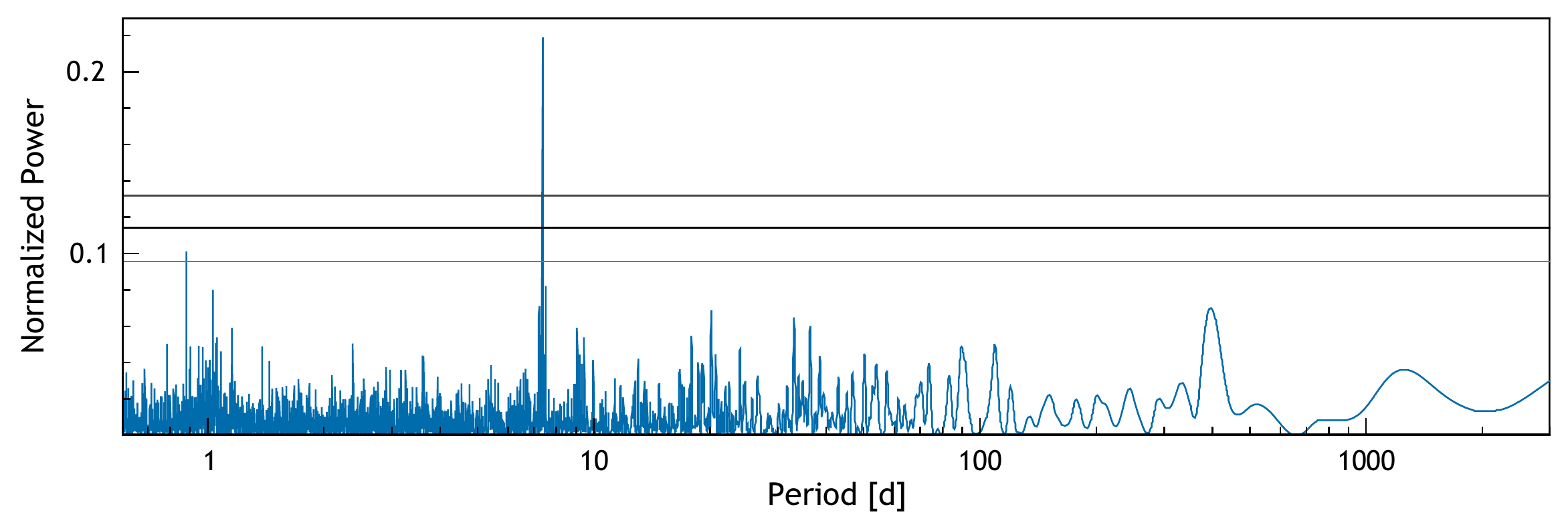}
            \put (21,29) { \tiny $7.34$~d $\displaystyle\rightarrow$ }
            \put (69,26) {\parbox{3cm} {\tiny After 13.18 d, 45.84 d \\ signal subtraction}}
        \end{overpic}
        \vspace{0.1cm}
        \begin{overpic}[width=0.49\textwidth]{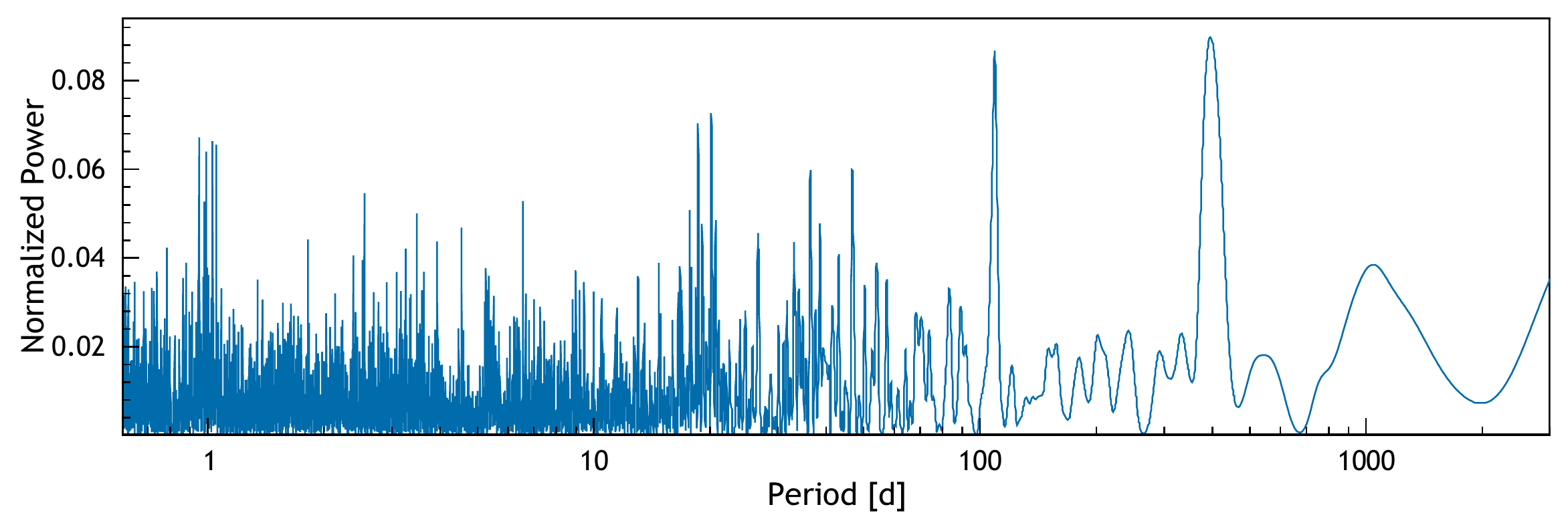}
            \put (13,27.7) {\parbox{3cm} {\colorbox{white}{\tiny After 13.18 d, 45.84 d, 7.34 d }}}
            \put (13,24) {\parbox{3cm} {\colorbox{white}{\tiny signal subtraction }}}
        \end{overpic}
    \end{multicols}
\end{minipage}
\end{figure*}

\newpage

\begin{figure*}[] 
\centering
\caption[]{\tabular[t]{p{0.93\textwidth}lp{0.95\textwidth}lp{0.95\textwidth}}\textit{Left:} Periodogram of the RV residuals of HD96700, after sequentially removing, from top to bottom, the instrumental RV offsets, the 8.12 d (resp. 103.5 d and 19.88 d) Keplerian signals. \\ \textit{Right:} Periodogram of the RV residuals of HD154088, after sequentially removing, from top to bottom, the instrumental RV offsets, the linear dependence with \rhk, and the 18.56 d Keplerian signal. \\ False alarm probability (FAP) thresholds are shown as horizontal lines for FAP=10$\%$, 1$\%$ and 0.1$\%$. \endtabular}

\label{fig:HD96700&HD154088-FAP}%
\begin{minipage}{\textwidth}%
    \centering
    \vspace{0.6cm}
    \begin{multicols}{2}%
        \begin{overpic}[width=0.49\textwidth]{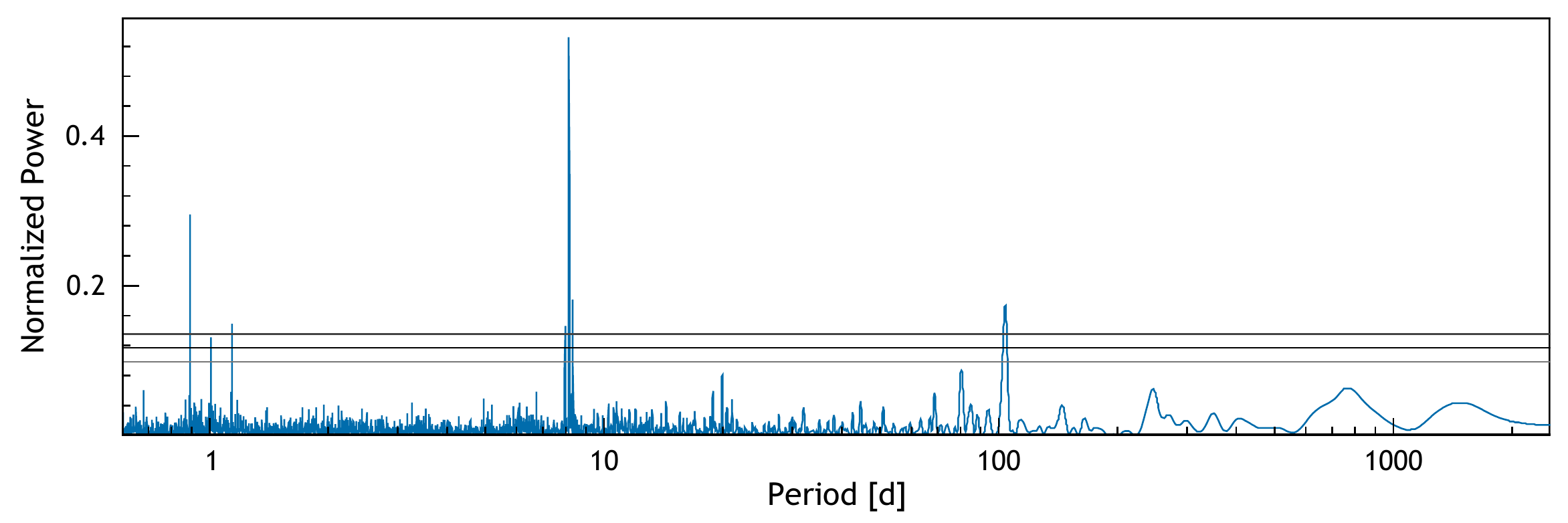}
            \put (45,35) {\bf HD\,96700 }
            \put (23,29) {\tiny $8.12$~d $\displaystyle\rightarrow$  }
            \put (68,28) {\tiny After offset subtraction}
        \end{overpic}\\
        \vspace{0.1cm}
        \begin{overpic}[width=0.49\textwidth]{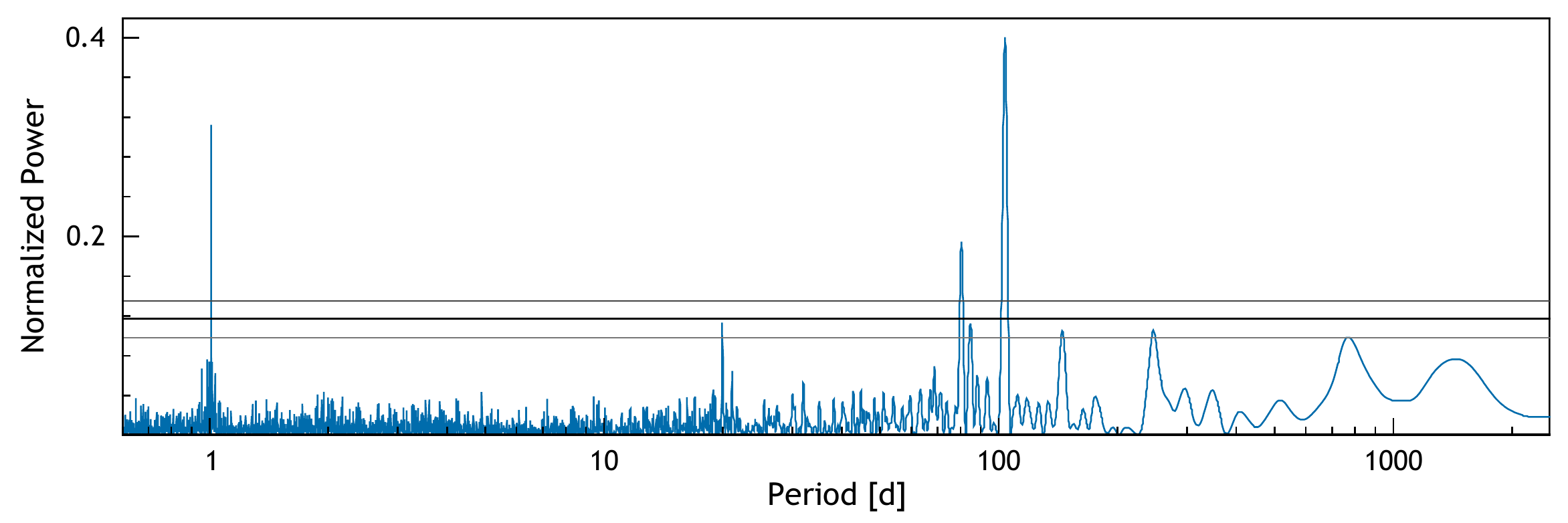}
            \put (50,29) {\tiny $103.5$~d $\displaystyle\rightarrow$ }
            \put (75,26) {\parbox{3cm} {\tiny After 8.12 d \\ signal subtraction}}
        \end{overpic}\\
        \vspace{0.1cm}
        \begin{overpic}[width=0.49\textwidth]{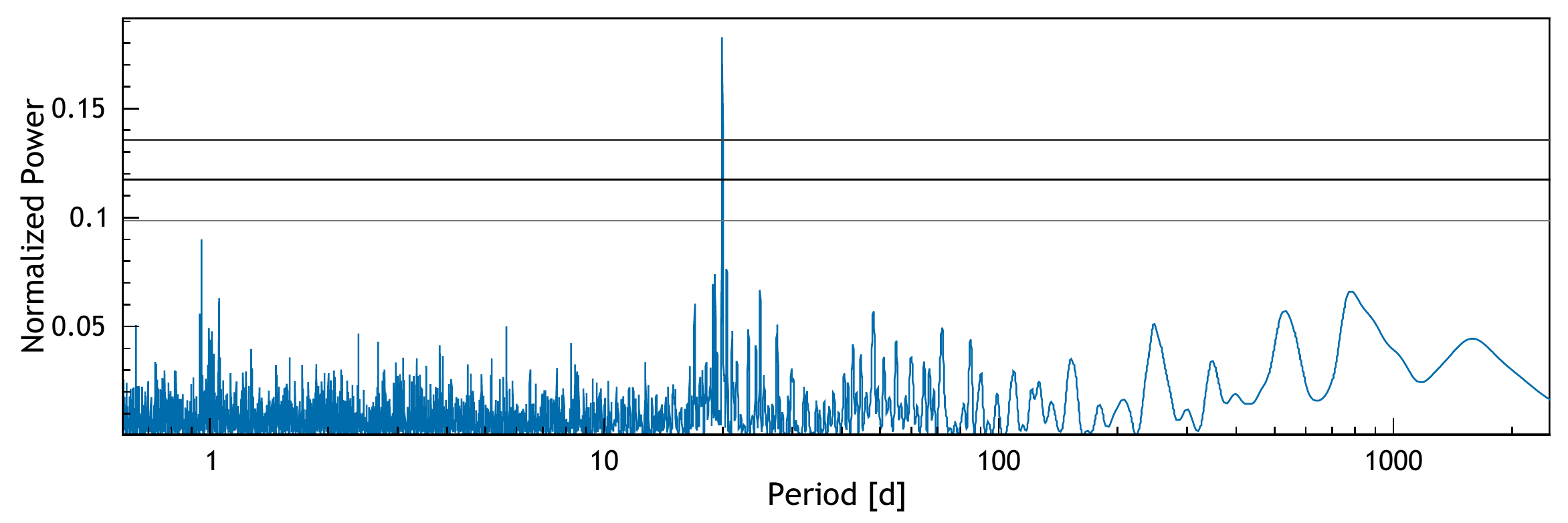}
            \put (30,29) { \tiny $19.88$~d $\displaystyle\rightarrow$ }
            \put (71,27) {\parbox{3cm} {\tiny After 8.12 d, 103.5 d \\ signal subtraction}}
        \end{overpic}
        \vspace{0.1cm}
        \begin{overpic}[width=0.49\textwidth]{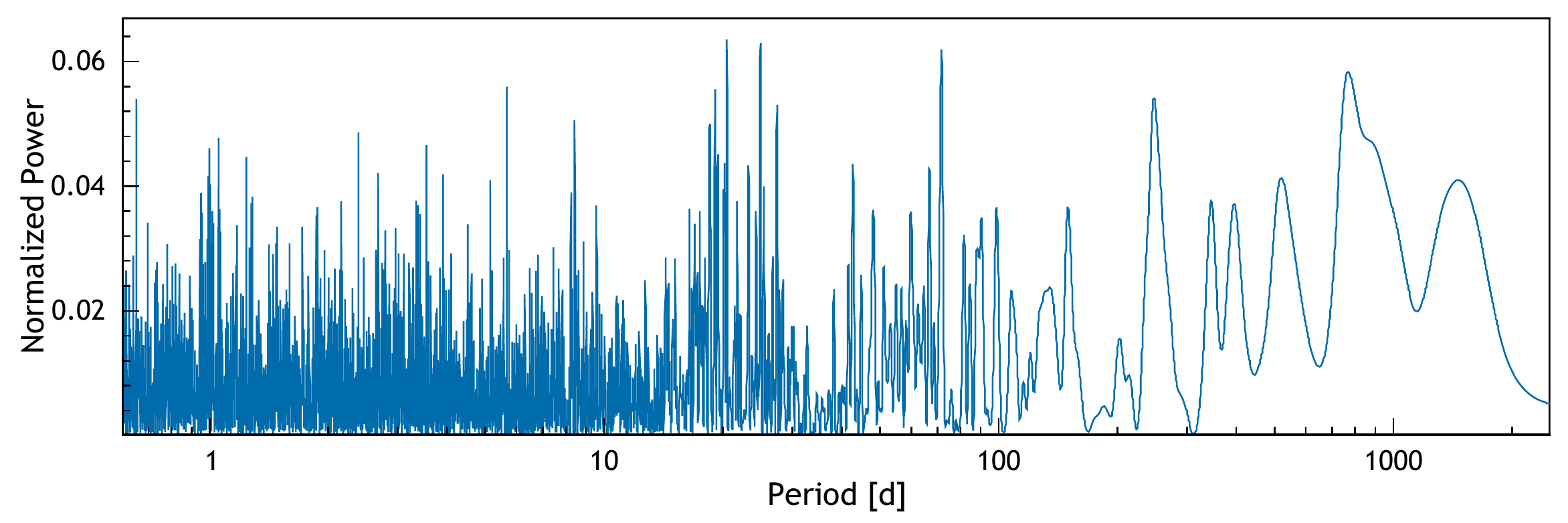}
            \put (10,27.7) {\parbox{3cm} {\colorbox{white}{\tiny After 8.12 d, 103.5 d, }}}
            \put (10,24) {\parbox{3cm} {\colorbox{white}{\tiny 19.88 d signal subtraction }}}
        \end{overpic}
        
        \columnbreak
        
        \begin{overpic}[width=0.49\textwidth]{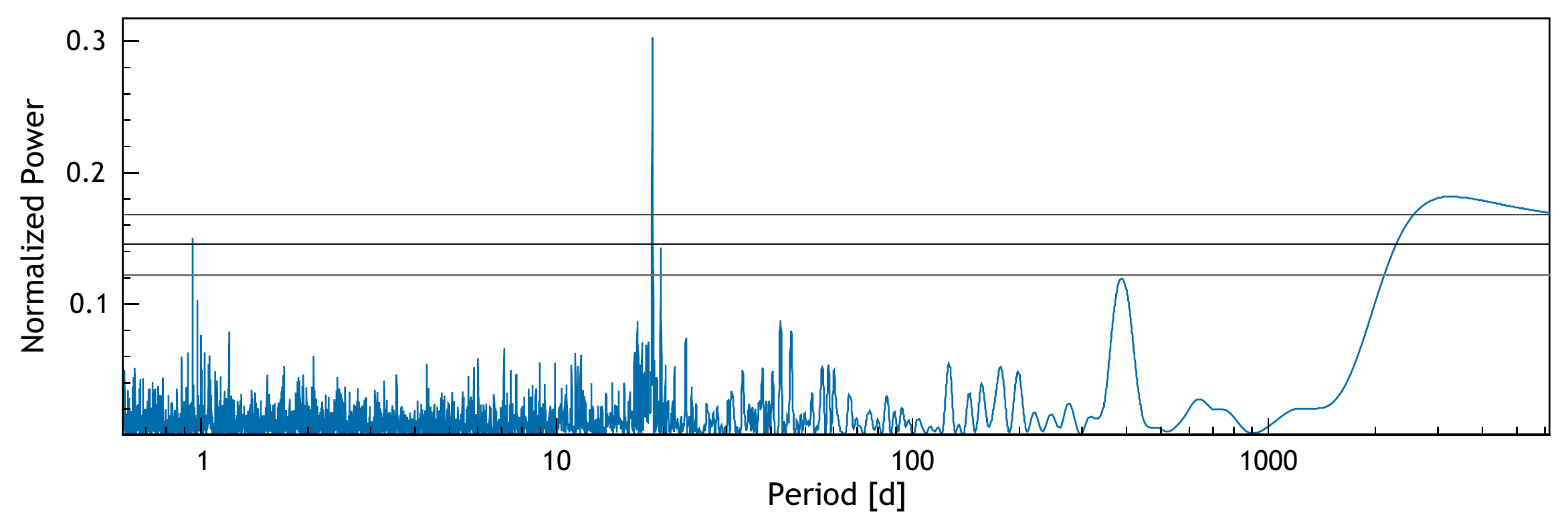}
            \put (45,35) {\bf HD\,154088 }
            \put (67,23) {\tiny Magnetic cycle}
            \put (87,21.5) {\tiny $\displaystyle\searrow$}
            \put (68,28) {\tiny After offset subtraction}
        \end{overpic}\\
        \vspace{0.1cm}
        \begin{overpic}[width=0.49\textwidth]{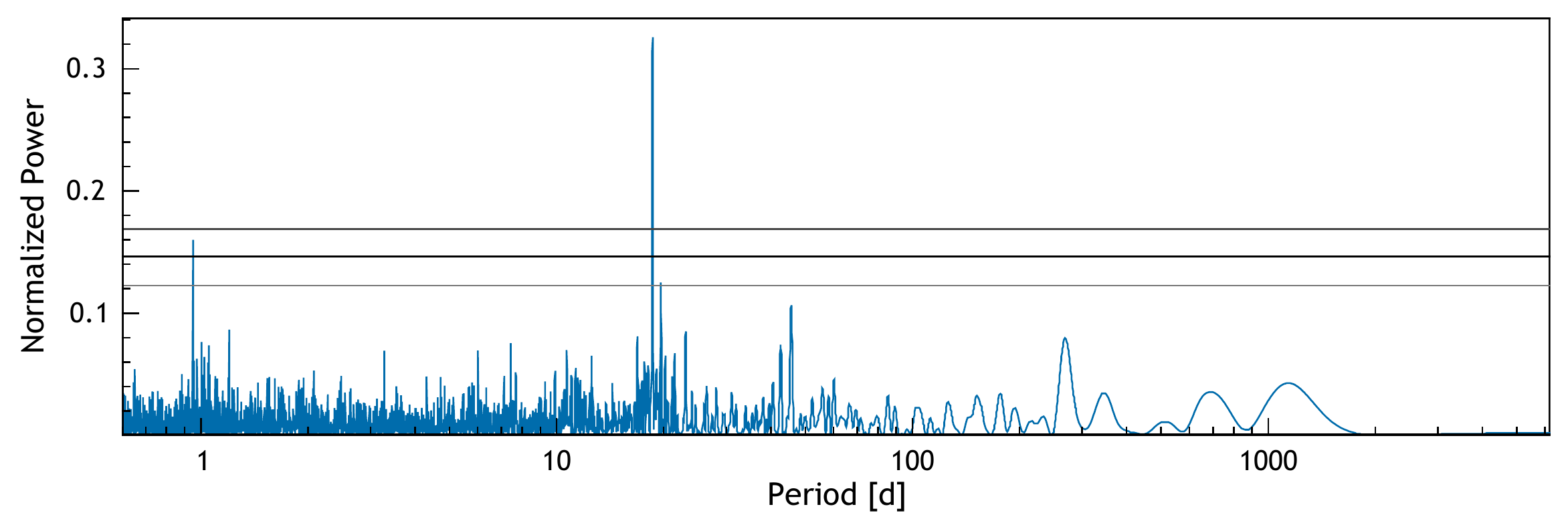}
            \put (28,29) {\tiny $18.56$~d $\displaystyle\rightarrow$ }
            \put (73,27) {\parbox{3cm} {\tiny After lin \rhk \\ subtraction}}
        \end{overpic}\\
        \vspace{0.1cm}
        \begin{overpic}[width=0.49\textwidth]{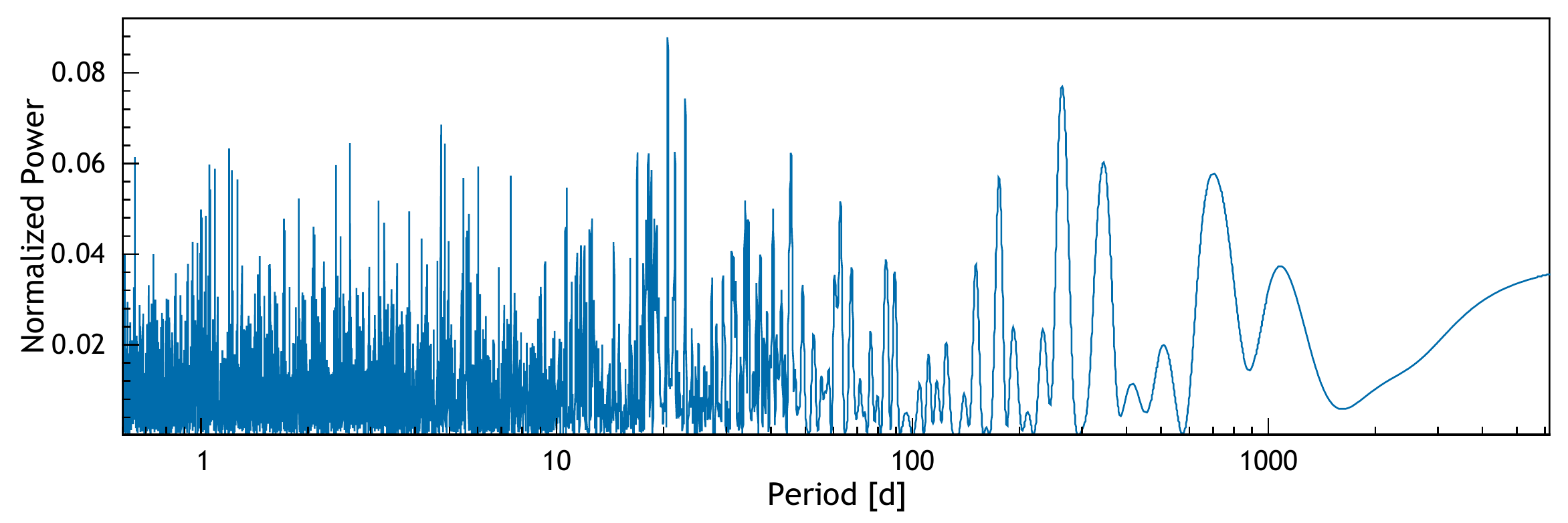}
            \put (64,27) {\colorbox{white}{\tiny After lin \rhk\ and}}
            \put (64,22.4) {\colorbox{white}{\tiny 18.56 d signal subtraction}}
        \end{overpic}
    \end{multicols}
\end{minipage}
\end{figure*}
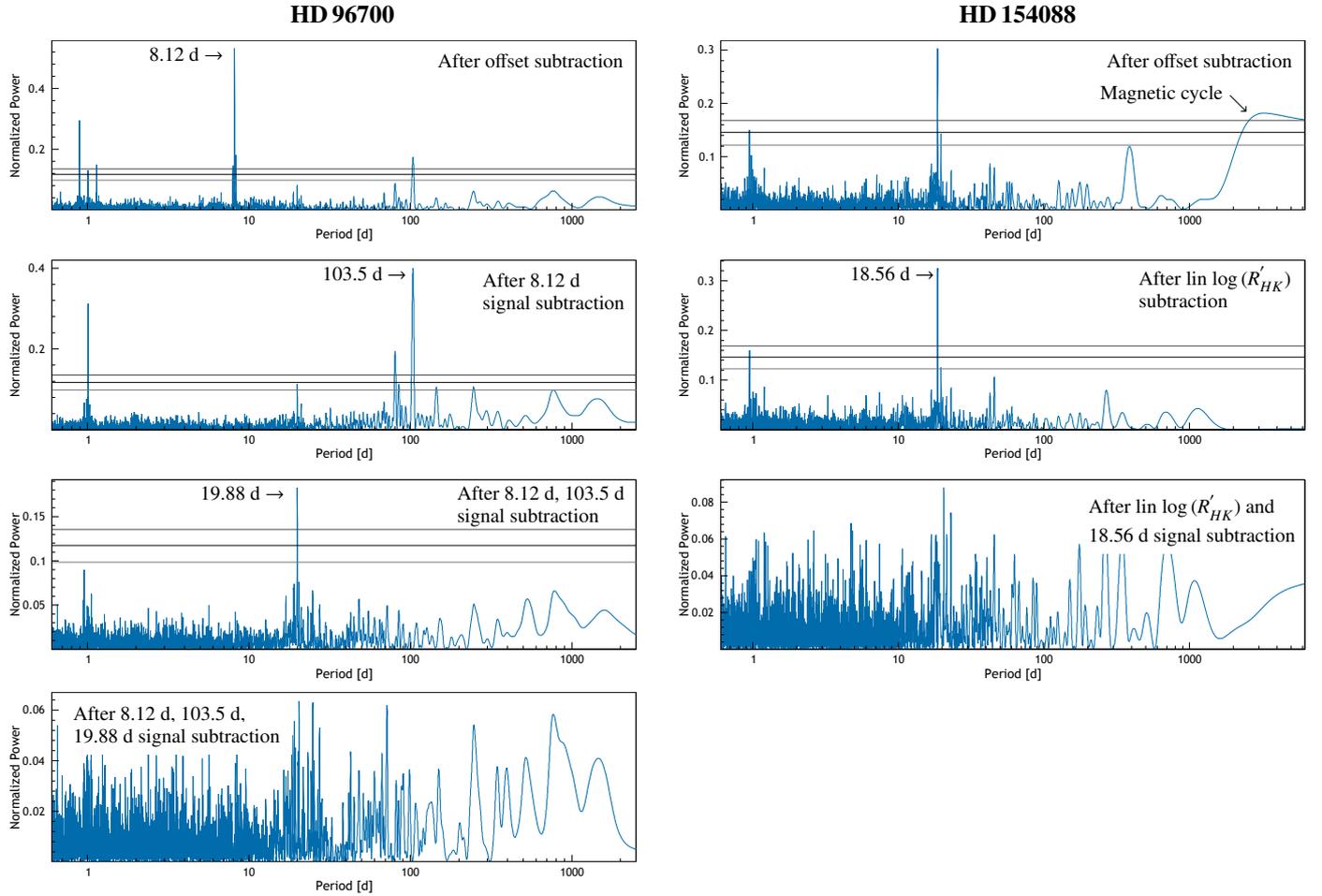

\newpage

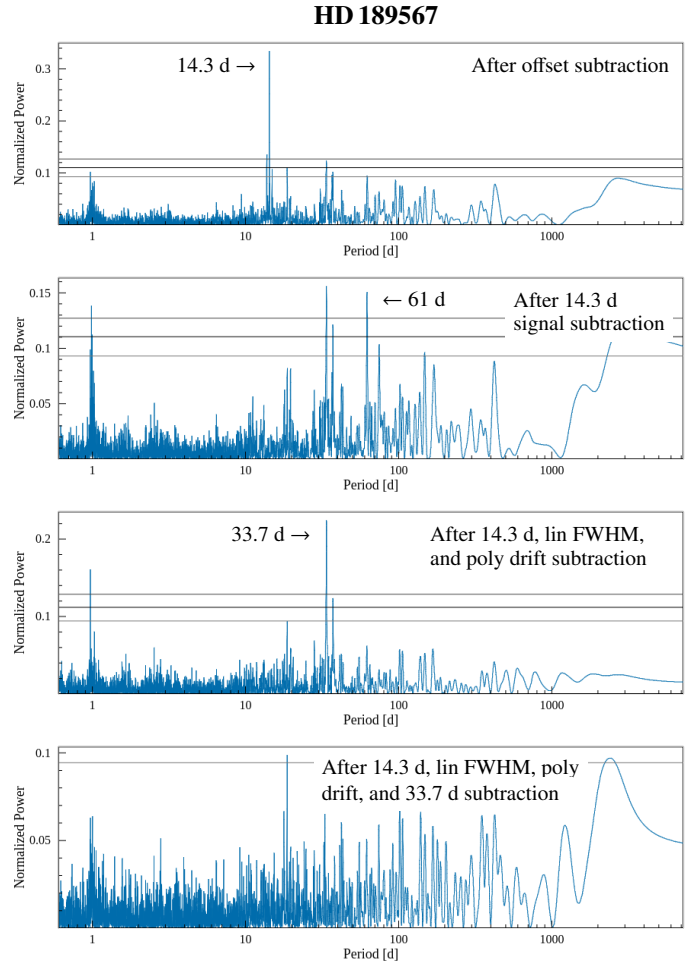
\begin{figure*}[] 
\centering
\caption[]{\tabular[t]{p{0.93\textwidth}lp{0.95\textwidth}lp{0.95\textwidth}}Periodogram of the RV residuals of HD189567, after sequentially removing, from top to bottom, the instrumental RV offsets, the 14.3 d Keplerian signal, a linear detrending with FWHM and polynomial drift, and lastly the 33.7 d Keplerian signal. \\ False alarm probability (FAP) thresholds are shown as horizontal lines for FAP=10$\%$, 1$\%$ and 0.1$\%$. \endtabular}

\label{fig:HD189567-FAP}%
\begin{minipage}{\textwidth}%
    \centering
    \vspace{0.6cm}
        \begin{overpic}[width=0.49\textwidth]{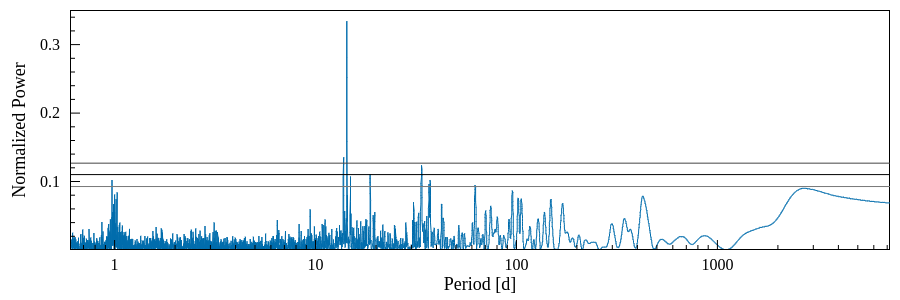}
            \put (45,35) {\bf HD\,189567 }
            \put (25,28) {\tiny $14.3$~d $\displaystyle\rightarrow$  }
            \put (68,28) {\tiny After offset subtraction}
        \end{overpic}\\
        \vspace{0.1cm}
        \begin{overpic}[width=0.49\textwidth]{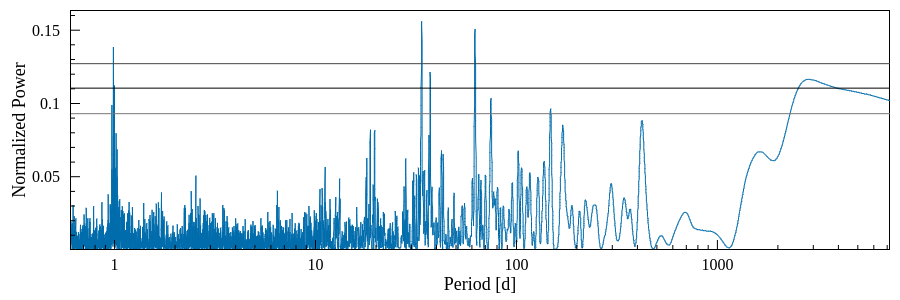}
            \put (55,28) {\tiny $\displaystyle\leftarrow$ $61$~d  }
            \put (73,28) {\parbox{3cm} {\colorbox{white}{\tiny After 14.3 d}}}
            \put (73,24.3) {\parbox{3cm} {\colorbox{white}{\tiny signal subtraction}}}
        \end{overpic}\\
        \vspace{0.1cm}
        \begin{overpic}[width=0.49\textwidth]{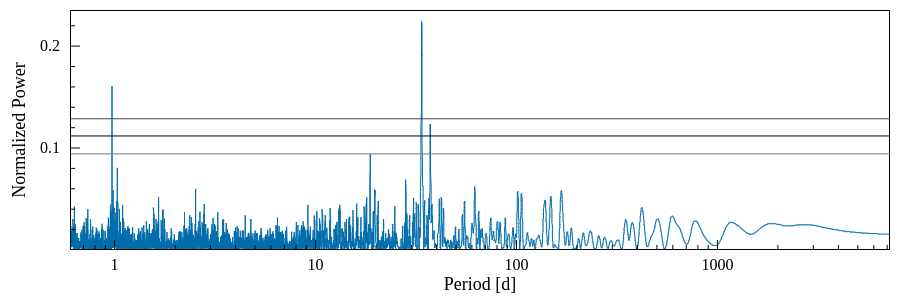}
            \put (32,28) { \tiny $33.7$~d $\displaystyle\rightarrow$ }
            \put (62,26) {\parbox{4cm} {\tiny After 14.3 d, lin FWHM, \\ and poly drift subtraction}}
        \end{overpic} \\
        \vspace{0.1cm}
        \begin{overpic}[width=0.49\textwidth]{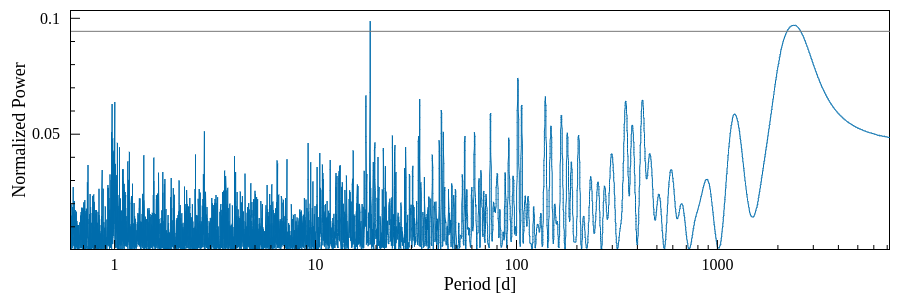}
            \put (45,28) {\parbox{3cm} {\colorbox{white}{\tiny After 14.3 d, lin FWHM, poly}}}
            \put (45,24.3) {\parbox{3cm} {\colorbox{white}{\tiny drift, and 33.7 d subtraction }}}
        \end{overpic}
\end{minipage}
\end{figure*}

\end{appendix}

\end{document}